\documentclass[
 aps,
 prc,
 amsmath,
 amssymb,
 superscriptaddress,
 reprint,
]{revtex4-2}

\usepackage{graphicx}
\usepackage{dcolumn}
\usepackage{bm}
\usepackage[mathlines]{lineno}
\usepackage[version=4]{mhchem}
\usepackage{hyperref}
\hypersetup{
  colorlinks = true,   
  citecolor  = blue,   
  linkcolor  = blue,  
  urlcolor   = blue    
}

\usepackage[utf8]{inputenc}
\usepackage[T1]{fontenc}

\begin{document}

\preprint{APS/123-QED}

\title{Production of Iodine Isotopes via Ultra-intense Laser Driven Photonuclear Reactions}

\author{Weifu Yin}
\affiliation{Shanghai Institute of Applied Physics, Chinese Academy of Sciences, Shanghai 201800, China}
\affiliation{University of Chinese Academy of Sciences, Beijing 100049, China}

\author{Tongjun Xu}\email{tjxu@siom.ac.cn}
\affiliation{State Key Laboratory of High Field Laser Physics and CAS Center for Excellence in Ultra-intense Laser Science, Shanghai Institute of Optics and Fine Mechanics (SIOM), Chinese Academy of Sciences (CAS), Shanghai 201800, China}

\author{Guoqiang Zhang}\email{zhangguoqiang@sari.ac.cn}
\affiliation{Shanghai Advanced Research Institute, Chinese Academy of Sciences, Shanghai 201210, China}
\affiliation{Shanghai Institute of Applied Physics, Chinese Academy of Sciences, Shanghai 201800, China}

\author{Putong Wang}
\affiliation{Shanghai Institute of Applied Physics, Chinese Academy of Sciences, Shanghai 201800, China}
\affiliation{University of Chinese Academy of Sciences, Beijing 100049, China}

\author{Yihang Zhang}
\author{Yufeng Dong}
\affiliation{Beijing National Laboratory for Condensed Matter Physics, Institute of Physics, Chinese Academy of Sciences, Beijing 100190, China}

\author{Xiangai Deng}
\author{Youjing Wang}
\author{Zhiguo Ma}
\author{Changbo Fu}
\author{Kai Zhao}
\affiliation{Key Laboratory of Nuclear Physics and Ion-Beam Application (MOE), Institute of Modern Physics, Fudan University, Shanghai 200433, China}
\author{Fenghua Qiao}
\affiliation{School of Physical Science and Technology, ShanghaiTech University, Shanghai 201210, China}

\author{Lulin Fan}
\author{Yingzi Dai}
\author{Bowen Zhang}
\author{Hui Zhang}
\author{Chenyu Qin}
\author{Dirui Xu}
\author{Jing Wang}
\affiliation{State Key Laboratory of High Field Laser Physics and CAS Center for Excellence in Ultra-intense Laser Science, Shanghai Institute of Optics and Fine Mechanics (SIOM), Chinese Academy of Sciences (CAS), Shanghai 201800, China}

\author{Jishao Xu}
\affiliation{Key Laboratory of Nuclear Physics and Ion-Beam Application (MOE), Institute of Modern Physics, Fudan University, Shanghai 200433, China}
\author{Wanqing Su}
\affiliation{Shanghai Advanced Research Institute, Chinese Academy of Sciences, Shanghai 201210, China}
\affiliation{College of Physics, Henan Normal University, Xinxiang 453007, China}
\affiliation{Shanghai Institute of Applied Physics, Chinese Academy of Sciences, Shanghai 201800, China}

\author{Lianghong Yu}
\author{Xiaoyan Liang}
\author{Liangliang Ji}\email{jill@siom.ac.cn}
\affiliation{State Key Laboratory of High Field Laser Physics and CAS Center for Excellence in Ultra-intense Laser Science, Shanghai Institute of Optics and Fine Mechanics (SIOM), Chinese Academy of Sciences (CAS), Shanghai 201800, China}

\author{Ruxin Li}
\affiliation{State Key Laboratory of High Field Laser Physics and CAS Center for Excellence in Ultra-intense Laser Science, Shanghai Institute of Optics and Fine Mechanics (SIOM), Chinese Academy of Sciences (CAS), Shanghai 201800, China}

\author{Yu-Gang Ma}\email{mayugang@fudan.edu.cn}
\affiliation{Key Laboratory of Nuclear Physics and Ion-Beam Application (MOE), Institute of Modern Physics, Fudan University, Shanghai 200433, China}
\affiliation{Shanghai Institute of Applied Physics, Chinese Academy of Sciences, Shanghai 201800, China}

\date{\today}
\begin{abstract}

The investigation and production of proton-rich iodine isotopes predominantly rely on conventional accelerator-based methods, typically requiring prolonged irradiation periods to measure or achieve quantifiable yields for isotopic isolation. Bremsstrahlung radiation sources generated by high-power laser-plasma-accelerated electron beams with ultrahigh charge (tens of nanocoulombs) bombarding high-Z targets demonstrate extraordinary photon flux characteristics. An electron beam with a total charge of approximately 47.7 nC (E$_e$ \textgreater 10.4 MeV) was generated in our experiment by focusing a ultra-intense laser pulse onto a deuterium gas jet. Laser-driven bremsstrahlung was employed to induce 
\ce{^{127}I}$(\gamma,xn)$ ($x$ = 1,3,4,6-8), and the product yields and the corresponding flux-weighted average cross sections are reported. Our results demonstrate production of medical isotopes, with average yields of \ce{^{124}I} and \ce{^{123}I} at approximately $9.83\pm0.45\times10^{5}$/shot and $2.81\pm0.11\times10^{5}$/shot, respectively. This method, utilizing high-power lasers to generate bremsstrahlung radiation, shows significant potential for medical applications and opens new avenues for studying photonuclear processes in astrophysical contexts.
\end{abstract}
\keywords{high-power laser, Laser-driven multi-MeV bremsstrahlung, \ce{^{127}I}$(\gamma,xn)(x=1-8)$, medical iodine isotopes.}

\maketitle

\section{INTRODUCTION}

Due to the development of chirped pulse amplification(CPA) and optical parametric CPA (OPCPA) technology~\cite{b23,b64,b24}, ultra-intense ultra-fast lasers can reach a focal intensity of $10^{22}$ W/cm${^2}$~\cite{b24,b63}, and their interaction with the plasma produces an accelerating field with a large acceleration gradient (hundreds of GV m$^{-1}$) and a small spatial scale (a length scale of centimeters)~\cite{b97}, which in turn drive the acceleration of electrons and produce electron beams with a high current intensity (hundreds of kiloamperes)~\cite{b25,b28}. New options for laser-accelerated electrons~\cite{b96} and various application for these high-current electron sources~\cite{b59,b92,b98} are being actively explored. For example, by adjusting the density of a clustered argon gas target, Shun Li \textit{et al}. used a femtosecond laser interacting with a near-critical-density plasma to obtain an electron beam of 15 nC(E$_e$ \textgreater 1.2 MeV) and produced bremsstrahlung with the photon-flux of $ 7 \times 10^{23}$ s$^{-1}$ by bombardment of a copper target~\cite{b26}. Certainly, the superposition of bremsstrahlung spectra, each with distinct endpoint energies, closely resembles the blackbody Planck spectrum at a specific temperature~\cite{b13}. This resemblance is particularly relevant in the context of investigating explosive nucleosynthesis phenomena in novae and supernovae~\cite{b31,b2}, including the $\gamma$-process and the $\nu$-process. Therefore, the ultra-intense laser-driven bremsstrahlung radiation serves as an ideal $\gamma$-ray source for replicating explosive nucleosynthesis processes in controlled laboratory environments. Furthermore, due to the ultrafast and high fluence nature of laser-driven bremsstrahlung, many researchers are applying it to diverse fields, including the study of photonuclear reaction cross sections~\cite{b27}, excitation of nuclear isomers~\cite{b28,b56}, transmutation of nuclear waste~\cite{b30,b58}, ultrashort pulsed photoeutron sources~\cite{b65,b66}, isotope production~\cite{b29,b57,b89}, and many other fields of application with ultra-high efficiencies.

Isotopes of iodine play a very important role in nuclear astrophysics~\cite{b60,b61}, environmental protection~\cite{b62}, medical diagnostics~\cite{b18,b19,b48,b53,b55} and cancer therapy~\cite{b20}. Although \ce{^{131}I} (iodine-131) pioneered clinical applications in thyroid disorder diagnosis and cancer theranostics~\cite{b86}, its utility is constrained by elevated radiation doses and suboptimal imaging characteristics due to predominant beta emission. The proton-rich isotopes \ce{^{120-125}I} are predominantly utilized in diagnostic imaging techniques, such as single photon emission computed tomography (SPECT) and positron emission tomography (PET)~\cite{b18,b53,b55}, within the realm of nuclear medicine diagnostics~\cite{b19,b48}. Specifically, \ce{^{123}I} is regarded as the most appropriate isotope of iodine for the diagnostic evaluation of thyroid disorders~\cite{b21,b51}, owing to its favorable half-life, low dose rate, and superior imaging resolution. 

The proton-rich isotopes do not necessitate a reactor-based production method. After the advent of accelerators, researchers conducted many theoretical and experimental studies to produce \ce{^{122-126}I} through charged particle-induced nuclear reactions~\cite{b32,b33,b49,b50,b52,b54}, as well as applying the bremsstrahlung photons produced by the electron beam to produce \ce{^{123}I} for a wide range of applications in nuclear medicine~\cite{b34}. For the quantitative prediction of proton-rich iodine isotopes for nuclear medicine applications production using bremsstrahlung photon beams, understanding the flux-weighted average photoneutron reaction cross-section of \ce{^{127}I} is of critical importance. Therefore, Naik \textit{et al}. investigated the flux-weighted average cross section of \ce{^{127}I}$(\gamma,xn;x=1-6)$ using bremsstrahlung radiation generated by electron linac accelerator with 50 MeV and 70 MeV endpoint energies~\cite{b10}. For iodine isotopes produced via photonuclear reactions, those located farther from the stability line require a greater photon flux above the threshold energy for experimental detection due to their greater total neutron separation energies. This explains why the average cross sections measurement of the \ce{^{127}I}$(\gamma,xn; x=6-9)$ reaction with the highest multiplicity of photoneutrons was achieved in the bremsstrahlung with an endpoint energy of 900 MeV experiment conducted by Jonsson and Lindgren (1969) at the 1.2 GeV electron synchrotron~\cite{b22}. Both large-scale and commercial electron accelerators exhibit accelerating gradients constrained to approximately 100 MV/m~\cite{b79} due to material limitations in radio-frequency accelerating cavities. This inherent limitation results in bremsstrahlung sources with insufficient photon flux to achieve statistically significant measurements within practical experimental durations\cite{b91}, while also presenting suboptimal irradiation efficiency for industrial applications. The utilization of high-flux above-threshold bremsstrahlung generated by ultra-intense and ultra-short laser pulses represents a novel experimental methodology for investigating photonuclear reactions and a revolutionary efficient tool for industrial irradiation processes.

In the present work, we systematically optimized the electron beam parameters and beam-targeting scheme to produce a series of proton-rich iodine isotopes. This paper is structured as follows: Section II details the experimental methodology and the simulation, including (A) laser and target parameters of experiment, (B) energy and absolute detection efficiency calibration of a high-purity germanium (HPGe) $\gamma$-spectrometry and (C) Fluent and PIC-simulation setup. Section III presents (A) optimized electron parameters and their resulting bremsstrahlung versus the energy window, (B) analysis of $\gamma$ spectrum, calculation of nuclide yields and identification of nuclear reaction channels, and (C) is the yield of medical iodine isotopes (e.g., \ce{^{124}I} and \ce{^{123}I}). Section IV summarizes the experiments and looks at some applications of laser-driven bremsstrahlung sources.

\section{Experiment and Simulation}
\subsection{Experimental setup}
\begin{figure*}
\includegraphics[scale=0.5]{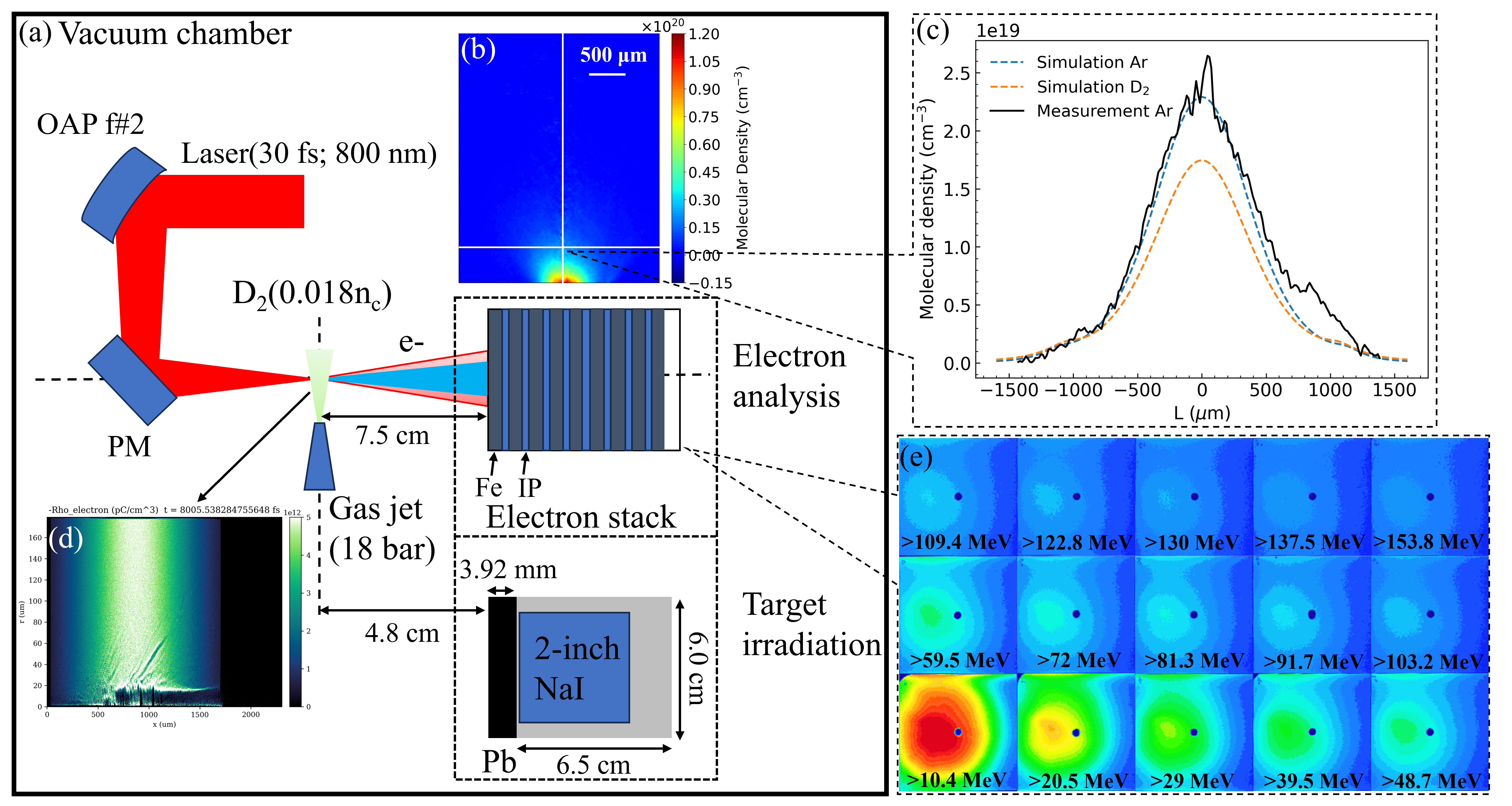}
\caption{\label{fig1}(a) Simple schematic of the layout of the experimental setup. (b) Ar density profile at 20 bar back pressure. (c) Comparison of simulated density profiles for Argon (\ce{Ar}) and Deuterium (\ce{D_{2}}) with the measured Ar profile at the same backing pressure, illustrating the gas density used in the experiment. (d) Electron density distribution during particle-in-cell (PIC) simulation. (e) Raw signals of the laser-accelerated electron beam recorded in the IPs.}
\end{figure*}
The experiment was conducted on the 10 PW laser beamline at the Shanghai Superintense Ultrafast Laser Facility (SULF), using a single-shot operational mode with a repetition interval of approximately 15 minutes between shots. Prior to the compressor stage, the laser energy was approximately 120 J, which was reduced to about 55 J upon reaching the target, attributed to an energy transmission efficiency of approximately 46\% from the compressor to the target ~\cite{b35}. The focal spot achieved a full width at half maximum (FWHM) of 2.1 $\mu$m. The fractional laser energy within the focus spot was approximately 16\% at FWHM, resulting in a peak intensity of 5.8 $\times$ 10$^{21}$ W/cm$^{2}$ and a corresponding normalized amplitude of a$_{0} \approx$ 52. Furthermore, the central wavelength of the laser was 800 nm, with a pulse duration of approximately 30 fs (FWHM). Fig.~\ref{fig1}(a) presents a schematic of the experimental setup. The laser pulses were focused by an f/2 off-axis parabolic (OAP) mirror and a plasma mirror(PM) onto a deuterium(\ce{D_{2}}) gas ejected from a nozzle with adjustable back pressure, which featured a Gaussian density distribution. Fig.~\ref{fig1}(b) was argon(\ce{Ar}) density profile measured at 20 bar back pressure. The preliminary energy spectrum, charge, and divergence angle of the electrons were characterized using a stack positioned approximately 7.5 cm from the nozzle's center along the laser's optical axis. The stack measured 50 mm $\times$ 50 mm, and iron sheets of varying thicknesses were inserted into different image plates (IPs) to attenuate the electron energy. The electronic parameters were ascertained using the electron stacks capable of the cutoff energy of 150 MeV and 200 MeV. Raw signals of the laser-accelerated electron beam are presented in Fig.~\ref{fig1}(e). Subsequently, a lead disk measured 51 mm $\times$ 51 mm with a thickness of 3.92 mm (as measured by a micrometer) was positioned approximately 4.8 cm from the nozzle's center along the laser's optical axis, serving as a converter for the bremsstrahlung radiation.

To achieve a sufficiently intense bremsstrahlung radiation, a 2-inch cylindrical sodium iodide (\ce{NaI(Tl)}) crystal with a diameter of 5.08 cm and a thickness of 5.08 cm, positioned immediately adjacent to the bremsstrahlung converter, was irradiated under the optimal electronic parameters. To maintain the stability of the laser parameters, the interval between each laser pulse was set to approximately 15 minutes. Owing to the constraints imposed by the plasma mirrors, it was necessary to de-vacuum the target chamber and replace the plasma mirrors after every 5 shots. Upon re-establishing vacuum conditions within the target chamber, it was essential to reconfirm the laser focal spot and other critical parameters. Consequently, the interval between the last five laser pulses and the initial five pulses spanned approximately 3 hours, with each pulse being separated by about 15 minutes.

\subsection{HPGe detector calibrations}

Following a 10-minute deflation of the target chamber, the irradiated \ce{NaI(Tl)} crystals were extracted. Subsequently, the $\gamma$-rays emitted from the radioactive decay of the photodisintegration products of the \ce{^{127}I}$(\gamma,\,xn;\,x\,=\,1,3,4,6-8)$ reaction, including \ce{^{126}I}, were detected using a high-purity germanium (HPGe) detector. This HPGe detector is a member of the N-type coaxial GMX series (model number GMX50P4-83) from ORTEC, featuring a carbon fiber endcap above the Beryllium window designed to minimize background noise~\cite{b39}. This specific model boasts a relative detection efficiency of 50\% and an endcap diameter of 83 mm. The relative photopeak efficiency is defined as the efficiency at the 1.33 MeV photopeak of a Co-60 point source positioned 25 cm from the surface of a standard $\Phi$3 inch $\times$ 3 inch cylindrical \ce{NaI(Tl)} detector~\cite{b40}. The output signal from the HPGe detector is captured by the fully integrated $\gamma$ spectrometer DSPEC 50~\cite{b41} and the computer software GammaVision~\cite{b42}, both developed by ORTEC. In list mode, both the signal size and timestamps are logged, enabling the acquisition of the $\gamma$-ray energy spectrum of the sample at any given time during measurement. This capability allows for the subsequent calculation of the half-life corresponding to a single full energy peak.

\begin{figure}
\includegraphics[scale=0.41]{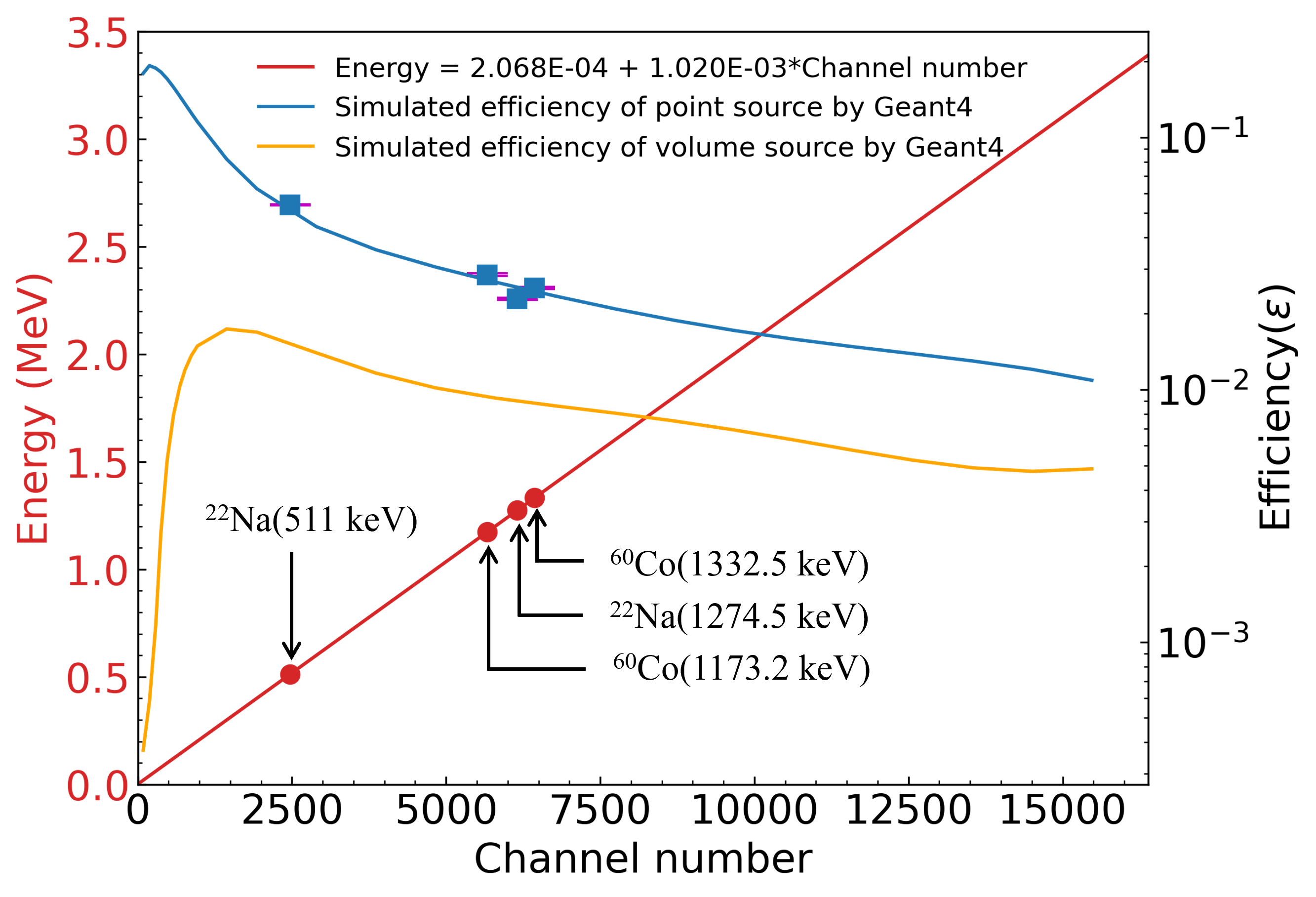}
\caption{\label{fig2}HPGe detector energy calibration and absolute detection efficiency calibration for point and volumetric sources.}
\end{figure}

The energy linearity of the entire detection system, calibrated with Co-60 and Na-22 standard sources, along with the absolute detection efficiency of the full energy peaks combined with Geant4~\cite{b80,b37,b38} simulations, is illustrated in Fig.~\ref{fig2}, where the volume source is defined based on Geant4~\cite{b37,b38} simulations of experimental beam-target conditions to obtain the distribution of \ce{^{126}I} in \ce{NaI(Tl)} crystals. The absolute efficiency curves of the HPGe detector for point and volumetric sources were obtained by fitting the absolute detection efficiencies obtained at different energy points of the Geant4 simulation by a polynomial function:
\begin{equation}
\epsilon(E_\gamma) = \sum_{i=0}^{8}p_{i}\cdot log_{10}^{i}(E_\gamma) 
\label{eq1}
\end{equation}
where $p_{i}$ were free parameters to be determined. Additionally, the energy resolution is characterized by a full width at half maximum (FWHM) of 2.02 keV at the \ce{^{60}Co} peak of 1332.5 keV. Details of the HPGe detector and Geant4~\cite{b37,b38} simulations are provided in the Supplementary Material\cite{b99}.

\subsection{Simulated setup}

Given that the initial gas density measurements were obtained using argon, we conducted computational fluid dynamics (CFD) simulations using the commercial software ANSYS FLUENT~\cite{b84}, which employs numerical discretization techniques to solve the governing Navier-Stokes equations on a discretized computational mesh~\cite{b85,b90}, to establish the corrected gas density for deuterium in our experimental system. The density distributions of both argon and deuterium under 20 bar inlet pressure conditions were numerically resolved using a density-based solver implemented on a 2D axisymmetric computational mesh with 5 $\mu$m characteristic cell size.

A particle-in-cell (PIC) simulations were performed to find suitable electron peak density for experiment using the Smilei~\cite{b81} code in cylindrical geometry with azimuthal mode decomposition~\cite{b82}, incorporating a laser envelope model~\cite{b83}. The computational grids are set to $L_{x}$=2304 $\mu$m $\times$ $L_{r}$=179.2 $\mu$m, with corresponding spatial resolutions of 11520 cells along the x direction and 448 cells along the r direction. Per-cell initialization employed 16 macro-particles (8 electron + 8 D$^{+}$ ions) possessing matching Gaussian density distributions centered at x$_{0}$ = 870 $\mu$m with FWHM = 946 $\mu$m. A y-polarized Gaussian beam ($\lambda_{L}$ = 800 nm) propagates in +x direction, focused at x$_{f}$ = x$_{0}$ - 100 $\mu$m. The laser pulse features a Gaussian transverse profile with waist $\omega_{0}$ = 1.8 $\mu$m at the focal spot and a longitudinal temporal envelope of 30 fs FWHM, corresponding to a normalized vector potential of a$_{0}$ = 52. A plane screen diagnostic was positioned at x$_{\mathrm{screen}}$= $L_{x}$ - 590 $\mu$m to detect electrons traversing the plane along the positive x-direction. Fig.~\ref{fig1}(d) is the electron density distribution during PIC simulation.

\section{RESULT AND DISCUSSION}
\subsection{Laser driven bremsstrahlung}

The molecular density of deuterium at the target position (500 $\mu$m downstream from the nozzle) demonstrates a reduction compared to argon under 20 bar back pressure conditions, attributable to the combined effects of its lower molecular weight(4.032 g/mol for \ce{D_{2}}, 39.948 g/mol for \ce{Ar}) and distinct transport properties including heat capacity, thermal conductivity, and viscosity\cite{b93}. The argon molecular density profile simulated by ANSYS Fluent under 20 bar back pressure demonstrates strong congruence with experimental measurements and the difference in density between argon and deuterium gases under the same back pressure condition at the targeting position, as show in Fig.~\ref{fig1}(c).

\begin{figure}
\includegraphics[scale=0.42]{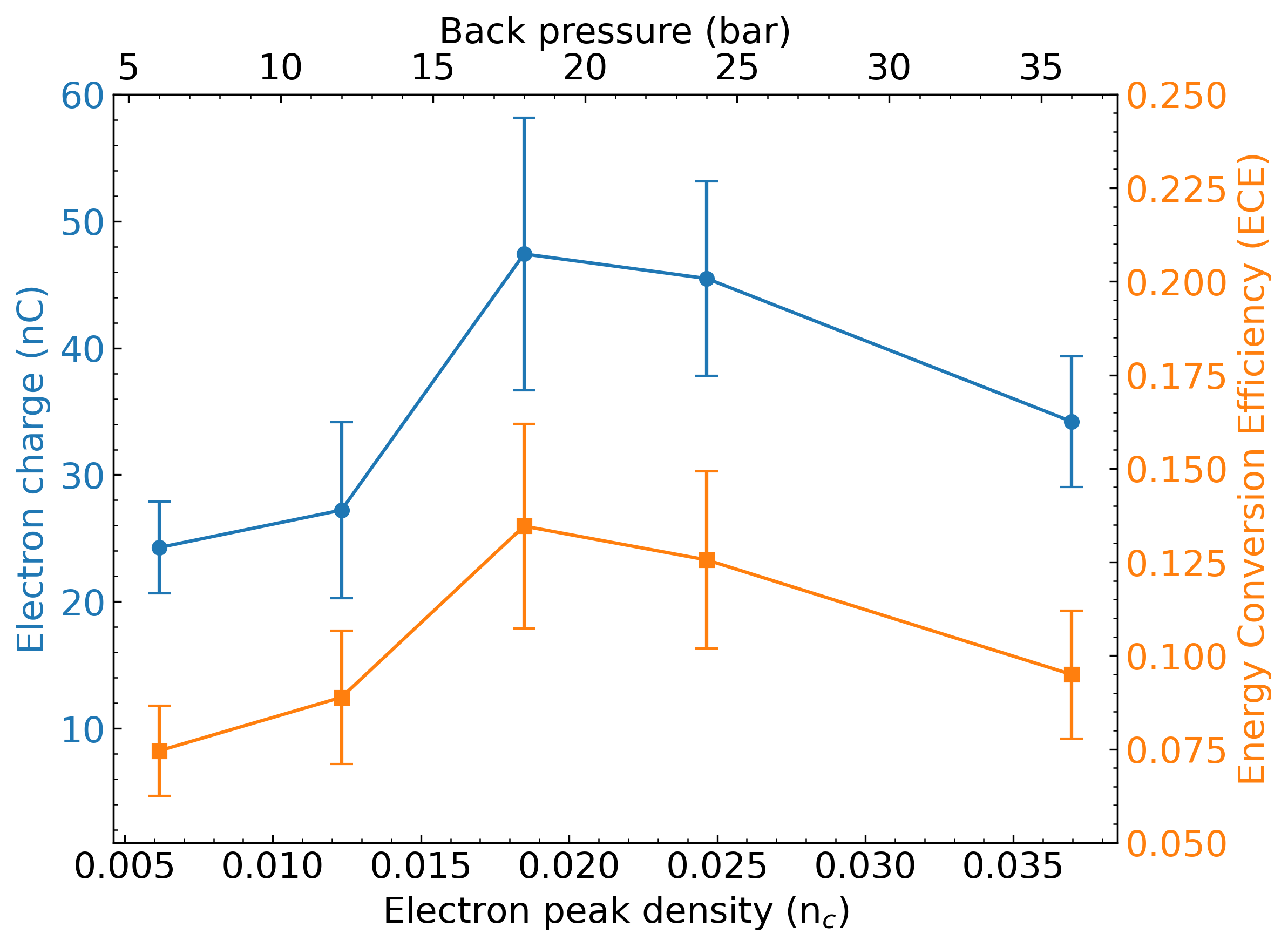}
\caption{\label{fig3}Blue symbols and lines denote the total charge (in nC) measured experimentally across varying peak electron densities. Similarly, yellow symbols and lines indicate the energy conversion efficiency from laser energy to total electron energy ($E_{e}\geq$ 10 MeV) measured experimentally. Vertical error bars for experimental data reflect measurement uncertainty in the electron stack diagnostics.}
\end{figure}

Prior to the electron irradiation of the bremsstrahlung converter, the optimal electron parameters, specifically the charge and energy, were determined by adjusting the gas pressure of the deuterium gas ejected from the nozzle. Our PIC simulations demonstrate that the optimal electron peak density matching the laser parameters lies within 0.012$n_{c}$ to 0.023$n_{c}$ in where the $n_{c}=\frac{\epsilon_{0}m_{e}\omega^{2}}{e^{2}}$, namely, $\sim1.7 \times 10^{21}$ cm$^{-3}$ for 800 nm laser wavelength. Since the simulations do not exactly match the experimental conditions, experimental scanning of the electron density parameters is required to determine the optimal targeting conditions. Experimental verification through analysis of electron spectrometer measurements reveals an optimal electron peak density of 0.018$n_{c}$ for laser-plasma electron acceleration, which falls within the simulated range. Owing to the relativistic self-focusing power P$_{f}$=P$_{0}(n_{c}/n_{e})$(where the natural power P$_{0}$=8.7 GW), the Gaussian laser beam undergoes strong self-focusing within the plasma under our experimental conditions. As illustrated in Fig~\ref{fig1}(d), PIC simulations reveal that the laser ponderomotively expels plasma electrons from the optical axis, forming a plasma channel. The interaction between the laser field and the nonlinear focusing forces within this channel leads to electron acceleration~\cite{b94}. Since the laser power significantly exceeds the relativistic self-focusing threshold, the laser beam splits into multiple sub-beams that focus independently, resulting in a multi-filament structure~\cite{b95}. The experimental results are presented in Fig.~\ref{fig3}. The experimental uncertainties originate from four primary sources: statistical fluctuations in the original photo-stimulated luminescence (PSL) measurements, inaccuracies in the saturation correction factor, errors associated with the temporal decay factor(7$\%$~\cite{b88}), and uncertainties in the electron number conversion process from PSL values(20$\%$~\cite{b88}). Quantitative analysis of electron spectrometer data (Fig.~\ref{fig1}(e)) indicates that laser-plasma acceleration at 0.018$n_{c}$ achieves: (1) a total electron charge of 47.4 nC(E$_e$ \textgreater 10.4 MeV), (2) an energy conversion efficiency of 13.5$\%$, and (3) transverse divergence angles of $17.95^\circ$  (x-direction) and $18.33^\circ$  (y-direction), representing optimal beam quality within the tested parameter space. The experimentally measured electron energy spectra at other back pressures are provided in the Supplementary Material\cite{b99}.

\begin{figure}
\includegraphics[scale=0.43]{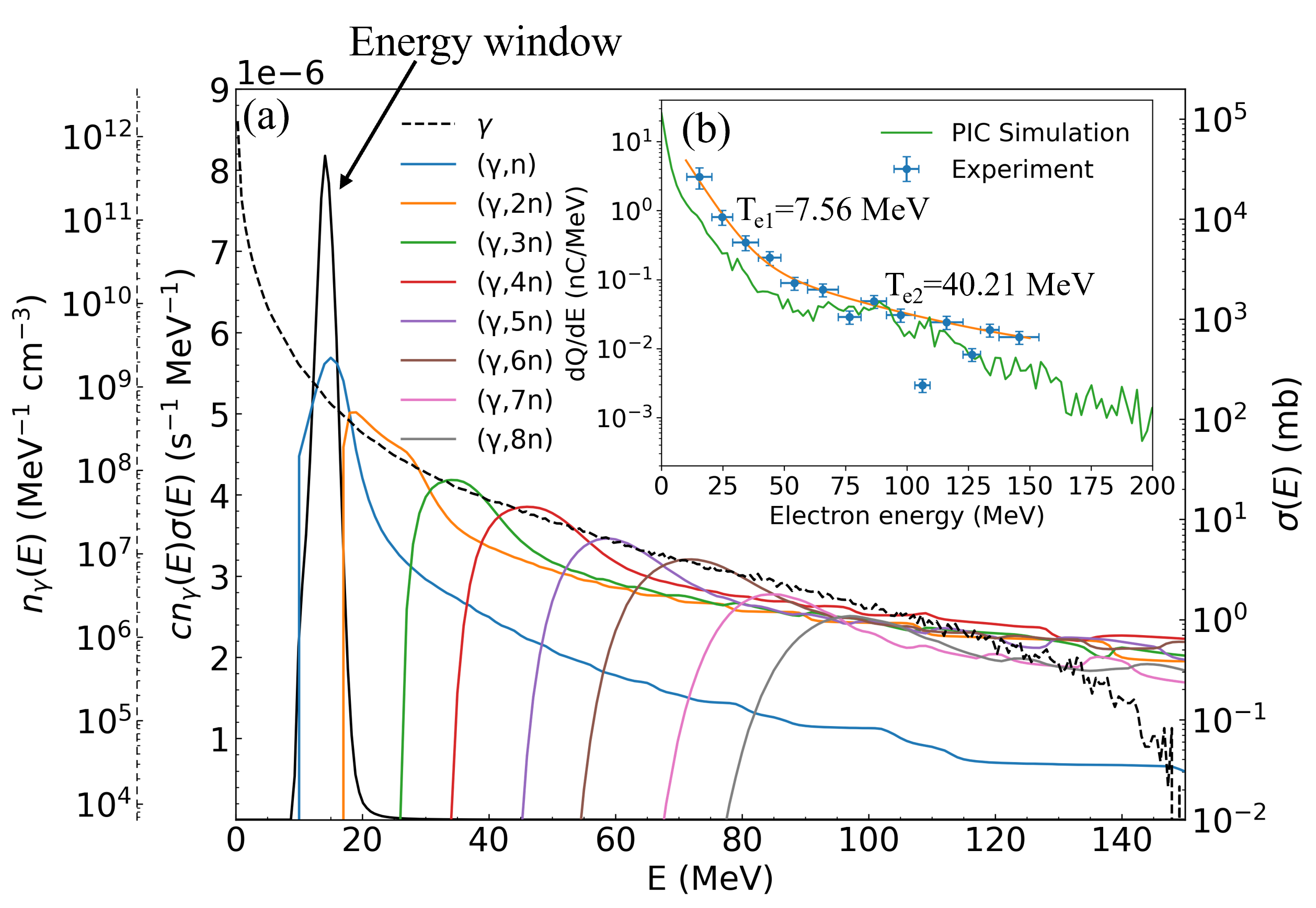}
\caption{\label{fig4}(a)Bremsstrahlung spectrum (Geant4~\cite{b37,b38}), \ce{^{127}I}$(\gamma,xn)$ cross sections (TALYS 2.0\cite{b36}), and energy window for \ce{^{127}I(\gamma,n)^{126}I}.The curve labeled 'Energy window' represents the product of the photon flux and the reaction cross section, highlighting the most effective energy range for the photonuclear reaction, analogous to the Gamow window in charged-particle-induced reactions. (b) Electron energy spectra obtained from electron stack measurements (blue data points) were fitted using a dual-temperature model ($T_{e1}$ and $T_{e2}$). Horizontal error bars indicate energy bin intervals, while vertical error bars represent measurement uncertainties in the electron stack. The green curve corresponds to the simulated energy spectrum derived from Particle-in-Cell (PIC) simulations.}
\end{figure}

The Gaussian laser beam, focused by a low f-number OAP mirror, achieves a high peak power density. However, its Rayleigh length $x_R = \pi \omega_0^{2} / \lambda_{L}$ = 12.7 $\mu$m (for a beam waist $\omega_{0}$ = 1.8 $\mu$m) indicates that plasma waves is not the dominant acceleration mechanism. As shown in Fig.~\ref{fig1}(c), the electron density exhibits a Gaussian profile. During laser propagation in low-density plasma, the laser pulse excites plasma waves, but electron injection is suppressed because the laser spot size $\omega(x)=\omega_{0}\sqrt{1+(\frac{x-x_{0}}{x_{R}})^{2}}$ significantly exceeds the plasma wavelength $\lambda_{p}(\mu m)\approx3.3\times10^{10}/\sqrt{n_{e}(cm^{-3})}$. At higher electron densities, intense laser self-focusing occurs prior to the geometric focus. The ponderomotive force of laser field that expels background electrons transversely, forming a plasma channel. Within this channel, the laser's transverse electric field accelerates electron and the laser's magnetic field collimates it along the propagation axis. Consequently, the electron energy spectra exhibit a broad energy spread characteristic of laser-plasma interactions, as shown in both the PIC-simulated and experimental results in Fig.~\ref{fig4}(b) at the electron peak density of 0.018n$_{C}$. The experimental data are fitted with a bi-exponential function corresponding to a two-temperature distribution (T$_{e1}$ = 7.56 MeV; T$_{e2}$ = 40.21 MeV). The discrepancies between the experimental and simulated energy spectra primarily arise from the following factors: (1) the experimental setup lacks perfect cylindrical symmetry; (2) intense forward-directed X-rays and $\gamma$-rays generated during direct laser acceleration of electrons Compton-scatter off the iron foil stack, producing secondary electrons; and (3) strong laser self-focusing in the plasma introduces differences not fully captured by the envelope approximation used in the PIC simulations. Utilizing the aforementioned electron energy spectrum and the dimensions of the lead target, the bremsstrahlung radiation spectrum was simulated using Geant4~\cite{b37,b38}, with the results depicted in Fig.~\ref{fig4}(a).

Referencec~\cite{b87} investigated the nucleosynthesis of proton-rich heavy nuclei in stellar photon baths under astrophysical $\gamma$-process temperatures, where the photon bath was modeled by superimposing bremsstrahlung spectra with six endpoint energies. As illustrated in Fig.~\ref{fig4}(a), the ultra-intense femtosecond bremsstrahlung radiation generated in our laser-driven experiment originates from the superposition of bremsstrahlung produced by multi-energy electrons interacting with the lead target. This radiation spectrum at the low temperature end exhibits remarkable similarities to the $X$-ray/$\gamma$-ray spectral profiles observed in astrophysical photodisintegration reactions. It should also be noted that we have computed the $(\gamma,xn;x=1-8)$ nuclear reaction cross-section for a \ce{^{127}I} target utilizing the default parameters of the Talys 2.0~\cite{b36} code. By integrating Bremsstrahlung spectrum analysis with nuclear reaction cross-section calculations within a unified theoretical framework, our computational methodology reveals the emergence of a energy window for photon-induced nuclear reactions that is strikingly analogous to the classical Gamow peak observed in charged-particle-induced processes under equivalent astrophysical conditions. Such experimental realization establishes a valuable platform for investigating the formation mechanisms of proton-rich nuclei in nuclear astrophysical environments, particularly regarding the nucleosynthesis processes of p-nuclei. In our experimental setup, proton-rich iodine isotopes are synthesized through bremsstrahlung irradiation, demonstrating a novel approach to probe nucleosynthesis pathways in proton-rich nuclei. 

\subsection{Laser driven nuclear reaction}

\begin{table*}
\renewcommand{\arraystretch}{1.2} 
\caption{\label{tab1}Correlated decay data for radionuclides from sodium iodide activation reactions. The energies and branch ratios were given by NuDat 3.0 \cite{b43}}
\begin{ruledtabular}
\begin{tabular}{cccccccc}
 Nucleus & Threshold/MeV  & T$_{1/2}$ & $\gamma$ energy/keV & $\gamma$ branch & Detector efficiency $\epsilon_{v}$ & Net counts N$_{net}$ & Yields $\bar{Y}$\\
 \hline
 $^{128}$I & & 24.99 min & 442.901\footnotemark[1] & 12.61\%  &0.0163 &2125$\pm$76 &6.28$\pm$0.23 $\times$ 10$^{5}$\\
 $^{126}$I & 9.14 & 12.93 d & 666.331 & 32.9\% & & &\\
 &  &  & 753.819 & 4.15\% & \\
 &  &  & 1420.2 & 0.30386\% & \\
 &  &  & 388.633\footnotemark[1] & 35.6\% &0.0170 &26756$\pm$180 &1.03$\pm$0.0069 $\times$ 10$^{8}$ \\
 &  &  & 491.243 & 2.88\% & \\
 &  &  & 879.876 & 0.743\% & \\
 $^{124}$I & 25.84 & 4.176 d & 602.73\footnotemark[1] & 62.9\% &0.0139 &1127$\pm$52 &9.83$\pm$0.45 $\times$ 10$^{5}$\\
 &  &  & 1690.96 & 11.15\% & \\
 &  &  & 722.78 & 10.36\% & \\
 &  &  & 1509.36 & 3.25\% & \\
 &  &  & 1376.09 & 1.79\% & \\
 &  &  & 1325.52 & 1.578\% & \\
 $^{123}$I & 33.33 & 13.223 h & 159.00\footnotemark[1] & 83.60\% &0.0119 &2364$\pm$89 &2.81$\pm$0.11 $\times$ 10$^{5}$ \\
 $^{121}$I & 51.17 & 2.12 h & 212.20\footnotemark[1] & 84.3\% &0.0154 &617$\pm$55 &2.00$\pm$0.18 $\times$ 10$^{4}$ \\
 $^{120}$I & 61.75 & 81.6 min & 560.40\footnotemark[1] & 69.6\% &0.0146 &123$\pm$30 &4.87$\pm$1.19 $\times$ 10$^{3}$ \\
$^{119}$I & 69.80 & 19.1 min & 257.52\footnotemark[1] & 86.3\% &0.0169 &63$\pm$41 &3.42$\pm$2.23 $\times$ 10$^{3}$\\
$^{24}$Na & & 14.956 h & 1368.625\footnotemark[1] & 99.994\% &0.00871 &36 $\pm$15 &5.44$\pm$2.27 $\times$ 10$^{3}$  \\
 &  &  & 2754.008 & 99.867\% & \\
\end{tabular}
\footnotetext[1]{Those $\gamma$-rays were used to determine the reaction yields.}
\end{ruledtabular}
\end{table*}

The decay data for radionuclides resulting from the \ce{NaI(Tl)} activation reaction are detailed in Table~\ref{tab1}. The identification of the nuclide type produced in an activated target can be accomplished through the analysis of the positions of the full-energy peaks within the $\gamma$-ray energy spectrum. Additionally, the consistency of the ratio of the net areas of these peaks, adjusted for detection efficiency, with the branching ratios of each $\gamma$ decay provides essential information. 

\begin{figure*}
\includegraphics[scale=0.52]{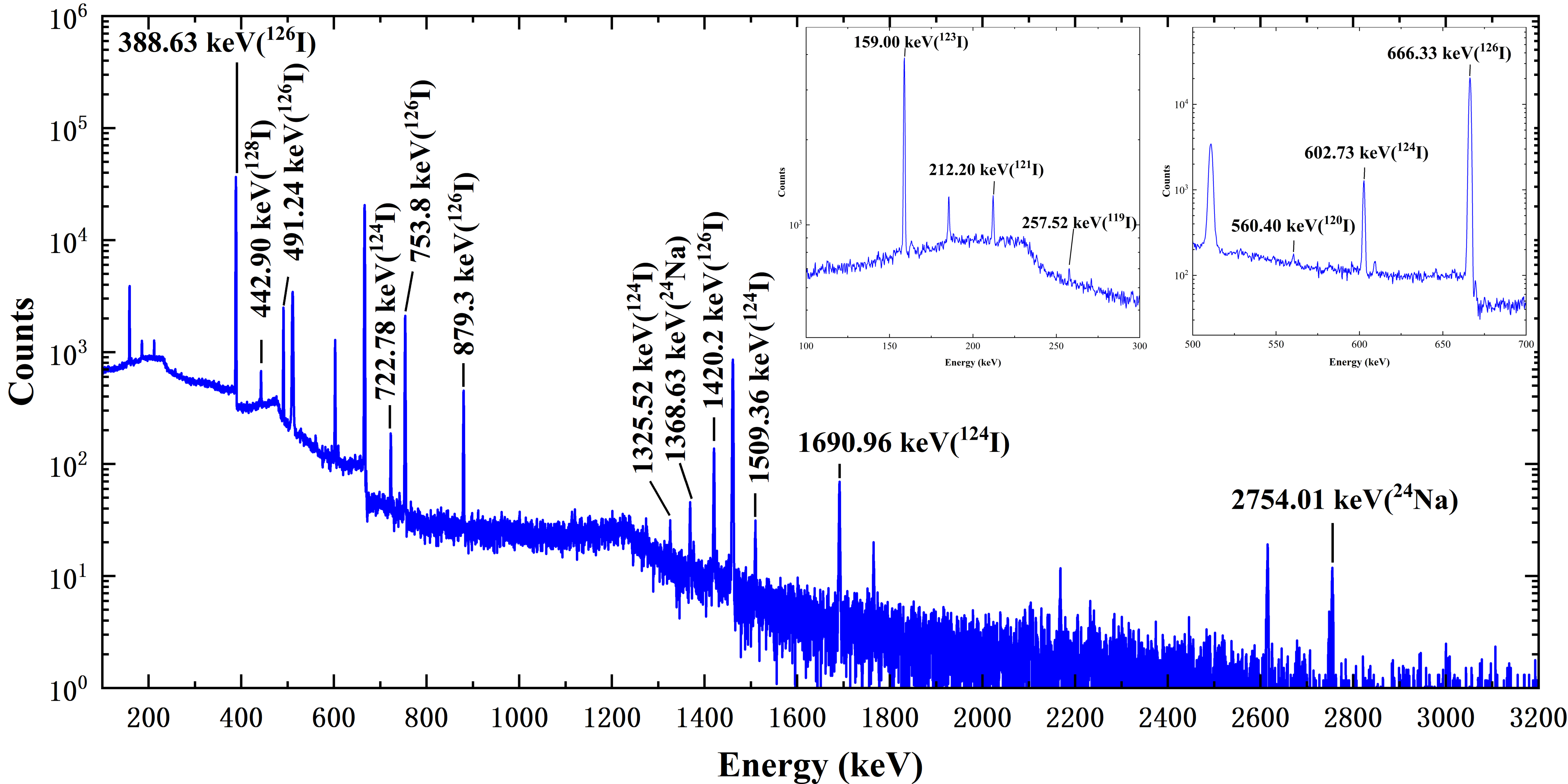}
\caption{\label{fig5} Typical $\gamma$ energy spectrum of \ce{NaI(Tl)} target activation signals measured by HPGe for 15.45 h.}
\end{figure*}

By analyzing the $\gamma$-ray energy spectrum of the activated target, the characteristic peaks of \ce{^{119-121,123,124,126,128}I} and \ce{^{24}Na} were clearly discernible above the background noise and Compton continuum, as depicted in Fig.~\ref{fig5}. As illustrated in Fig.~\ref{fig5}, the characteristic $\gamma$-rays of the nuclides listed in Table~\ref{tab1} manifest as full-energy peaks in the $\gamma$-ray energy spectra acquired from the measurements of the activated sodium iodide target using a high-purity germanium detector. Given that the measurement of the sodium iodide target commenced 10 minutes post-targeting and the data acquisition duration for the energy spectrum presented in Fig.~\ref{fig5} was 15.45 hours, the characteristic full-energy peaks for \ce{^{122}I}($T_{1/2}=3.63$ min) and \ce{^{125}I}($T_{1/2}=59.41$ d) are absent. In contrast to \ce{^{126}I} and \ce{^{124}I}, which have lower production thresholds, other nuclides further from the line of stability exhibit only the signature $\gamma$-ray with the largest branching ratio, showing a weak full-energy peak in the measured spectra due to their low yields and the significantly smaller branching ratio of the secondary signature $\gamma$-ray. As the radionuclide decays, the net count rate of the full-energy peak decreases exponentially at a rate defined by its decay constant, $\lambda$. The half-life of the peak, which corresponds to the half-life of the nuclide, can be determined by fitting the decay of the net count rate. The fitting function employed adheres to the standard formula for radioactive decay:

\begin{equation}
\frac{dN_{net}}{dt} = \lambda N_{0}\cdot exp(-\lambda t)
\label{eq2}
\end{equation}

\begin{figure*}
\includegraphics[scale=0.32]{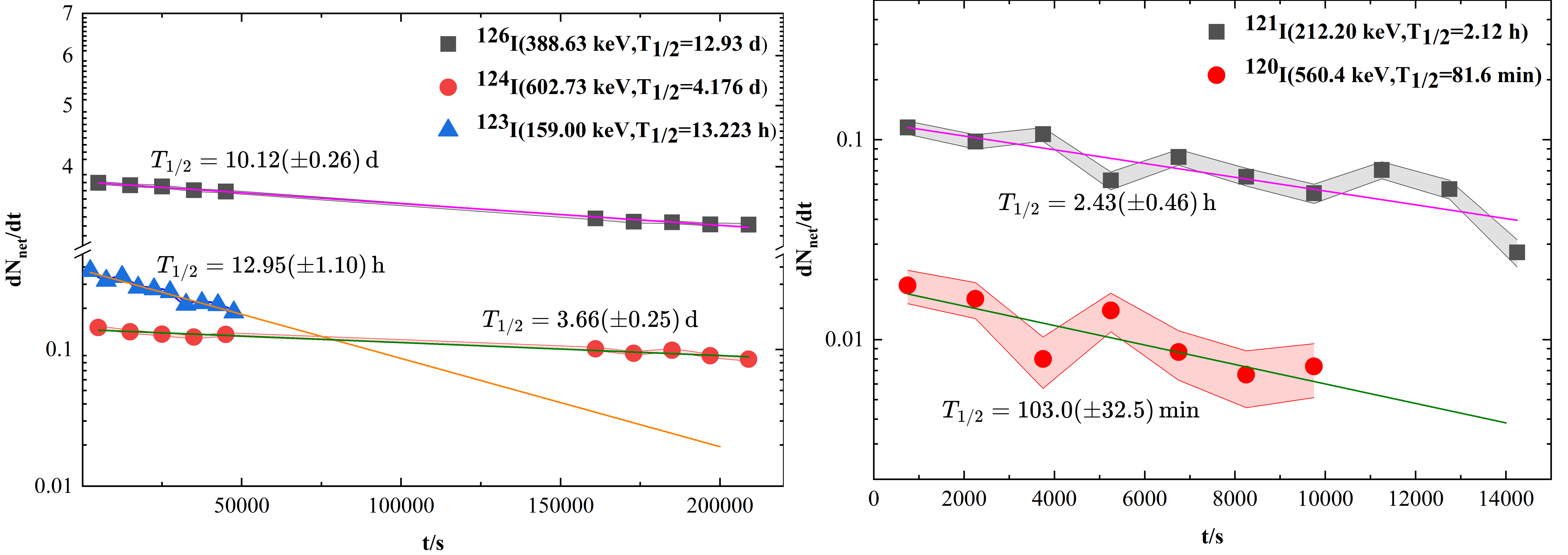}
\caption{\label{fig6} Typical decay curves for the isotopes \ce{^{126}I}, \ce{^{124}I}, \ce{^{123}I}, \ce{^{121}I}, \ce{^{120}I} obtained from an irradiation of \ce{NaI(Tl)} with bremsstrahlung.(The legend for the data points indicates the full energy peaks of nuclide used for the calculations and the half-life values for that nuclide from NuDat 3.0~\cite{b43})}
\end{figure*}

Consequently, by computing the net area of the distinctive all-energy peaks in the $\gamma$ energy spectrum at each timestamp and determining the net count rate, the uncertainty in which is obtained from one standard deviation of the increment in net counts, we can plot the variation in the net count rate of the characteristic peaks against time. This plot is then fitted to Equation ~\ref{eq2}, as illustrated in Fig~\ref{fig6}. The determined half-lives from the fittings are generally consistent with those provided by the NuDat 3.0~\cite{b43} database. The greater deviations observed for the longer-lived nuclides, \ce{^{126}I} and \ce{^{124}I}, can be attributed to the shorter measurement durations employed.
In instances where the signal is strong and the measurement duration exceeds the half-life of the nuclide, the discrepancy between our calculated half-life values and those reported by previous authors is less than 10\%. Conversely, when our measurements are conducted over a shorter period than the nuclides' lifetimes. In summary, by correlating the peak positions and half-lives, we can ascertain the nuclides generated by the experiment, with the notable exception of \ce{^{119}I}. The yield of \ce{^{119}I} was insufficient to perform a reliable half-life analysis due to poor statistics in the 257.81 keV full-energy peak. However, this energy peak is definitively assigned to \ce{^{119}I} based on the known reaction channel on the \ce{NaI} target and corroboration with the NuDat 3.0 database~\cite{b43}.

Subsequently, the final yield of the nuclide at the conclusion of irradiation can be deduced from the net area under the photo-peaks. This calculation is articulated as follows:
\begin{equation}
 \begin{split}
\bar{Y} &= \frac{N_{net}}{[\sum_{i=1}^{n}exp(-\lambda \cdot \Delta t_{i})]\cdot I_{\gamma}\cdot \epsilon_{v}\cdot[1-exp(-\lambda t)]}\\
&= N_{v}D\int_{E_{thr}}^{E_{max}}\sigma(E)\phi_{\gamma}(E)dE
\label{eq3}
\end{split}
\end{equation}
In the first term of equation~\ref{eq3}, \(\bar{Y}\) denotes the average reaction yield per laser pulse, \(N_{\text{net}}\) represents the net area under the photo-peaks, \(I_{\gamma}\) is the decay intensity of the characteristic $\gamma$-rays, \(t\) is the actual measurement time, \(\Delta t_{i}\) is the time difference between the \(i\)-th laser pulse and the start of the sample measurement, \(\lambda\) is the decay constant, and \(\epsilon_v\) is the peak detection efficiency of the HPGe detector for volumetric sources. Based on equation~\ref{eq3} and the parameters in Table~\ref{tab1}, we can calculate the average yield of nuclides at the end of per targeting. The error in the nuclear reaction yields comes from the error in calculating the net area of the Gaussian fit to the full energy peak. For continuous $\gamma$-spectra, the nuclear reaction yield can be theoretically calculated by the second term of equation~\ref{eq3}, where \(N_{v}\) is the number density of target nuclei, \(D\) is the target thickness, \(E_{thr}\) is the reaction threshold energy, \(E_{max}\) is the maximum bremsstrahlung photon energy, \(\sigma(E)\) is the energy-dependent reaction cross section, \(\phi(E)\) is the bremsstrahlung photon flux. In studies of bremsstrahlung-induced photonuclear reactions, the flux-weighted average cross section $\langle \sigma \rangle$ is commonly employed to compare experimental measurements with theoretical calculations, defined as:
\begin{equation}
\langle \sigma \rangle = \frac{\int_{E_{thr}}^{E_{max}}\sigma(E)\phi_{\gamma}(E)dE}{\int_{E_{thr}}^{E_{max}}\phi_{\gamma}(E)dE}
\label{eq4}
\end{equation}
The flux-weighted average cross section was experimentally derived from nuclear reaction product yields using the Geant4\cite{b37,b38}-simulated bremsstrahlung spectrum, while theoretically computed via TALYS 2.0 code\cite{b36}, with both results summarized in Table~\ref{tab2}. 

\begin{table}[!htb]
\renewcommand{\arraystretch}{1.5} 
\begin{ruledtabular}
\caption{The flux-weight average cross sections of \ce{^{127}I}$(\gamma,xn)$\ce{^{119-126}I} reactions and the theoretical values based on the TALYS 2.0 code\cite{b36}.}
\label{tab2}
\small
\begin{tabular}{@{}p{2cm}p{2cm}p{2cm}p{2cm}@{}}
\centering Nuclear reaction &\centering  $\gamma$ endpoint energy/MeV  & \multicolumn{2}{@{}p{4cm}@{}}{\centering Flux-weighted average cross section ($\langle \sigma \rangle$)/mb} \\
\cline{3-4}
& &  Experiment & TALYS 2.0 \\
\hline 
\ce{^{127}I}$(\gamma,n)$\ce{^{126}I} & \centering 150.0 &  102.76$\pm$0.69 & 68.66 \\
\ce{^{127}I}$(\gamma,2n)$\ce{^{125}I} & \centering 150.0 &  Not observed & 31.46 \\
\ce{^{127}I}$(\gamma,3n)$\ce{^{124}I} &\centering 150.0 & 6.54$\pm$0.30 & 6.27 \\
\ce{^{127}I}$(\gamma,4n)$\ce{^{123}I} &\centering 150.0 & 3.12$\pm$0.12 & 3.50 \\
\ce{^{127}I}$(\gamma,5n)$\ce{^{122}I} &\centering 150.0 & Not observed & 1.47 \\
\ce{^{127}I}$(\gamma,6n)$\ce{^{121}I} & \centering 150.0 & 0.57$\pm$0.051 & 0.88 \\
\ce{^{127}I}$(\gamma,7n)$\ce{^{120}I} & \centering 150.0 & 0.23$\pm$0.057 & 0.38 \\
\ce{^{127}I}$(\gamma,8n)$\ce{^{119}I} &\centering 150.0 & 0.24$\pm$0.15 & 0.22 \\
\end{tabular}
\end{ruledtabular}
\end{table}

\begin{figure}
\includegraphics[scale=0.45]{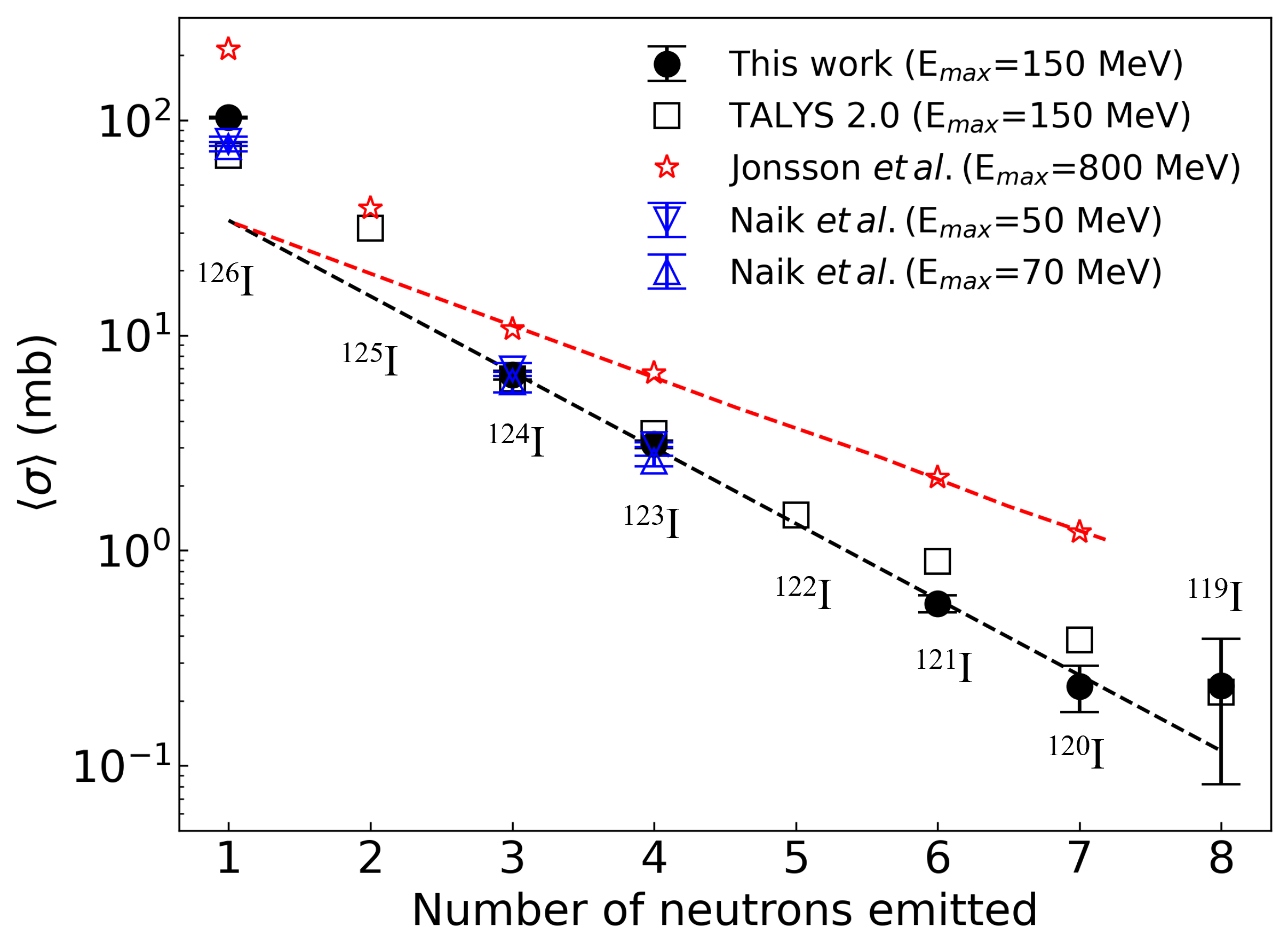}
\caption{\label{fig7} The experimentally measured flux-weighted average cross section as functions of number of emitted neutrons, Compared with theoretical values based on the TALYS 2.0 code\cite{b36}, the work of Jonsson \textit{et al}.~\cite{b8} and the work of Naik \textit{et al}.~\cite{b10}. The dashed lines represent exponential fit to the data.}
\end{figure}

As depicted in Fig~\ref{fig4}(a), The cross-section in photonuclear processes exhibits an inverse correlation with neutron evaporation multiplicity. Fig.~\ref{fig7} illustrates the experimentally measured flux-weighted average cross section as functions of number of emitted neutrons, Compared with theoretical values based on the TALYS 2.0 code\cite{b36}, the work of Jonsson \textit{et al}.~\cite{b8} and the work of Naik \textit{et al}.~\cite{b10}. The reaction channels leading to \ce{^{125}I} (x=2) and \ce{^{122}I} (x=5) were not experimentally observed, consistent with their respective long and short half-lives under our experimental conditions. However, their theoretical flux-weighted average cross sections, calculated using TALYS 2.0, are included in Table~\ref{tab2} and Figure~\ref{fig7} for a comprehensive comparison. The experimentally measured flux-weighted average cross section decreases approximately exponentially with the number of neutrons emitted, particularly when the number of emitted neutrons exceeds two. The predicted cross section for the ($\gamma$,2n) reaction is comparable to that of the ($\gamma$,3n) channel, while the ($\gamma$,5n) channel follows the established exponential decay trend. The exponential coefficients are 0.81 per neutron. As discussed by Jonsson and Forkman in Ref.\cite{b8}, when the number of neutrons exceeds two, the flux-weighted average cross section decreases nearly exponentially with respect to the number of emitted neutrons, exhibiting exponential coefficients of 0.56 per neutron at photon energies of 800 MeV, respectively. The steeper exponential decline observed in our work (0.81 per neutron) compared to that reported by Jonsson et al.\cite{b8} (0.56 per neutron)  may be attributed to the different endpoint energies of the bremsstrahlung sources (150 MeV vs. 800 MeV). The higher photon flux from bremsstrahlung driven by higher-energy electron beams, within the relevant nuclear energy window, promotes population of multiple neutron emission channels. This can lead to a flatter decreasing trend of the photoneutron cross section as a function of the emitted neutron number. This correlation arises because, with increasing energy, reaction channels such as emitted charged particles compete with neutron emission. Additionally, the yield of isotope \ce{^{126}I} might include contributions from the photonuclear reactions induced by Compton-scattered photons in thick targets and the reaction channel \ce{^{127}I}$(n,2n)$\ce{^{126}I}, which has a threshold of 9.217 MeV. Halpern \textit{et al}. observed in the Cu-As region that it is more facile to evaporate neutrons from stable nuclei with a higher atomic number A~\cite{b47}. For heavy nuclei like \ce{^{127}I}, Jonsson \textit{et al}.'s previous $(\gamma, xn)$ experiment~\cite{b22} utilized a monoenergetic electron beam converted to bremsstrahlung to irradiate iodine samples from reaction thresholds up to 900 MeV, measuring \ce{^{118}I}. In contrast, our experiment, employing a significantly lower energy but higher charge, measured a substantially lower nuclear reaction cross section for \ce{^{119}I}. As demonstrated in Fig.~\ref{fig4}(a), when $x$ is greater than or equal to 7, the cross-section reduces to below 1 mb. This suggests that laser-accelerated electrons produce a high-current electron beam over a short timescale, which is then converted into a high-flux bremsstrahlung radiation suitable for studying cross-section $(\gamma, xn)$ nuclear reactions.

As shown in Fig.~\ref{fig5} and Table~\ref{tab1}, the appearance of the 442.90 keV $\gamma$ line in the energy spectrum given by the reaction \ce{^{127}I}$(n,\gamma)$\ce{^{128}I} and the calculation of its half-life clarifies the production of \ce{^{128}I} with an average yield for a single shot of $6.28 \pm 0.23 \times 10^{5}$; the appearance of the 1368.63 keV $\gamma$ line given by the reaction \ce{^{23}Na}$(n,\gamma)$\ce{^{24}Na} in the energy spectrum and the calculation of its half-life clarified the production of \ce{^{24}Na} with an average yield of $5.44 \pm 2.27 \times 10^{3}$ for a single shot. The nuclear reaction yield of \ce{^{128}I} is more than that of \ce{^{24}Na} because the neutron capture cross section of \ce{^{127}I} is larger than that of \ce{^{23}Na}. This indicates that a significant flux of photoneutrons was generated, which in turn induced secondary neutron capture reactions on \ce{^{127}I} and \ce{^{23}Na} within the thick target. This further demonstrates that the neutrons generated through these photonuclear reactions can serve as an ultrashort-pulse neutron source for investigating fundamental phenomena in neutron physics.

\subsection{Yields of medical iodine isotopes}

Erwann \textit{et al}. conducted a comparison of the image quality between \ce{^{131}I} and \ce{^{123}I} for single-photon emission computed tomography (SPECT) imaging and \ce{^{124}I} for positron emission tomography (PET) imaging~\cite{b18}. Their findings indicated that \ce{^{124}I} exhibited the superior imaging performance, which can be attributed to the electron collimation and a short coincidence time window. The isotope of \ce{^{124}I} and \ce{^{123}I} is primarily produced utilizing charged particles beams. In our experimental setup, laser-accelerated electrons with energies ranging from 25.8 to 150.0 MeV produced a total charge of 16.18 nC/shot. This electron beam subsequently generated bremsstrahlung radiation, which induced a photonuclear reaction yielding 9.83 $\pm$ 0.45 $\times$ 10$^{5}$/shot of \ce{^{124}I}. In contrast to \ce{^{131}I}, which is previously extensively employed as a thyroid scanning agent, \ce{^{123}I} possesses a half-life of 13.223 hours, rendering it highly suitable for 24-hour iodine uptake tests~\cite{b21}. Our study achieved a \ce{^{123}I} yield of 2.81$\pm$0.11 $\times$ 10$^{5}$/shot within the energy range of 33.3-150.0 MeV. 

\begin{figure}
\includegraphics[scale=0.43]{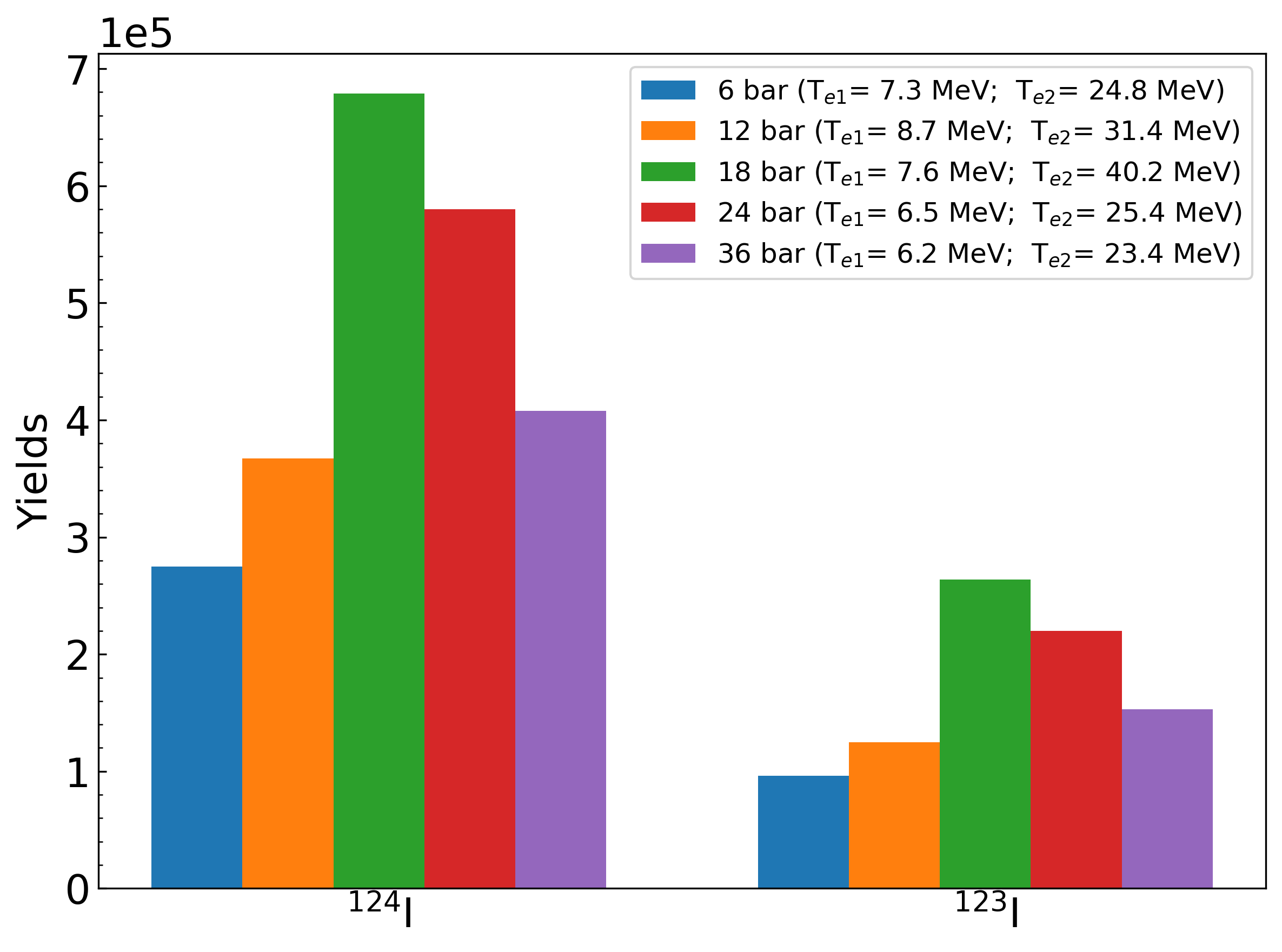}
\caption{\label{fig8} The \ce{^{124}I} and \ce{^{123}I} yields obtained from Geant4~\cite{b37,b38} simulations are plotted with different colors corresponding to distinct gas jet backing pressures, representing the yields under electron energy spectra at different effective temperatures.}
\end{figure}

To compare the effects of electron energy spectra at different effective temperatures on the production of \ce{^{124}I} and \ce{^{123}I}, we performed Geant4~\cite{b37,b38} simulations under experimental irradiation conditions. As shown in Fig.~\ref{fig8}, the maximum yields of \ce{^{124}I} and \ce{^{123}I} were achieved at a gas jet backing pressure of 18 bar(n$_{e}$=0.018n$_{c}$), corresponding to effective temperatures of T$_{e1}$ = 7.6 MeV and T$_{e2}$ = 40.2 MeV. The production of \ce{^{124}I} and \ce{^{123}I} via photonuclear reactions on \ce{^{127}I} requires the stripping of three and four neutrons from the nucleus, respectively. The reaction cross-sections govern the production yields of these isotopes. Two primary optimization approaches exist: (1) Enhancing the $\gamma$-ray flux of laser-driven bremsstrahlung sources above the reaction threshold, and (2) employment of photonuclear reaction channels with larger cross-sections, such as \ce{^{124}Xe}($\gamma, n$)\ce{^{123}Xe} $\rightarrow$ \ce{^{123}I}~\cite{b34}.

\section{Conclusion}

We have conducted experiments irradiating a sodium iodide target with high-flux bremsstrahlung radiation, generated by ultra-intense, ultra-fast laser-driven electron acceleration at the SULF-10PW beamline. Following a ten-shot irradiation, our investigation successfully produced a series of proton-rich iodine isotopes and calculated the corresponding flux-weighted average cross-section, notably including \ce{^{123}I} and \ce{^{124}I} that are extensively utilized in nuclear medicine for diagnostic and therapeutic applications. Particularly noteworthy is the identification of the $(\gamma,8n)$ reaction channel in the photonuclear reaction of \ce{^{127}I}, Additionally, we measured the characteristic peaks of \ce{^{128}I} and \ce{^{24}Na}, which were predominantly produced by secondary reactions triggered by photoneutrons. Furthermore, the experiments have verified that laser-driven bremsstrahlung is a promising alternative compared to traditional methods of photonuclear reaction investigation and medical iodine isotope production.

Utilizing the laser-driven bremsstrahlung source, we can delve into the nuclear reactions involving multiphoton absorption on a short time scale with a large $\gamma$ flux, specifically, the photon absorption reactions at certain short-lived energy levels within the nucleus. This approach is instrumental in generating and investigating the properties of proton-rich nuclides far from the valley of stability. This bremsstrahlung-generated photon bath with thermal spectral characteristics enables systematic studies of $\gamma$-induced reaction cross-sections and rates in astrophysically relevant energy regimes. Beyond fundamental physics, ultra-intense, ultra-fast lasers are poised to become an efficient tool in applications such as the production of medical isotopes.

\section{ACKNOWLEDGMENTS}

The authors thank the staff of the 10 PW laser beamline at the Shanghai Superintense Ultrafast Laser Facility for the excellent operation of the laser facility and their support during the experiment. This work was supported by the National Key R\&D Program of China (Nos.2022YFA1602200, 2022YFA1602400), the Strategic Priority Research Program of the CAS (No.XDB0890302), and the National Natural Science Foundation of China (Nos.12235003, 12388102).

\bibliography{ref}

\begin{thebibliography}{71}%
\makeatletter
\providecommand \@ifxundefined [1]{%
 \@ifx{#1\undefined}
}%
\providecommand \@ifnum [1]{%
 \ifnum #1\expandafter \@firstoftwo
 \else \expandafter \@secondoftwo
 \fi
}%
\providecommand \@ifx [1]{%
 \ifx #1\expandafter \@firstoftwo
 \else \expandafter \@secondoftwo
 \fi
}%
\providecommand \natexlab [1]{#1}%
\providecommand \enquote  [1]{``#1''}%
\providecommand \bibnamefont  [1]{#1}%
\providecommand \bibfnamefont [1]{#1}%
\providecommand \citenamefont [1]{#1}%
\providecommand \href@noop [0]{\@secondoftwo}%
\providecommand \href [0]{\begingroup \@sanitize@url \@href}%
\providecommand \@href[1]{\@@startlink{#1}\@@href}%
\providecommand \@@href[1]{\endgroup#1\@@endlink}%
\providecommand \@sanitize@url [0]{\catcode `\\12\catcode `\$12\catcode
  `\&12\catcode `\#12\catcode `\^12\catcode `\_12\catcode `\%12\relax}%
\providecommand \@@startlink[1]{}%
\providecommand \@@endlink[0]{}%
\providecommand \url  [0]{\begingroup\@sanitize@url \@url }%
\providecommand \@url [1]{\endgroup\@href {#1}{\urlprefix }}%
\providecommand \urlprefix  [0]{URL }%
\providecommand \Eprint [0]{\href }%
\providecommand \doibase [0]{https://doi.org/}%
\providecommand \selectlanguage [0]{\@gobble}%
\providecommand \bibinfo  [0]{\@secondoftwo}%
\providecommand \bibfield  [0]{\@secondoftwo}%
\providecommand \translation [1]{[#1]}%
\providecommand \BibitemOpen [0]{}%
\providecommand \bibitemStop [0]{}%
\providecommand \bibitemNoStop [0]{.\EOS\space}%
\providecommand \EOS [0]{\spacefactor3000\relax}%
\providecommand \BibitemShut  [1]{\csname bibitem#1\endcsname}%
\let\auto@bib@innerbib\@empty
\bibitem [{\citenamefont {Donna}\ and\ \citenamefont {Gerard}(1985)}]{b23}%
  \BibitemOpen
  \bibfield  {author} {\bibinfo {author} {\bibfnamefont {S.}~\bibnamefont
  {Donna}}\ and\ \bibinfo {author} {\bibfnamefont {M.}~\bibnamefont {Gerard}},\
  }\bibfield  {title} {\bibinfo {title} {Compression of amplified chirped
  optical pulses},\ }\href
  {https://doi.org/https://doi.org/10.1016/0030-4018(85)90120-8} {\bibfield
  {journal} {\bibinfo  {journal} {Optics Communications}\ }\textbf {\bibinfo
  {volume} {56}},\ \bibinfo {pages} {219} (\bibinfo {year} {1985})}\BibitemShut
  {NoStop}%
\bibitem [{\citenamefont {Mourou}\ \emph {et~al.}(2006)\citenamefont {Mourou},
  \citenamefont {Tajima},\ and\ \citenamefont {Bulanov}}]{b64}%
  \BibitemOpen
  \bibfield  {author} {\bibinfo {author} {\bibfnamefont {G.~A.}\ \bibnamefont
  {Mourou}}, \bibinfo {author} {\bibfnamefont {T.}~\bibnamefont {Tajima}},\
  and\ \bibinfo {author} {\bibfnamefont {S.~V.}\ \bibnamefont {Bulanov}},\
  }\bibfield  {title} {\bibinfo {title} {Optics in the relativistic regime},\
  }\href {https://doi.org/10.1103/RevModPhys.78.309} {\bibfield  {journal}
  {\bibinfo  {journal} {Rev. Mod. Phys.}\ }\textbf {\bibinfo {volume} {78}},\
  \bibinfo {pages} {309} (\bibinfo {year} {2006})}\BibitemShut {NoStop}%
\bibitem [{\citenamefont {Danson}\ \emph {et~al.}(2019)\citenamefont {Danson},
  \citenamefont {Haefner}, \citenamefont {Bromage}, \citenamefont {Butcher},
  \citenamefont {Chanteloup}, \citenamefont {Chowdhury}, \citenamefont
  {Galvanauskas}, \citenamefont {Gizzi}, \citenamefont {Hein}, \citenamefont
  {Hillier},\ and\ \citenamefont {\textit{et al.}}}]{b24}%
  \BibitemOpen
  \bibfield  {author} {\bibinfo {author} {\bibfnamefont {C.~N.}\ \bibnamefont
  {Danson}}, \bibinfo {author} {\bibfnamefont {C.}~\bibnamefont {Haefner}},
  \bibinfo {author} {\bibfnamefont {J.}~\bibnamefont {Bromage}}, \bibinfo
  {author} {\bibfnamefont {T.}~\bibnamefont {Butcher}}, \bibinfo {author}
  {\bibfnamefont {J.-C.~F.}\ \bibnamefont {Chanteloup}}, \bibinfo {author}
  {\bibfnamefont {E.~A.}\ \bibnamefont {Chowdhury}}, \bibinfo {author}
  {\bibfnamefont {A.}~\bibnamefont {Galvanauskas}}, \bibinfo {author}
  {\bibfnamefont {L.~A.}\ \bibnamefont {Gizzi}}, \bibinfo {author}
  {\bibfnamefont {J.}~\bibnamefont {Hein}}, \bibinfo {author} {\bibfnamefont
  {D.~I.}\ \bibnamefont {Hillier}},\ and\ \bibinfo {author} {\bibnamefont
  {\textit{et al.}}},\ }\bibfield  {title} {\bibinfo {title} {Petawatt and
  exawatt class lasers worldwide},\ }\href
  {https://doi.org/10.1017/hpl.2019.36} {\bibfield  {journal} {\bibinfo
  {journal} {High Power Laser Science and Engineering}\ }\textbf {\bibinfo
  {volume} {7}},\ \bibinfo {pages} {e54} (\bibinfo {year} {2019})}\BibitemShut
  {NoStop}%
\bibitem [{\citenamefont {Tanaka}\ \emph {et~al.}(2020)\citenamefont {Tanaka},
  \citenamefont {Spohr}, \citenamefont {Balabanski}, \citenamefont {Balascuta},
  \citenamefont {Capponi}, \citenamefont {Cernaianu}, \citenamefont {Cuciuc},
  \citenamefont {Cucoanes}, \citenamefont {Dancus}, \citenamefont {Dhal},
  \citenamefont {Diaconescu}, \citenamefont {Doria}, \citenamefont {Ghenuche},
  \citenamefont {Ghita}, \citenamefont {Kisyov}, \citenamefont {Nastasa},
  \citenamefont {Ong}, \citenamefont {Rotaru}, \citenamefont {Sangwan},
  \citenamefont {Söderström}, \citenamefont {Stutman}, \citenamefont
  {Suliman}, \citenamefont {Tesileanu}, \citenamefont {Tudor}, \citenamefont
  {Tsoneva}, \citenamefont {Ur}, \citenamefont {Ursescu},\ and\ \citenamefont
  {Zamfir}}]{b63}%
  \BibitemOpen
  \bibfield  {author} {\bibinfo {author} {\bibfnamefont {K.~A.}\ \bibnamefont
  {Tanaka}}, \bibinfo {author} {\bibfnamefont {K.~M.}\ \bibnamefont {Spohr}},
  \bibinfo {author} {\bibfnamefont {D.~L.}\ \bibnamefont {Balabanski}},
  \bibinfo {author} {\bibfnamefont {S.}~\bibnamefont {Balascuta}}, \bibinfo
  {author} {\bibfnamefont {L.}~\bibnamefont {Capponi}}, \bibinfo {author}
  {\bibfnamefont {M.~O.}\ \bibnamefont {Cernaianu}}, \bibinfo {author}
  {\bibfnamefont {M.}~\bibnamefont {Cuciuc}}, \bibinfo {author} {\bibfnamefont
  {A.}~\bibnamefont {Cucoanes}}, \bibinfo {author} {\bibfnamefont
  {I.}~\bibnamefont {Dancus}}, \bibinfo {author} {\bibfnamefont
  {A.}~\bibnamefont {Dhal}}, \bibinfo {author} {\bibfnamefont {B.}~\bibnamefont
  {Diaconescu}}, \bibinfo {author} {\bibfnamefont {D.}~\bibnamefont {Doria}},
  \bibinfo {author} {\bibfnamefont {P.}~\bibnamefont {Ghenuche}}, \bibinfo
  {author} {\bibfnamefont {D.~G.}\ \bibnamefont {Ghita}}, \bibinfo {author}
  {\bibfnamefont {S.}~\bibnamefont {Kisyov}}, \bibinfo {author} {\bibfnamefont
  {V.}~\bibnamefont {Nastasa}}, \bibinfo {author} {\bibfnamefont {J.~F.}\
  \bibnamefont {Ong}}, \bibinfo {author} {\bibfnamefont {F.}~\bibnamefont
  {Rotaru}}, \bibinfo {author} {\bibfnamefont {D.}~\bibnamefont {Sangwan}},
  \bibinfo {author} {\bibfnamefont {P.~A.}\ \bibnamefont {Söderström}},
  \bibinfo {author} {\bibfnamefont {D.}~\bibnamefont {Stutman}}, \bibinfo
  {author} {\bibfnamefont {G.}~\bibnamefont {Suliman}}, \bibinfo {author}
  {\bibfnamefont {O.}~\bibnamefont {Tesileanu}}, \bibinfo {author}
  {\bibfnamefont {L.}~\bibnamefont {Tudor}}, \bibinfo {author} {\bibfnamefont
  {N.}~\bibnamefont {Tsoneva}}, \bibinfo {author} {\bibfnamefont {C.~A.}\
  \bibnamefont {Ur}}, \bibinfo {author} {\bibfnamefont {D.}~\bibnamefont
  {Ursescu}},\ and\ \bibinfo {author} {\bibfnamefont {N.~V.}\ \bibnamefont
  {Zamfir}},\ }\bibfield  {title} {\bibinfo {title} {Current status and
  highlights of the eli-np research program},\ }\href
  {https://doi.org/10.1063/1.5093535} {\bibfield  {journal} {\bibinfo
  {journal} {Matter and Radiation at Extremes}\ }\textbf {\bibinfo {volume}
  {5}},\ \bibinfo {pages} {024402} (\bibinfo {year} {2020})}\BibitemShut
  {NoStop}%
\bibitem [{\citenamefont {Esarey}\ \emph {et~al.}(2009)\citenamefont {Esarey},
  \citenamefont {Schroeder},\ and\ \citenamefont {Leemans}}]{b97}%
  \BibitemOpen
  \bibfield  {author} {\bibinfo {author} {\bibfnamefont {E.}~\bibnamefont
  {Esarey}}, \bibinfo {author} {\bibfnamefont {C.~B.}\ \bibnamefont
  {Schroeder}},\ and\ \bibinfo {author} {\bibfnamefont {W.~P.}\ \bibnamefont
  {Leemans}},\ }\bibfield  {title} {\bibinfo {title} {Physics of laser-driven
  plasma-based electron accelerators},\ }\href
  {https://doi.org/10.1103/RevModPhys.81.1229} {\bibfield  {journal} {\bibinfo
  {journal} {Rev. Mod. Phys.}\ }\textbf {\bibinfo {volume} {81}},\ \bibinfo
  {pages} {1229} (\bibinfo {year} {2009})}\BibitemShut {NoStop}%
\bibitem [{\citenamefont {Tajima}\ and\ \citenamefont {Dawson}(1979)}]{b25}%
  \BibitemOpen
  \bibfield  {author} {\bibinfo {author} {\bibfnamefont {T.}~\bibnamefont
  {Tajima}}\ and\ \bibinfo {author} {\bibfnamefont {J.~M.}\ \bibnamefont
  {Dawson}},\ }\bibfield  {title} {\bibinfo {title} {Laser electron
  accelerator},\ }\href {https://doi.org/10.1103/PhysRevLett.43.267} {\bibfield
   {journal} {\bibinfo  {journal} {Phys. Rev. Lett.}\ }\textbf {\bibinfo
  {volume} {43}},\ \bibinfo {pages} {267} (\bibinfo {year} {1979})}\BibitemShut
  {NoStop}%
\bibitem [{\citenamefont {Feng}\ \emph {et~al.}(2023)\citenamefont {Feng},
  \citenamefont {Li}, \citenamefont {Tan}, \citenamefont {Wang}, \citenamefont
  {Li}, \citenamefont {Zhang}, \citenamefont {Meng}, \citenamefont {Ge},
  \citenamefont {Liu}, \citenamefont {Yan}, \citenamefont {Fu}, \citenamefont
  {Chen},\ and\ \citenamefont {Zhang}}]{b28}%
  \BibitemOpen
  \bibfield  {author} {\bibinfo {author} {\bibfnamefont {J.}~\bibnamefont
  {Feng}}, \bibinfo {author} {\bibfnamefont {Y.~J.}\ \bibnamefont {Li}},
  \bibinfo {author} {\bibfnamefont {J.~H.}\ \bibnamefont {Tan}}, \bibinfo
  {author} {\bibfnamefont {W.~Z.}\ \bibnamefont {Wang}}, \bibinfo {author}
  {\bibfnamefont {Y.~F.}\ \bibnamefont {Li}}, \bibinfo {author} {\bibfnamefont
  {X.~P.}\ \bibnamefont {Zhang}}, \bibinfo {author} {\bibfnamefont
  {Y.}~\bibnamefont {Meng}}, \bibinfo {author} {\bibfnamefont {X.~L.}\
  \bibnamefont {Ge}}, \bibinfo {author} {\bibfnamefont {F.}~\bibnamefont
  {Liu}}, \bibinfo {author} {\bibfnamefont {W.~C.}\ \bibnamefont {Yan}},
  \bibinfo {author} {\bibfnamefont {C.~B.}\ \bibnamefont {Fu}}, \bibinfo
  {author} {\bibfnamefont {L.~M.}\ \bibnamefont {Chen}},\ and\ \bibinfo
  {author} {\bibfnamefont {J.}~\bibnamefont {Zhang}},\ }\bibfield  {title}
  {\bibinfo {title} {Laser plasma-accelerated ultra-intense electron beam for
  efficiently exciting nuclear isomers},\ }\href
  {https://doi.org/https://doi.org/10.1002/lpor.202300514} {\bibfield
  {journal} {\bibinfo  {journal} {Laser \& Photonics Reviews}\ }\textbf
  {\bibinfo {volume} {17}},\ \bibinfo {pages} {2300514} (\bibinfo {year}
  {2023})}\BibitemShut {NoStop}%
\bibitem [{\citenamefont {Shaw}\ \emph {et~al.}(2021)\citenamefont {Shaw},
  \citenamefont {Romo-Gonzalez}, \citenamefont {Lemos}, \citenamefont {King},
  \citenamefont {Bruhaug}, \citenamefont {Miller}, \citenamefont {Dorrer},
  \citenamefont {Kruschwitz}, \citenamefont {Waxer}, \citenamefont {Williams},
  \citenamefont {Ambat}, \citenamefont {McKie}, \citenamefont {Sinclair},
  \citenamefont {Mori}, \citenamefont {Joshi}, \citenamefont {Chen},
  \citenamefont {Palastro}, \citenamefont {Albert},\ and\ \citenamefont
  {Froula}}]{b96}%
  \BibitemOpen
  \bibfield  {author} {\bibinfo {author} {\bibfnamefont {J.~L.}\ \bibnamefont
  {Shaw}}, \bibinfo {author} {\bibfnamefont {M.~A.}\ \bibnamefont
  {Romo-Gonzalez}}, \bibinfo {author} {\bibfnamefont {N.}~\bibnamefont
  {Lemos}}, \bibinfo {author} {\bibfnamefont {P.~M.}\ \bibnamefont {King}},
  \bibinfo {author} {\bibfnamefont {G.}~\bibnamefont {Bruhaug}}, \bibinfo
  {author} {\bibfnamefont {K.~G.}\ \bibnamefont {Miller}}, \bibinfo {author}
  {\bibfnamefont {C.}~\bibnamefont {Dorrer}}, \bibinfo {author} {\bibfnamefont
  {B.}~\bibnamefont {Kruschwitz}}, \bibinfo {author} {\bibfnamefont
  {L.}~\bibnamefont {Waxer}}, \bibinfo {author} {\bibfnamefont {G.~J.}\
  \bibnamefont {Williams}}, \bibinfo {author} {\bibfnamefont {M.~V.}\
  \bibnamefont {Ambat}}, \bibinfo {author} {\bibfnamefont {M.~M.}\ \bibnamefont
  {McKie}}, \bibinfo {author} {\bibfnamefont {M.~D.}\ \bibnamefont {Sinclair}},
  \bibinfo {author} {\bibfnamefont {W.~B.}\ \bibnamefont {Mori}}, \bibinfo
  {author} {\bibfnamefont {C.}~\bibnamefont {Joshi}}, \bibinfo {author}
  {\bibfnamefont {H.}~\bibnamefont {Chen}}, \bibinfo {author} {\bibfnamefont
  {J.~P.}\ \bibnamefont {Palastro}}, \bibinfo {author} {\bibfnamefont
  {F.}~\bibnamefont {Albert}},\ and\ \bibinfo {author} {\bibfnamefont {D.~H.}\
  \bibnamefont {Froula}},\ }\bibfield  {title} {\bibinfo {title} {Microcoulomb
  (0.7$\pm$$\frac{0.4}{0.2}$$\mu${C}) laser plasma accelerator on {OMEGA EP}},\
  }\href {https://doi.org/10.1038/s41598-021-86523-5} {\bibfield  {journal}
  {\bibinfo  {journal} {Scientific Reports}\ }\textbf {\bibinfo {volume}
  {11}},\ \bibinfo {pages} {7498} (\bibinfo {year} {2021})}\BibitemShut
  {NoStop}%
\bibitem [{\citenamefont {Mirani}\ \emph {et~al.}(2021)\citenamefont {Mirani},
  \citenamefont {Calzolari}, \citenamefont {Formenti},\ and\ \citenamefont
  {Passoni}}]{b59}%
  \BibitemOpen
  \bibfield  {author} {\bibinfo {author} {\bibfnamefont {F.}~\bibnamefont
  {Mirani}}, \bibinfo {author} {\bibfnamefont {D.}~\bibnamefont {Calzolari}},
  \bibinfo {author} {\bibfnamefont {A.}~\bibnamefont {Formenti}},\ and\
  \bibinfo {author} {\bibfnamefont {M.}~\bibnamefont {Passoni}},\ }\bibfield
  {title} {\bibinfo {title} {Superintense laser-driven photon activation
  analysis},\ }\href
  {https://doi.org/https://doi.org/10.1038/s42005-021-00685-2} {\bibfield
  {journal} {\bibinfo  {journal} {Communications Physics}\ }\textbf {\bibinfo
  {volume} {4}},\ \bibinfo {pages} {185} (\bibinfo {year} {2021})}\BibitemShut
  {NoStop}%
\bibitem [{\citenamefont {Li}\ \emph {et~al.}(2023)\citenamefont {Li},
  \citenamefont {Wang}, \citenamefont {Luo}, \citenamefont {Yang},
  \citenamefont {Cao}, \citenamefont {Li}, \citenamefont {Yuan}, \citenamefont
  {Zhao},\ and\ \citenamefont {Fan}}]{b92}%
  \BibitemOpen
  \bibfield  {author} {\bibinfo {author} {\bibfnamefont {Z.~C.}\ \bibnamefont
  {Li}}, \bibinfo {author} {\bibfnamefont {H.~W.}\ \bibnamefont {Wang}},
  \bibinfo {author} {\bibfnamefont {W.}~\bibnamefont {Luo}}, \bibinfo {author}
  {\bibfnamefont {Y.}~\bibnamefont {Yang}}, \bibinfo {author} {\bibfnamefont
  {Z.~W.}\ \bibnamefont {Cao}}, \bibinfo {author} {\bibfnamefont {X.~X.}\
  \bibnamefont {Li}}, \bibinfo {author} {\bibfnamefont {Y.}~\bibnamefont
  {Yuan}}, \bibinfo {author} {\bibfnamefont {Z.~Q.}\ \bibnamefont {Zhao}},\
  and\ \bibinfo {author} {\bibfnamefont {G.~T.}\ \bibnamefont {Fan}},\
  }\bibfield  {title} {\bibinfo {title} {Effective extraction of photoneutron
  cross-section distribution using gamma activation and reaction yield ratio
  method},\ }\href {https://doi.org/10.1007/s41365-023-01330-z} {\bibfield
  {journal} {\bibinfo  {journal} {Nuclear Science and Techniques}\ }\textbf
  {\bibinfo {volume} {34}},\ \bibinfo {pages} {170} (\bibinfo {year}
  {2023})}\BibitemShut {NoStop}%
\bibitem [{\citenamefont {Fan}\ \emph {et~al.}(2023)\citenamefont {Fan},
  \citenamefont {Xu}, \citenamefont {Li}, \citenamefont {Xu}, \citenamefont
  {Xu}, \citenamefont {Zhu}, \citenamefont {Shen},\ and\ \citenamefont
  {Ji}}]{b98}%
  \BibitemOpen
  \bibfield  {author} {\bibinfo {author} {\bibfnamefont {L.}~\bibnamefont
  {Fan}}, \bibinfo {author} {\bibfnamefont {T.}~\bibnamefont {Xu}}, \bibinfo
  {author} {\bibfnamefont {S.}~\bibnamefont {Li}}, \bibinfo {author}
  {\bibfnamefont {Z.}~\bibnamefont {Xu}}, \bibinfo {author} {\bibfnamefont
  {J.}~\bibnamefont {Xu}}, \bibinfo {author} {\bibfnamefont {J.}~\bibnamefont
  {Zhu}}, \bibinfo {author} {\bibfnamefont {B.}~\bibnamefont {Shen}},\ and\
  \bibinfo {author} {\bibfnamefont {L.}~\bibnamefont {Ji}},\ }\bibfield
  {title} {\bibinfo {title} {Collimated gamma beams with high peak flux driven
  by laser-accelerated electrons},\ }\href
  {https://doi.org/10.1017/hpl.2023.25} {\bibfield  {journal} {\bibinfo
  {journal} {High Power Laser Science and Engineering}\ }\textbf {\bibinfo
  {volume} {11}},\ \bibinfo {pages} {e26} (\bibinfo {year} {2023})}\BibitemShut
  {NoStop}%
\bibitem [{\citenamefont {Li}\ \emph {et~al.}(2017)\citenamefont {Li},
  \citenamefont {Shen}, \citenamefont {Xu}, \citenamefont {Xu}, \citenamefont
  {Yu}, \citenamefont {Li}, \citenamefont {Lu}, \citenamefont {Wang},
  \citenamefont {Wang}, \citenamefont {Liang}, \citenamefont {Leng},
  \citenamefont {Li},\ and\ \citenamefont {Xu}}]{b26}%
  \BibitemOpen
  \bibfield  {author} {\bibinfo {author} {\bibfnamefont {S.}~\bibnamefont
  {Li}}, \bibinfo {author} {\bibfnamefont {B.~F.}\ \bibnamefont {Shen}},
  \bibinfo {author} {\bibfnamefont {J.~C.}\ \bibnamefont {Xu}}, \bibinfo
  {author} {\bibfnamefont {T.~J.}\ \bibnamefont {Xu}}, \bibinfo {author}
  {\bibfnamefont {Y.}~\bibnamefont {Yu}}, \bibinfo {author} {\bibfnamefont
  {J.~F.}\ \bibnamefont {Li}}, \bibinfo {author} {\bibfnamefont {X.~M.}\
  \bibnamefont {Lu}}, \bibinfo {author} {\bibfnamefont {C.}~\bibnamefont
  {Wang}}, \bibinfo {author} {\bibfnamefont {X.~L.}\ \bibnamefont {Wang}},
  \bibinfo {author} {\bibfnamefont {X.~Y.}\ \bibnamefont {Liang}}, \bibinfo
  {author} {\bibfnamefont {Y.~X.}\ \bibnamefont {Leng}}, \bibinfo {author}
  {\bibfnamefont {R.~X.}\ \bibnamefont {Li}},\ and\ \bibinfo {author}
  {\bibfnamefont {Z.~Z.}\ \bibnamefont {Xu}},\ }\bibfield  {title} {\bibinfo
  {title} {Ultrafast multi-mev gamma-ray beam produced by laser-accelerated
  electrons},\ }\href {https://doi.org/10.1063/1.4996020} {\bibfield  {journal}
  {\bibinfo  {journal} {Physics of Plasmas}\ }\textbf {\bibinfo {volume}
  {24}},\ \bibinfo {pages} {093104} (\bibinfo {year} {2017})}\BibitemShut
  {NoStop}%
\bibitem [{\citenamefont {Arnould}\ and\ \citenamefont {Goriely}(2003)}]{b13}%
  \BibitemOpen
  \bibfield  {author} {\bibinfo {author} {\bibfnamefont {M.}~\bibnamefont
  {Arnould}}\ and\ \bibinfo {author} {\bibfnamefont {S.}~\bibnamefont
  {Goriely}},\ }\bibfield  {title} {\bibinfo {title} {The p-process of stellar
  nucleosynthesis: astrophysics and nuclear physics status},\ }\href
  {https://doi.org/https://doi.org/10.1016/S0370-1573(03)00242-4} {\bibfield
  {journal} {\bibinfo  {journal} {Physics Reports}\ }\textbf {\bibinfo {volume}
  {384}},\ \bibinfo {pages} {1} (\bibinfo {year} {2003})}\BibitemShut {NoStop}%
\bibitem [{\citenamefont {Hayakawa}\ \emph {et~al.}(2017)\citenamefont
  {Hayakawa}, \citenamefont {Nakamura}, \citenamefont {Kotaki}, \citenamefont
  {Kando},\ and\ \citenamefont {Kajino}}]{b31}%
  \BibitemOpen
  \bibfield  {author} {\bibinfo {author} {\bibfnamefont {T.}~\bibnamefont
  {Hayakawa}}, \bibinfo {author} {\bibfnamefont {T.}~\bibnamefont {Nakamura}},
  \bibinfo {author} {\bibfnamefont {H.}~\bibnamefont {Kotaki}}, \bibinfo
  {author} {\bibfnamefont {M.}~\bibnamefont {Kando}},\ and\ \bibinfo {author}
  {\bibfnamefont {T.}~\bibnamefont {Kajino}},\ }\bibfield  {title} {\bibinfo
  {title} {Explosive nucleosynthesis study using laser driven $\gamma$-ray
  pulses},\ }\href {https://doi.org/https://doi.org/10.3390/qubs1010003}
  {\bibfield  {journal} {\bibinfo  {journal} {Quantum Beam Science}\ }\textbf
  {\bibinfo {volume} {1}} (\bibinfo {year} {2017})}\BibitemShut {NoStop}%
\bibitem [{\citenamefont {Zilges}\ \emph {et~al.}(2022)\citenamefont {Zilges},
  \citenamefont {Balabanski}, \citenamefont {Isaak},\ and\ \citenamefont
  {Pietralla}}]{b2}%
  \BibitemOpen
  \bibfield  {author} {\bibinfo {author} {\bibfnamefont {A.}~\bibnamefont
  {Zilges}}, \bibinfo {author} {\bibfnamefont {D.~L.}\ \bibnamefont
  {Balabanski}}, \bibinfo {author} {\bibfnamefont {J.}~\bibnamefont {Isaak}},\
  and\ \bibinfo {author} {\bibfnamefont {N.}~\bibnamefont {Pietralla}},\
  }\bibfield  {title} {\bibinfo {title} {Photonuclear reactions—from basic
  research to applications},\ }\href
  {https://doi.org/https://doi.org/10.1016/j.ppnp.2021.103903} {\bibfield
  {journal} {\bibinfo  {journal} {Progress in Particle and Nuclear Physics}\
  }\textbf {\bibinfo {volume} {122}},\ \bibinfo {pages} {103903} (\bibinfo
  {year} {2022})}\BibitemShut {NoStop}%
\bibitem [{\citenamefont {Wu}\ \emph {et~al.}(2023)\citenamefont {Wu},
  \citenamefont {Lan}, \citenamefont {Zhang}, \citenamefont {Liu},
  \citenamefont {Lu}, \citenamefont {Lv}, \citenamefont {Wu}, \citenamefont
  {Zhang}, \citenamefont {Cai}, \citenamefont {Ma}, \citenamefont {Xia},
  \citenamefont {Wang}, \citenamefont {Wang}, \citenamefont {Yang},
  \citenamefont {Xu}, \citenamefont {Geng}, \citenamefont {Zhao}, \citenamefont
  {Lin}, \citenamefont {Ma}, \citenamefont {Yu}, \citenamefont {Wang},
  \citenamefont {Liu}, \citenamefont {He}, \citenamefont {Guo}, \citenamefont
  {Zhu}, \citenamefont {Zhang}, \citenamefont {Wang}, \citenamefont {Ma},\ and\
  \citenamefont {Yan}}]{b27}%
  \BibitemOpen
  \bibfield  {author} {\bibinfo {author} {\bibfnamefont {D.}~\bibnamefont
  {Wu}}, \bibinfo {author} {\bibfnamefont {H.~Y.}\ \bibnamefont {Lan}},
  \bibinfo {author} {\bibfnamefont {J.~Y.}\ \bibnamefont {Zhang}}, \bibinfo
  {author} {\bibfnamefont {J.~X.}\ \bibnamefont {Liu}}, \bibinfo {author}
  {\bibfnamefont {H.~G.}\ \bibnamefont {Lu}}, \bibinfo {author} {\bibfnamefont
  {J.~F.}\ \bibnamefont {Lv}}, \bibinfo {author} {\bibfnamefont {X.~Z.}\
  \bibnamefont {Wu}}, \bibinfo {author} {\bibfnamefont {H.}~\bibnamefont
  {Zhang}}, \bibinfo {author} {\bibfnamefont {J.}~\bibnamefont {Cai}}, \bibinfo
  {author} {\bibfnamefont {Q.~Y.}\ \bibnamefont {Ma}}, \bibinfo {author}
  {\bibfnamefont {Y.~H.}\ \bibnamefont {Xia}}, \bibinfo {author} {\bibfnamefont
  {Z.~N.}\ \bibnamefont {Wang}}, \bibinfo {author} {\bibfnamefont {M.~Z.}\
  \bibnamefont {Wang}}, \bibinfo {author} {\bibfnamefont {Z.~Y.}\ \bibnamefont
  {Yang}}, \bibinfo {author} {\bibfnamefont {X.~L.}\ \bibnamefont {Xu}},
  \bibinfo {author} {\bibfnamefont {Y.~X.}\ \bibnamefont {Geng}}, \bibinfo
  {author} {\bibfnamefont {Y.~Y.}\ \bibnamefont {Zhao}}, \bibinfo {author}
  {\bibfnamefont {C.}~\bibnamefont {Lin}}, \bibinfo {author} {\bibfnamefont
  {W.~J.}\ \bibnamefont {Ma}}, \bibinfo {author} {\bibfnamefont {J.~Q.}\
  \bibnamefont {Yu}}, \bibinfo {author} {\bibfnamefont {H.~R.}\ \bibnamefont
  {Wang}}, \bibinfo {author} {\bibfnamefont {F.~L.}\ \bibnamefont {Liu}},
  \bibinfo {author} {\bibfnamefont {C.~Y.}\ \bibnamefont {He}}, \bibinfo
  {author} {\bibfnamefont {B.}~\bibnamefont {Guo}}, \bibinfo {author}
  {\bibfnamefont {P.}~\bibnamefont {Zhu}}, \bibinfo {author} {\bibfnamefont
  {G.~Q.}\ \bibnamefont {Zhang}}, \bibinfo {author} {\bibfnamefont {N.~Y.}\
  \bibnamefont {Wang}}, \bibinfo {author} {\bibfnamefont {Y.~G.}\ \bibnamefont
  {Ma}},\ and\ \bibinfo {author} {\bibfnamefont {X.~Q.}\ \bibnamefont {Yan}},\
  }\bibfield  {title} {\bibinfo {title} {\ce{^{197}Au}$(\gamma,xn;x=1\sim9)$
  reaction cross section measurements using laser-driven ultra-intense
  $\gamma$-ray source},\ }\href
  {https://doi.org/https://doi.org/10.48550/arXiv.2209.13947} {\bibfield
  {journal} {\bibinfo  {journal} {arXiv:2209.13947 [nucl-ex, physics:physics]}\
  } (\bibinfo {year} {2023})}\BibitemShut {NoStop}%
\bibitem [{\citenamefont {Lan}\ \emph {et~al.}(2023)\citenamefont {Lan},
  \citenamefont {Wu}, \citenamefont {Liu}, \citenamefont {Zhang}, \citenamefont
  {Lu}, \citenamefont {Lv}, \citenamefont {Wu}, \citenamefont {Luo},\ and\
  \citenamefont {Yan}}]{b56}%
  \BibitemOpen
  \bibfield  {author} {\bibinfo {author} {\bibfnamefont {H.~Y.}\ \bibnamefont
  {Lan}}, \bibinfo {author} {\bibfnamefont {D.}~\bibnamefont {Wu}}, \bibinfo
  {author} {\bibfnamefont {J.~X.}\ \bibnamefont {Liu}}, \bibinfo {author}
  {\bibfnamefont {J.~Y.}\ \bibnamefont {Zhang}}, \bibinfo {author}
  {\bibfnamefont {H.~G.}\ \bibnamefont {Lu}}, \bibinfo {author} {\bibfnamefont
  {J.~F.}\ \bibnamefont {Lv}}, \bibinfo {author} {\bibfnamefont {X.~Z.}\
  \bibnamefont {Wu}}, \bibinfo {author} {\bibfnamefont {W.}~\bibnamefont
  {Luo}},\ and\ \bibinfo {author} {\bibfnamefont {X.~Q.}\ \bibnamefont {Yan}},\
  }\bibfield  {title} {\bibinfo {title} {Photonuclear production of nuclear
  isomers using bremsstrahlung induced by laser-wakefield electrons},\ }\href
  {https://doi.org/10.1007/s41365-023-01219-x} {\bibfield  {journal} {\bibinfo
  {journal} {Nuclear Science and Techniques}\ }\textbf {\bibinfo {volume}
  {34}},\ \bibinfo {pages} {74} (\bibinfo {year} {2023})}\BibitemShut {NoStop}%
\bibitem [{\citenamefont {Magill}\ \emph {et~al.}(2003)\citenamefont {Magill},
  \citenamefont {Schwoerer}, \citenamefont {Ewald}, \citenamefont {Galy},
  \citenamefont {Schenkel},\ and\ \citenamefont {Sauerbrey}}]{b30}%
  \BibitemOpen
  \bibfield  {author} {\bibinfo {author} {\bibfnamefont {J.}~\bibnamefont
  {Magill}}, \bibinfo {author} {\bibfnamefont {H.}~\bibnamefont {Schwoerer}},
  \bibinfo {author} {\bibfnamefont {F.}~\bibnamefont {Ewald}}, \bibinfo
  {author} {\bibfnamefont {J.}~\bibnamefont {Galy}}, \bibinfo {author}
  {\bibfnamefont {R.}~\bibnamefont {Schenkel}},\ and\ \bibinfo {author}
  {\bibfnamefont {R.}~\bibnamefont {Sauerbrey}},\ }\bibfield  {title} {\bibinfo
  {title} {Laser transmutation of iodine-129},\ }\href
  {https://doi.org/https://doi.org/10.1007/s00340-003-1306-4} {\bibfield
  {journal} {\bibinfo  {journal} {Applied Physics B}\ }\textbf {\bibinfo
  {volume} {77}},\ \bibinfo {pages} {387} (\bibinfo {year} {2003})}\BibitemShut
  {NoStop}%
\bibitem [{\citenamefont {Takashima}\ \emph {et~al.}(2004)\citenamefont
  {Takashima}, \citenamefont {Hasegawa}, \citenamefont {Nemoto},\ and\
  \citenamefont {Kato}}]{b58}%
  \BibitemOpen
  \bibfield  {author} {\bibinfo {author} {\bibfnamefont {R.}~\bibnamefont
  {Takashima}}, \bibinfo {author} {\bibfnamefont {S.}~\bibnamefont {Hasegawa}},
  \bibinfo {author} {\bibfnamefont {K.}~\bibnamefont {Nemoto}},\ and\ \bibinfo
  {author} {\bibfnamefont {K.}~\bibnamefont {Kato}},\ }\bibfield  {title}
  {\bibinfo {title} {Possibility of transmutation of \ce{^{135}Cs} by
  ultraintense laser},\ }\href {https://doi.org/10.1063/1.1847715} {\bibfield
  {journal} {\bibinfo  {journal} {Applied Physics Letters}\ }\textbf {\bibinfo
  {volume} {86}},\ \bibinfo {pages} {011501} (\bibinfo {year}
  {2004})}\BibitemShut {NoStop}%
\bibitem [{\citenamefont {Pomerantz}\ \emph {et~al.}(2014)\citenamefont
  {Pomerantz}, \citenamefont {McCary}, \citenamefont {Meadows}, \citenamefont
  {Arefiev}, \citenamefont {Bernstein}, \citenamefont {Chester}, \citenamefont
  {Cortez}, \citenamefont {Donovan}, \citenamefont {Dyer}, \citenamefont
  {Gaul}, \citenamefont {Hamilton}, \citenamefont {Kuk}, \citenamefont
  {Lestrade}, \citenamefont {Wang}, \citenamefont {Ditmire},\ and\
  \citenamefont {Hegelich}}]{b65}%
  \BibitemOpen
  \bibfield  {author} {\bibinfo {author} {\bibfnamefont {I.}~\bibnamefont
  {Pomerantz}}, \bibinfo {author} {\bibfnamefont {E.}~\bibnamefont {McCary}},
  \bibinfo {author} {\bibfnamefont {A.~R.}\ \bibnamefont {Meadows}}, \bibinfo
  {author} {\bibfnamefont {A.}~\bibnamefont {Arefiev}}, \bibinfo {author}
  {\bibfnamefont {A.~C.}\ \bibnamefont {Bernstein}}, \bibinfo {author}
  {\bibfnamefont {C.}~\bibnamefont {Chester}}, \bibinfo {author} {\bibfnamefont
  {J.}~\bibnamefont {Cortez}}, \bibinfo {author} {\bibfnamefont {M.~E.}\
  \bibnamefont {Donovan}}, \bibinfo {author} {\bibfnamefont {G.}~\bibnamefont
  {Dyer}}, \bibinfo {author} {\bibfnamefont {E.~W.}\ \bibnamefont {Gaul}},
  \bibinfo {author} {\bibfnamefont {D.}~\bibnamefont {Hamilton}}, \bibinfo
  {author} {\bibfnamefont {D.}~\bibnamefont {Kuk}}, \bibinfo {author}
  {\bibfnamefont {A.~C.}\ \bibnamefont {Lestrade}}, \bibinfo {author}
  {\bibfnamefont {C.}~\bibnamefont {Wang}}, \bibinfo {author} {\bibfnamefont
  {T.}~\bibnamefont {Ditmire}},\ and\ \bibinfo {author} {\bibfnamefont {B.~M.}\
  \bibnamefont {Hegelich}},\ }\bibfield  {title} {\bibinfo {title} {Ultrashort
  pulsed neutron source},\ }\href
  {https://doi.org/10.1103/PhysRevLett.113.184801} {\bibfield  {journal}
  {\bibinfo  {journal} {Phys. Rev. Lett.}\ }\textbf {\bibinfo {volume} {113}},\
  \bibinfo {pages} {184801} (\bibinfo {year} {2014})}\BibitemShut {NoStop}%
\bibitem [{\citenamefont {Günther}\ \emph {et~al.}(2022)\citenamefont
  {Günther}, \citenamefont {Rosmej}, \citenamefont {Tavana}, \citenamefont
  {Gyrdymov}, \citenamefont {Skobliakov}, \citenamefont {Kantsyrev},
  \citenamefont {Zähter}, \citenamefont {Borisenko}, \citenamefont {Pukhov},\
  and\ \citenamefont {Andreev}}]{b66}%
  \BibitemOpen
  \bibfield  {author} {\bibinfo {author} {\bibfnamefont {M.~M.}\ \bibnamefont
  {Günther}}, \bibinfo {author} {\bibfnamefont {O.~N.}\ \bibnamefont
  {Rosmej}}, \bibinfo {author} {\bibfnamefont {P.}~\bibnamefont {Tavana}},
  \bibinfo {author} {\bibfnamefont {M.}~\bibnamefont {Gyrdymov}}, \bibinfo
  {author} {\bibfnamefont {A.}~\bibnamefont {Skobliakov}}, \bibinfo {author}
  {\bibfnamefont {A.}~\bibnamefont {Kantsyrev}}, \bibinfo {author}
  {\bibfnamefont {{\c{S}}.}~\bibnamefont {Zähter}}, \bibinfo {author}
  {\bibfnamefont {N.~G.}\ \bibnamefont {Borisenko}}, \bibinfo {author}
  {\bibfnamefont {A.}~\bibnamefont {Pukhov}},\ and\ \bibinfo {author}
  {\bibfnamefont {N.~E.}\ \bibnamefont {Andreev}},\ }\bibfield  {title}
  {\bibinfo {title} {Forward-looking insights in laser-generated ultra-intense
  $\gamma$-ray and neutron sources for nuclear application and science},\
  }\href {https://doi.org/10.1038/s41467-021-27694-7} {\bibfield  {journal}
  {\bibinfo  {journal} {Nature Communications}\ }\textbf {\bibinfo {volume}
  {13}},\ \bibinfo {pages} {170} (\bibinfo {year} {2022})}\BibitemShut
  {NoStop}%
\bibitem [{\citenamefont {Tavana}\ \emph {et~al.}(2023)\citenamefont {Tavana},
  \citenamefont {Bukharskii}, \citenamefont {Gyrdymov}, \citenamefont
  {Spillmann}, \citenamefont {Zähter}, \citenamefont {Cikhardt}, \citenamefont
  {Borisenko}, \citenamefont {Korneev}, \citenamefont {Jacoby}, \citenamefont
  {Spielmann}, \citenamefont {Andreev}, \citenamefont {Günther},\ and\
  \citenamefont {Rosmej}}]{b29}%
  \BibitemOpen
  \bibfield  {author} {\bibinfo {author} {\bibfnamefont {P.}~\bibnamefont
  {Tavana}}, \bibinfo {author} {\bibfnamefont {N.}~\bibnamefont {Bukharskii}},
  \bibinfo {author} {\bibfnamefont {M.}~\bibnamefont {Gyrdymov}}, \bibinfo
  {author} {\bibfnamefont {U.}~\bibnamefont {Spillmann}}, \bibinfo {author}
  {\bibfnamefont {{\c{S}}.}~\bibnamefont {Zähter}}, \bibinfo {author}
  {\bibfnamefont {J.}~\bibnamefont {Cikhardt}}, \bibinfo {author}
  {\bibfnamefont {N.~G.}\ \bibnamefont {Borisenko}}, \bibinfo {author}
  {\bibfnamefont {P.}~\bibnamefont {Korneev}}, \bibinfo {author} {\bibfnamefont
  {J.}~\bibnamefont {Jacoby}}, \bibinfo {author} {\bibfnamefont
  {C.}~\bibnamefont {Spielmann}}, \bibinfo {author} {\bibfnamefont {N.~E.}\
  \bibnamefont {Andreev}}, \bibinfo {author} {\bibfnamefont {M.~M.}\
  \bibnamefont {Günther}},\ and\ \bibinfo {author} {\bibfnamefont {O.~N.}\
  \bibnamefont {Rosmej}},\ }\bibfield  {title} {\bibinfo {title} {Ultra-high
  efficiency bremsstrahlung production in the interaction of direct
  laser-accelerated electrons with high-z material},\ }\href
  {https://doi.org/10.3389/fphy.2023.1178967} {\bibfield  {journal} {\bibinfo
  {journal} {Frontiers in Physics}\ }\textbf {\bibinfo {volume} {11}} (\bibinfo
  {year} {2023})}\BibitemShut {NoStop}%
\bibitem [{\citenamefont {Ledingham}\ \emph {et~al.}(2000)\citenamefont
  {Ledingham}, \citenamefont {Spencer}, \citenamefont {McCanny}, \citenamefont
  {Singhal}, \citenamefont {Santala}, \citenamefont {Clark}, \citenamefont
  {Watts}, \citenamefont {Beg}, \citenamefont {Zepf}, \citenamefont
  {Krushelnick}, \citenamefont {Tatarakis}, \citenamefont {Dangor},
  \citenamefont {Norreys}, \citenamefont {Allott}, \citenamefont {Neely},
  \citenamefont {Clark}, \citenamefont {Machacek}, \citenamefont {Wark},
  \citenamefont {Cresswell}, \citenamefont {Sanderson},\ and\ \citenamefont
  {Magill}}]{b57}%
  \BibitemOpen
  \bibfield  {author} {\bibinfo {author} {\bibfnamefont {K.~W.~D.}\
  \bibnamefont {Ledingham}}, \bibinfo {author} {\bibfnamefont {I.}~\bibnamefont
  {Spencer}}, \bibinfo {author} {\bibfnamefont {T.}~\bibnamefont {McCanny}},
  \bibinfo {author} {\bibfnamefont {R.~P.}\ \bibnamefont {Singhal}}, \bibinfo
  {author} {\bibfnamefont {M.~I.~K.}\ \bibnamefont {Santala}}, \bibinfo
  {author} {\bibfnamefont {E.}~\bibnamefont {Clark}}, \bibinfo {author}
  {\bibfnamefont {I.}~\bibnamefont {Watts}}, \bibinfo {author} {\bibfnamefont
  {F.~N.}\ \bibnamefont {Beg}}, \bibinfo {author} {\bibfnamefont
  {M.}~\bibnamefont {Zepf}}, \bibinfo {author} {\bibfnamefont {K.}~\bibnamefont
  {Krushelnick}}, \bibinfo {author} {\bibfnamefont {M.}~\bibnamefont
  {Tatarakis}}, \bibinfo {author} {\bibfnamefont {A.~E.}\ \bibnamefont
  {Dangor}}, \bibinfo {author} {\bibfnamefont {P.~A.}\ \bibnamefont {Norreys}},
  \bibinfo {author} {\bibfnamefont {R.}~\bibnamefont {Allott}}, \bibinfo
  {author} {\bibfnamefont {D.}~\bibnamefont {Neely}}, \bibinfo {author}
  {\bibfnamefont {R.~J.}\ \bibnamefont {Clark}}, \bibinfo {author}
  {\bibfnamefont {A.~C.}\ \bibnamefont {Machacek}}, \bibinfo {author}
  {\bibfnamefont {J.~S.}\ \bibnamefont {Wark}}, \bibinfo {author}
  {\bibfnamefont {A.~J.}\ \bibnamefont {Cresswell}}, \bibinfo {author}
  {\bibfnamefont {D.~C.~W.}\ \bibnamefont {Sanderson}},\ and\ \bibinfo {author}
  {\bibfnamefont {J.}~\bibnamefont {Magill}},\ }\bibfield  {title} {\bibinfo
  {title} {Photonuclear physics when a multiterawatt laser pulse interacts with
  solid targets},\ }\href {https://doi.org/10.1103/PhysRevLett.84.899}
  {\bibfield  {journal} {\bibinfo  {journal} {Phys. Rev. Lett.}\ }\textbf
  {\bibinfo {volume} {84}},\ \bibinfo {pages} {899} (\bibinfo {year}
  {2000})}\BibitemShut {NoStop}%
\bibitem [{\citenamefont {Zhang}\ \emph {et~al.}(2024)\citenamefont {Zhang},
  \citenamefont {Wu}, \citenamefont {Lan}, \citenamefont {Lu}, \citenamefont
  {Liu}, \citenamefont {Lv}, \citenamefont {Wang},\ and\ \citenamefont
  {Yan}}]{b89}%
  \BibitemOpen
  \bibfield  {author} {\bibinfo {author} {\bibfnamefont {J.~Y.}\ \bibnamefont
  {Zhang}}, \bibinfo {author} {\bibfnamefont {D.}~\bibnamefont {Wu}}, \bibinfo
  {author} {\bibfnamefont {H.~Y.}\ \bibnamefont {Lan}}, \bibinfo {author}
  {\bibfnamefont {H.~G.}\ \bibnamefont {Lu}}, \bibinfo {author} {\bibfnamefont
  {J.~X.}\ \bibnamefont {Liu}}, \bibinfo {author} {\bibfnamefont {J.~F.}\
  \bibnamefont {Lv}}, \bibinfo {author} {\bibfnamefont {M.~Z.}\ \bibnamefont
  {Wang}},\ and\ \bibinfo {author} {\bibfnamefont {X.~Q.}\ \bibnamefont
  {Yan}},\ }\bibfield  {title} {\bibinfo {title} {Generation of medical
  isotopes \ce{^{47}Sc}, \ce{^{67}Cu} through laser-induced ($\gamma, p$)
  reaction},\ }\href {https://doi.org/10.1007/s41365-024-01550-x} {\bibfield
  {journal} {\bibinfo  {journal} {Nuclear Science and Techniques}\ }\textbf
  {\bibinfo {volume} {35}},\ \bibinfo {pages} {206} (\bibinfo {year}
  {2024})}\BibitemShut {NoStop}%
\bibitem [{\citenamefont {Brazzle}\ \emph {et~al.}(1999)\citenamefont
  {Brazzle}, \citenamefont {Pravdivtseva}, \citenamefont {Meshik},\ and\
  \citenamefont {Hohenberg}}]{b60}%
  \BibitemOpen
  \bibfield  {author} {\bibinfo {author} {\bibfnamefont {R.~H.}\ \bibnamefont
  {Brazzle}}, \bibinfo {author} {\bibfnamefont {O.~V.}\ \bibnamefont
  {Pravdivtseva}}, \bibinfo {author} {\bibfnamefont {A.~P.}\ \bibnamefont
  {Meshik}},\ and\ \bibinfo {author} {\bibfnamefont {C.~M.}\ \bibnamefont
  {Hohenberg}},\ }\bibfield  {title} {\bibinfo {title} {Verification and
  interpretation of the i-xe chronometer},\ }\href
  {https://doi.org/https://doi.org/10.1016/S0016-7037(98)00314-7} {\bibfield
  {journal} {\bibinfo  {journal} {Geochimica et Cosmochimica Acta}\ }\textbf
  {\bibinfo {volume} {63}},\ \bibinfo {pages} {739} (\bibinfo {year}
  {1999})}\BibitemShut {NoStop}%
\bibitem [{\citenamefont {Gary}\ \emph {et~al.}(2009)\citenamefont {Gary},
  \citenamefont {Bradley}, \citenamefont {Gopalan}, \citenamefont {Jitendra},\
  and\ \citenamefont {Sandeep}}]{b61}%
  \BibitemOpen
  \bibfield  {author} {\bibinfo {author} {\bibfnamefont {R.~H.}\ \bibnamefont
  {Gary}}, \bibinfo {author} {\bibfnamefont {S.~M.}\ \bibnamefont {Bradley}},
  \bibinfo {author} {\bibfnamefont {S.}~\bibnamefont {Gopalan}}, \bibinfo
  {author} {\bibfnamefont {N.~G.}\ \bibnamefont {Jitendra}},\ and\ \bibinfo
  {author} {\bibfnamefont {S.}~\bibnamefont {Sandeep}},\ }\bibfield  {title}
  {\bibinfo {title} {Stellar sources of the short-lived radionuclides in the
  early solar system},\ }\href
  {https://doi.org/https://doi.org/10.1016/j.gca.2009.01.039} {\bibfield
  {journal} {\bibinfo  {journal} {Geochimica et Cosmochimica Acta}\ }\textbf
  {\bibinfo {volume} {73}},\ \bibinfo {pages} {4922} (\bibinfo {year}
  {2009})},\ \bibinfo {note} {the Chronology of Meteorites and the Early Solar
  System}\BibitemShut {NoStop}%
\bibitem [{\citenamefont {Hou}\ \emph {et~al.}(2009)\citenamefont {Hou},
  \citenamefont {Hansen}, \citenamefont {Aldahan}, \citenamefont {Possnert},
  \citenamefont {Lind},\ and\ \citenamefont {Lujaniene}}]{b62}%
  \BibitemOpen
  \bibfield  {author} {\bibinfo {author} {\bibfnamefont {X.~L.}\ \bibnamefont
  {Hou}}, \bibinfo {author} {\bibfnamefont {V.}~\bibnamefont {Hansen}},
  \bibinfo {author} {\bibfnamefont {A.}~\bibnamefont {Aldahan}}, \bibinfo
  {author} {\bibfnamefont {G.}~\bibnamefont {Possnert}}, \bibinfo {author}
  {\bibfnamefont {O.~C.}\ \bibnamefont {Lind}},\ and\ \bibinfo {author}
  {\bibfnamefont {G.}~\bibnamefont {Lujaniene}},\ }\bibfield  {title} {\bibinfo
  {title} {A review on speciation of iodine-129 in the environmental and
  biological samples},\ }\href
  {https://doi.org/https://doi.org/10.1016/j.aca.2008.11.013} {\bibfield
  {journal} {\bibinfo  {journal} {Analytica Chimica Acta}\ }\textbf {\bibinfo
  {volume} {632}},\ \bibinfo {pages} {181} (\bibinfo {year}
  {2009})}\BibitemShut {NoStop}%
\bibitem [{\citenamefont {Rault}\ \emph {et~al.}(2007)\citenamefont {Rault},
  \citenamefont {Vandenberghe}, \citenamefont {Van~Holen}, \citenamefont
  {De~Beenhouwer}, \citenamefont {Staelens},\ and\ \citenamefont
  {Lemahieu}}]{b18}%
  \BibitemOpen
  \bibfield  {author} {\bibinfo {author} {\bibfnamefont {E.}~\bibnamefont
  {Rault}}, \bibinfo {author} {\bibfnamefont {S.}~\bibnamefont {Vandenberghe}},
  \bibinfo {author} {\bibfnamefont {R.}~\bibnamefont {Van~Holen}}, \bibinfo
  {author} {\bibfnamefont {J.}~\bibnamefont {De~Beenhouwer}}, \bibinfo {author}
  {\bibfnamefont {S.}~\bibnamefont {Staelens}},\ and\ \bibinfo {author}
  {\bibfnamefont {I.}~\bibnamefont {Lemahieu}},\ }\bibfield  {title} {\bibinfo
  {title} {Comparison of image quality of different iodine isotopes
  (\ce{^{123}I}, \ce{^{124}I}, and \ce{^{131}I})},\ }\href
  {https://doi.org/10.1089/cbr.2006.323} {\bibfield  {journal} {\bibinfo
  {journal} {Cancer Biotherapy and Radiopharmaceuticals}\ }\textbf {\bibinfo
  {volume} {22}},\ \bibinfo {pages} {423} (\bibinfo {year} {2007})}\BibitemShut
  {NoStop}%
\bibitem [{\citenamefont {Hertz}\ and\ \citenamefont {Schuller}(2010)}]{b19}%
  \BibitemOpen
  \bibfield  {author} {\bibinfo {author} {\bibfnamefont {B.~E.}\ \bibnamefont
  {Hertz}}\ and\ \bibinfo {author} {\bibfnamefont {K.~E.}\ \bibnamefont
  {Schuller}},\ }\bibfield  {title} {\bibinfo {title} {Saul hertz, md
  (1905-1950): A pioneer in the use of radioactive iodine},\ }\href
  {https://doi.org/10.4158/EP10065.CO} {\bibfield  {journal} {\bibinfo
  {journal} {Endocrine Practice}\ }\textbf {\bibinfo {volume} {16}},\ \bibinfo
  {pages} {713} (\bibinfo {year} {2010})}\BibitemShut {NoStop}%
\bibitem [{\citenamefont {Green}\ \emph {et~al.}(2017)\citenamefont {Green},
  \citenamefont {Lowe}, \citenamefont {Kadirvel}, \citenamefont {McMahon},
  \citenamefont {Westwood}, \citenamefont {Chua},\ and\ \citenamefont
  {Brown}}]{b48}%
  \BibitemOpen
  \bibfield  {author} {\bibinfo {author} {\bibfnamefont {M.}~\bibnamefont
  {Green}}, \bibinfo {author} {\bibfnamefont {J.}~\bibnamefont {Lowe}},
  \bibinfo {author} {\bibfnamefont {M.}~\bibnamefont {Kadirvel}}, \bibinfo
  {author} {\bibfnamefont {A.}~\bibnamefont {McMahon}}, \bibinfo {author}
  {\bibfnamefont {N.}~\bibnamefont {Westwood}}, \bibinfo {author}
  {\bibfnamefont {S.}~\bibnamefont {Chua}},\ and\ \bibinfo {author}
  {\bibfnamefont {G.}~\bibnamefont {Brown}},\ }\bibfield  {title} {\bibinfo
  {title} {Radiosynthesis of no-carrier-added \textit{meta}-[\ce{^{124}I}]
  iodobenzylguanidine for pet imaging of metastatic neuroblastoma},\ }\href
  {https://doi.org/10.1007/s10967-016-5073-1} {\bibfield  {journal} {\bibinfo
  {journal} {Journal of Radioanalytical and Nuclear Chemistry}\ }\textbf
  {\bibinfo {volume} {311}},\ \bibinfo {pages} {727} (\bibinfo {year}
  {2017})}\BibitemShut {NoStop}%
\bibitem [{\citenamefont {Herzog}\ \emph {et~al.}(2002)\citenamefont {Herzog},
  \citenamefont {Tellmann}, \citenamefont {Qaim}, \citenamefont {Spellerberg},
  \citenamefont {Schmid},\ and\ \citenamefont {Coenen}}]{b53}%
  \BibitemOpen
  \bibfield  {author} {\bibinfo {author} {\bibfnamefont {H.}~\bibnamefont
  {Herzog}}, \bibinfo {author} {\bibfnamefont {L.}~\bibnamefont {Tellmann}},
  \bibinfo {author} {\bibfnamefont {S.~M.}\ \bibnamefont {Qaim}}, \bibinfo
  {author} {\bibfnamefont {S.}~\bibnamefont {Spellerberg}}, \bibinfo {author}
  {\bibfnamefont {A.}~\bibnamefont {Schmid}},\ and\ \bibinfo {author}
  {\bibfnamefont {H.~H.}\ \bibnamefont {Coenen}},\ }\bibfield  {title}
  {\bibinfo {title} {{PET} quantitation and imaging of the non-pure
  positron-emitting iodine isotope \ce{^{124}I}},\ }\href
  {https://doi.org/https://doi.org/10.1016/S0969-8043(01)00283-4} {\bibfield
  {journal} {\bibinfo  {journal} {Applied Radiation and Isotopes}\ }\textbf
  {\bibinfo {volume} {56}},\ \bibinfo {pages} {673} (\bibinfo {year}
  {2002})}\BibitemShut {NoStop}%
\bibitem [{\citenamefont {Schmitz}(2011)}]{b55}%
  \BibitemOpen
  \bibfield  {author} {\bibinfo {author} {\bibfnamefont {J.}~\bibnamefont
  {Schmitz}},\ }\bibfield  {title} {\bibinfo {title} {The production of
  [\ce{^{124}I}]iodine and [\ce{^{86}Y}]yttrium},\ }\href
  {https://doi.org/10.1007/s00259-011-1782-4} {\bibfield  {journal} {\bibinfo
  {journal} {European Journal of Nuclear Medicine and Molecular Imaging}\
  }\textbf {\bibinfo {volume} {38}},\ \bibinfo {pages} {4} (\bibinfo {year}
  {2011})}\BibitemShut {NoStop}%
\bibitem [{\citenamefont {Robbins}\ and\ \citenamefont
  {Schneider}(2000)}]{b20}%
  \BibitemOpen
  \bibfield  {author} {\bibinfo {author} {\bibfnamefont {J.}~\bibnamefont
  {Robbins}}\ and\ \bibinfo {author} {\bibfnamefont {A.~B.}\ \bibnamefont
  {Schneider}},\ }\bibfield  {title} {\bibinfo {title} {Thyroid cancer
  following exposure to radioactive iodine},\ }\href
  {https://doi.org/10.1023/A:1010031115233} {\bibfield  {journal} {\bibinfo
  {journal} {Reviews in Endocrine and Metabolic Disorders}\ }\textbf {\bibinfo
  {volume} {1}},\ \bibinfo {pages} {197} (\bibinfo {year} {2000})}\BibitemShut
  {NoStop}%
\bibitem [{\citenamefont {Ferris}\ \emph {et~al.}(2021)\citenamefont {Ferris},
  \citenamefont {Carroll}, \citenamefont {Jenner},\ and\ \citenamefont
  {Aboagye}}]{b86}%
  \BibitemOpen
  \bibfield  {author} {\bibinfo {author} {\bibfnamefont {T.}~\bibnamefont
  {Ferris}}, \bibinfo {author} {\bibfnamefont {L.}~\bibnamefont {Carroll}},
  \bibinfo {author} {\bibfnamefont {S.}~\bibnamefont {Jenner}},\ and\ \bibinfo
  {author} {\bibfnamefont {E.~O.}\ \bibnamefont {Aboagye}},\ }\bibfield
  {title} {\bibinfo {title} {Use of radioiodine in nuclear medicine—a brief
  overview},\ }\href {https://doi.org/https://doi.org/10.1002/jlcr.3891}
  {\bibfield  {journal} {\bibinfo  {journal} {Journal of Labelled Compounds and
  Radiopharmaceuticals}\ }\textbf {\bibinfo {volume} {64}},\ \bibinfo {pages}
  {92} (\bibinfo {year} {2021})}\BibitemShut {NoStop}%
\bibitem [{\citenamefont {Park}(2002)}]{b21}%
  \BibitemOpen
  \bibfield  {author} {\bibinfo {author} {\bibfnamefont {H.~M.}\ \bibnamefont
  {Park}},\ }\bibfield  {title} {\bibinfo {title} {\ce{^{123}I}: Almost a
  designer radioiodine for thyroid scanning},\ }\href
  {https://jnm.snmjournals.org/content/43/1/77} {\bibfield  {journal} {\bibinfo
   {journal} {Journal of Nuclear Medicine}\ }\textbf {\bibinfo {volume} {43}},\
  \bibinfo {pages} {77} (\bibinfo {year} {2002})}\BibitemShut {NoStop}%
\bibitem [{\citenamefont {GERARD}\ and\ \citenamefont {CAVALIERI}(2002)}]{b51}%
  \BibitemOpen
  \bibfield  {author} {\bibinfo {author} {\bibfnamefont {S.~K.}\ \bibnamefont
  {GERARD}}\ and\ \bibinfo {author} {\bibfnamefont {R.~R.}\ \bibnamefont
  {CAVALIERI}},\ }\bibfield  {title} {\bibinfo {title} {I-123 diagnostic
  thyroid tumor whole-body scanning with imaging at 6, 24, and 48 hours},\
  }\href {https://doi.org/10.1097/00003072-200201000-00001} {\bibfield
  {journal} {\bibinfo  {journal} {Clinical Nuclear Medicinee}\ }\textbf
  {\bibinfo {volume} {27}},\ \bibinfo {pages} {1} (\bibinfo {year}
  {2002})}\BibitemShut {NoStop}%
\bibitem [{\citenamefont {Qaim}(2004)}]{b32}%
  \BibitemOpen
  \bibfield  {author} {\bibinfo {author} {\bibfnamefont {S.}~\bibnamefont
  {Qaim}},\ }\bibfield  {title} {\bibinfo {title} {Use of cyclotrons in
  medicine},\ }\href
  {https://doi.org/https://doi.org/10.1016/j.radphyschem.2004.04.124}
  {\bibfield  {journal} {\bibinfo  {journal} {Radiation Physics and Chemistry}\
  }\textbf {\bibinfo {volume} {71}},\ \bibinfo {pages} {917} (\bibinfo {year}
  {2004})}\BibitemShut {NoStop}%
\bibitem [{\citenamefont {Artun}\ and\ \citenamefont {Aytekin}(2015)}]{b33}%
  \BibitemOpen
  \bibfield  {author} {\bibinfo {author} {\bibfnamefont {O.}~\bibnamefont
  {Artun}}\ and\ \bibinfo {author} {\bibfnamefont {H.}~\bibnamefont
  {Aytekin}},\ }\bibfield  {title} {\bibinfo {title} {Calculation of excitation
  functions of proton, alpha and deuteron induced reactions for production of
  medical radioisotopes \ce{^{122–125}I}},\ }\href
  {https://doi.org/https://doi.org/10.1016/j.nimb.2014.12.029} {\bibfield
  {journal} {\bibinfo  {journal} {Nuclear Instruments and Methods in Physics
  Research Section B: Beam Interactions with Materials and Atoms}\ }\textbf
  {\bibinfo {volume} {345}},\ \bibinfo {pages} {1} (\bibinfo {year}
  {2015})}\BibitemShut {NoStop}%
\bibitem [{\citenamefont {Canbula}\ and\ \citenamefont {Canbula}(2023)}]{b49}%
  \BibitemOpen
  \bibfield  {author} {\bibinfo {author} {\bibfnamefont {D.}~\bibnamefont
  {Canbula}}\ and\ \bibinfo {author} {\bibfnamefont {B.}~\bibnamefont
  {Canbula}},\ }\bibfield  {title} {\bibinfo {title} {Cross-section
  calculations for the production of \ce{^{123}I} and \ce{^{124}I}
  radioisotopes via (p,n) and (p,2n) reactions using collective nuclear level
  density model},\ }\href {https://doi.org/10.1080/00295450.2022.2163802}
  {\bibfield  {journal} {\bibinfo  {journal} {Nuclear Technology}\ }\textbf
  {\bibinfo {volume} {209}},\ \bibinfo {pages} {895} (\bibinfo {year}
  {2023})}\BibitemShut {NoStop}%
\bibitem [{\citenamefont {Bzowski}\ \emph {et~al.}(2022)\citenamefont
  {Bzowski}, \citenamefont {Borys}, \citenamefont {Gorczewski}, \citenamefont
  {Chmura}, \citenamefont {Daszewska}, \citenamefont {Gorczewska},
  \citenamefont {Kastelik-Hryniewiecka}, \citenamefont {Szydło}, \citenamefont
  {d’Amico},\ and\ \citenamefont {Sokół}}]{b50}%
  \BibitemOpen
  \bibfield  {author} {\bibinfo {author} {\bibfnamefont {P.}~\bibnamefont
  {Bzowski}}, \bibinfo {author} {\bibfnamefont {D.}~\bibnamefont {Borys}},
  \bibinfo {author} {\bibfnamefont {K.}~\bibnamefont {Gorczewski}}, \bibinfo
  {author} {\bibfnamefont {A.}~\bibnamefont {Chmura}}, \bibinfo {author}
  {\bibfnamefont {K.}~\bibnamefont {Daszewska}}, \bibinfo {author}
  {\bibfnamefont {I.}~\bibnamefont {Gorczewska}}, \bibinfo {author}
  {\bibfnamefont {A.}~\bibnamefont {Kastelik-Hryniewiecka}}, \bibinfo {author}
  {\bibfnamefont {M.}~\bibnamefont {Szydło}}, \bibinfo {author} {\bibfnamefont
  {A.}~\bibnamefont {d’Amico}},\ and\ \bibinfo {author} {\bibfnamefont
  {M.}~\bibnamefont {Sokół}},\ }\bibfield  {title} {\bibinfo {title}
  {Efficiency of \ce{^{124}I} radioisotope production from natural and enriched
  tellurium dioxide using \ce{^{124}Te}(p,xn)\ce{^{124}I} reaction},\ }\href
  {https://doi.org/10.1186/s40658-022-00471-1} {\bibfield  {journal} {\bibinfo
  {journal} {EJNMMI Physics}\ }\textbf {\bibinfo {volume} {9}},\ \bibinfo
  {pages} {41} (\bibinfo {year} {2022})}\BibitemShut {NoStop}%
\bibitem [{\citenamefont {Sheh}\ \emph {et~al.}(2000)\citenamefont {Sheh},
  \citenamefont {Koziorowski}, \citenamefont {Balatoni}, \citenamefont {Lom},
  \citenamefont {Dahl},\ and\ \citenamefont {Finn}}]{b52}%
  \BibitemOpen
  \bibfield  {author} {\bibinfo {author} {\bibfnamefont {Y.}~\bibnamefont
  {Sheh}}, \bibinfo {author} {\bibfnamefont {J.}~\bibnamefont {Koziorowski}},
  \bibinfo {author} {\bibfnamefont {J.}~\bibnamefont {Balatoni}}, \bibinfo
  {author} {\bibfnamefont {C.}~\bibnamefont {Lom}}, \bibinfo {author}
  {\bibfnamefont {J.~R.}\ \bibnamefont {Dahl}},\ and\ \bibinfo {author}
  {\bibfnamefont {R.~D.}\ \bibnamefont {Finn}},\ }\bibfield  {title} {\bibinfo
  {title} {Low energy cyclotron production and chemical separation of "no
  carrier added" iodine-124 from a reusable, enriched tellurium-124
  dioxide/aluminum oxide solid solution target},\ }\href
  {https://doi.org/doi:10.1524/ract.2000.88.3-4.169} {\bibfield  {journal}
  {\bibinfo  {journal} {Radiochimica Acta}\ }\textbf {\bibinfo {volume} {88}},\
  \bibinfo {pages} {169} (\bibinfo {year} {2000})}\BibitemShut {NoStop}%
\bibitem [{\citenamefont {Wellman}\ \emph {et~al.}(1971)\citenamefont
  {Wellman}, \citenamefont {Sodd},\ and\ \citenamefont {Mack}}]{b54}%
  \BibitemOpen
  \bibfield  {author} {\bibinfo {author} {\bibfnamefont {H.~N.}\ \bibnamefont
  {Wellman}}, \bibinfo {author} {\bibfnamefont {V.~J.}\ \bibnamefont {Sodd}},\
  and\ \bibinfo {author} {\bibfnamefont {J.~F.}\ \bibnamefont {Mack}},\
  }\bibfield  {title} {\bibinfo {title} {Production and clinical development of
  a new ideal radioisotope of iodine\textemdash iodine-123},\ }\href
  {https://doi.org/10.1007/978-3-642-65092-5_3} {\bibfield  {journal} {\bibinfo
   {journal} {Frontiers of Nuclear Medicine/Aktuelle Nuklearmedizin}\ ,\
  \bibinfo {pages} {19}} (\bibinfo {year} {1971})}\BibitemShut {NoStop}%
\bibitem [{\citenamefont {Avetisyan}\ \emph {et~al.}(2017)\citenamefont
  {Avetisyan}, \citenamefont {Avagyan}, \citenamefont {Dallakyan},
  \citenamefont {Avdalyan}, \citenamefont {Dobrovolsky}, \citenamefont
  {Gavalyan}, \citenamefont {Kerobyan},\ and\ \citenamefont
  {Harutyunyan}}]{b34}%
  \BibitemOpen
  \bibfield  {author} {\bibinfo {author} {\bibfnamefont {A.}~\bibnamefont
  {Avetisyan}}, \bibinfo {author} {\bibfnamefont {R.}~\bibnamefont {Avagyan}},
  \bibinfo {author} {\bibfnamefont {R.}~\bibnamefont {Dallakyan}}, \bibinfo
  {author} {\bibfnamefont {G.}~\bibnamefont {Avdalyan}}, \bibinfo {author}
  {\bibfnamefont {N.}~\bibnamefont {Dobrovolsky}}, \bibinfo {author}
  {\bibfnamefont {V.}~\bibnamefont {Gavalyan}}, \bibinfo {author}
  {\bibfnamefont {I.}~\bibnamefont {Kerobyan}},\ and\ \bibinfo {author}
  {\bibfnamefont {G.}~\bibnamefont {Harutyunyan}},\ }\bibfield  {title}
  {\bibinfo {title} {Investigation of \ce{^{123}I} production using electron
  accelerator},\ }\href
  {https://doi.org/https://doi.org/10.1016/j.nucmedbio.2016.12.007} {\bibfield
  {journal} {\bibinfo  {journal} {Nuclear Medicine and Biology}\ }\textbf
  {\bibinfo {volume} {47}},\ \bibinfo {pages} {44} (\bibinfo {year}
  {2017})}\BibitemShut {NoStop}%
\bibitem [{\citenamefont {Naik}\ \emph {et~al.}(2020)\citenamefont {Naik},
  \citenamefont {Kim}, \citenamefont {Kim}, \citenamefont {Zaman},\ and\
  \citenamefont {Nguyen}}]{b10}%
  \BibitemOpen
  \bibfield  {author} {\bibinfo {author} {\bibfnamefont {H.}~\bibnamefont
  {Naik}}, \bibinfo {author} {\bibfnamefont {G.}~\bibnamefont {Kim}}, \bibinfo
  {author} {\bibfnamefont {K.}~\bibnamefont {Kim}}, \bibinfo {author}
  {\bibfnamefont {M.}~\bibnamefont {Zaman}},\ and\ \bibinfo {author}
  {\bibfnamefont {T.~H.}\ \bibnamefont {Nguyen}},\ }\bibfield  {title}
  {\bibinfo {title} {Flux-weighted average cross sections of medical isotopes
  in the \ce{^{127}I}$(\gamma,xn)$ reactions with the bremsstrahlung end-point
  energies of 50 and 70 {MeV}},\ }\href
  {https://doi.org/https://doi.org/10.1016/j.apradiso.2019.108842} {\bibfield
  {journal} {\bibinfo  {journal} {Applied Radiation and Isotopes}\ }\textbf
  {\bibinfo {volume} {156}},\ \bibinfo {pages} {108842} (\bibinfo {year}
  {2020})}\BibitemShut {NoStop}%
\bibitem [{\citenamefont {Jonsson}\ and\ \citenamefont {Lindgren}(1970)}]{b22}%
  \BibitemOpen
  \bibfield  {author} {\bibinfo {author} {\bibfnamefont {G.}~\bibnamefont
  {Jonsson}}\ and\ \bibinfo {author} {\bibfnamefont {K.}~\bibnamefont
  {Lindgren}},\ }\bibfield  {title} {\bibinfo {title} {Pion effects in
  \ce{^{127}I} $(\gamma, xn)$ reactions of high multiplicity},\ }\href
  {https://doi.org/https://doi.org/10.1016/0375-9474(70)90851-1} {\bibfield
  {journal} {\bibinfo  {journal} {Nuclear Physics A}\ }\textbf {\bibinfo
  {volume} {141}},\ \bibinfo {pages} {355} (\bibinfo {year}
  {1970})}\BibitemShut {NoStop}%
\bibitem [{\citenamefont {Allen}\ \emph {et~al.}(1989)\citenamefont {Allen},
  \citenamefont {Boyd}, \citenamefont {Callin}, \citenamefont {Deruyter},
  \citenamefont {Eppley}, \citenamefont {Fant}, \citenamefont {Fowkes},
  \citenamefont {Haimson}, \citenamefont {Hoag}, \citenamefont {Hopkins},
  \citenamefont {Houck}, \citenamefont {Koontz}, \citenamefont {Lavine},
  \citenamefont {Loew}, \citenamefont {Mecklenburg}, \citenamefont {Miller},
  \citenamefont {Ruth}, \citenamefont {Ryne}, \citenamefont {Sessler},
  \citenamefont {Vlieks}, \citenamefont {Wang}, \citenamefont {Westenskow},\
  and\ \citenamefont {Yu}}]{b79}%
  \BibitemOpen
  \bibfield  {author} {\bibinfo {author} {\bibfnamefont {M.~A.}\ \bibnamefont
  {Allen}}, \bibinfo {author} {\bibfnamefont {J.~K.}\ \bibnamefont {Boyd}},
  \bibinfo {author} {\bibfnamefont {R.~S.}\ \bibnamefont {Callin}}, \bibinfo
  {author} {\bibfnamefont {H.}~\bibnamefont {Deruyter}}, \bibinfo {author}
  {\bibfnamefont {K.~R.}\ \bibnamefont {Eppley}}, \bibinfo {author}
  {\bibfnamefont {K.~S.}\ \bibnamefont {Fant}}, \bibinfo {author}
  {\bibfnamefont {W.~R.}\ \bibnamefont {Fowkes}}, \bibinfo {author}
  {\bibfnamefont {J.}~\bibnamefont {Haimson}}, \bibinfo {author} {\bibfnamefont
  {H.~A.}\ \bibnamefont {Hoag}}, \bibinfo {author} {\bibfnamefont {D.~B.}\
  \bibnamefont {Hopkins}}, \bibinfo {author} {\bibfnamefont {T.}~\bibnamefont
  {Houck}}, \bibinfo {author} {\bibfnamefont {R.~F.}\ \bibnamefont {Koontz}},
  \bibinfo {author} {\bibfnamefont {T.~L.}\ \bibnamefont {Lavine}}, \bibinfo
  {author} {\bibfnamefont {G.~A.}\ \bibnamefont {Loew}}, \bibinfo {author}
  {\bibfnamefont {B.}~\bibnamefont {Mecklenburg}}, \bibinfo {author}
  {\bibfnamefont {R.~H.}\ \bibnamefont {Miller}}, \bibinfo {author}
  {\bibfnamefont {R.~D.}\ \bibnamefont {Ruth}}, \bibinfo {author}
  {\bibfnamefont {R.~D.}\ \bibnamefont {Ryne}}, \bibinfo {author}
  {\bibfnamefont {A.~M.}\ \bibnamefont {Sessler}}, \bibinfo {author}
  {\bibfnamefont {A.~E.}\ \bibnamefont {Vlieks}}, \bibinfo {author}
  {\bibfnamefont {J.~W.}\ \bibnamefont {Wang}}, \bibinfo {author}
  {\bibfnamefont {G.~A.}\ \bibnamefont {Westenskow}},\ and\ \bibinfo {author}
  {\bibfnamefont {S.~S.}\ \bibnamefont {Yu}},\ }\bibfield  {title} {\bibinfo
  {title} {High-gradient electron accelerator powered by a relativisitic
  klystron},\ }\href {https://doi.org/10.1103/PhysRevLett.63.2472} {\bibfield
  {journal} {\bibinfo  {journal} {Phys. Rev. Lett.}\ }\textbf {\bibinfo
  {volume} {63}},\ \bibinfo {pages} {2472} (\bibinfo {year}
  {1989})}\BibitemShut {NoStop}%
\bibitem [{\citenamefont {Bayram}\ and\ \citenamefont {Akkoyun}(2018)}]{b91}%
  \BibitemOpen
  \bibfield  {author} {\bibinfo {author} {\bibfnamefont {T.}~\bibnamefont
  {Bayram}}\ and\ \bibinfo {author} {\bibfnamefont {S.}~\bibnamefont
  {Akkoyun}},\ }\bibfield  {title} {\bibinfo {title} {Determinations of
  \ce{^{171}Er} half-life and some \ce{^{171}Tm} transition energies},\ }\href
  {https://doi.org/10.1007/s41365-018-0378-0} {\bibfield  {journal} {\bibinfo
  {journal} {Nuclear Science and Techniques}\ }\textbf {\bibinfo {volume}
  {29}},\ \bibinfo {pages} {39} (\bibinfo {year} {2018})}\BibitemShut {NoStop}%
\bibitem [{\citenamefont {Li}\ \emph {et~al.}(2022)\citenamefont {Li},
  \citenamefont {Qin}, \citenamefont {Zhang}, \citenamefont {Li}, \citenamefont
  {Fan}, \citenamefont {Wang}, \citenamefont {Xu}, \citenamefont {Wang},
  \citenamefont {Yu}, \citenamefont {Xu},\ and\ \citenamefont {\textit{et
  al.}}}]{b35}%
  \BibitemOpen
  \bibfield  {author} {\bibinfo {author} {\bibfnamefont {A.~X.}\ \bibnamefont
  {Li}}, \bibinfo {author} {\bibfnamefont {C.~Y.}\ \bibnamefont {Qin}},
  \bibinfo {author} {\bibfnamefont {H.}~\bibnamefont {Zhang}}, \bibinfo
  {author} {\bibfnamefont {S.}~\bibnamefont {Li}}, \bibinfo {author}
  {\bibfnamefont {L.~L.}\ \bibnamefont {Fan}}, \bibinfo {author} {\bibfnamefont
  {Q.~S.}\ \bibnamefont {Wang}}, \bibinfo {author} {\bibfnamefont {T.~J.}\
  \bibnamefont {Xu}}, \bibinfo {author} {\bibfnamefont {N.~W.}\ \bibnamefont
  {Wang}}, \bibinfo {author} {\bibfnamefont {L.~H.}\ \bibnamefont {Yu}},
  \bibinfo {author} {\bibfnamefont {Y.}~\bibnamefont {Xu}},\ and\ \bibinfo
  {author} {\bibnamefont {\textit{et al.}}},\ }\bibfield  {title} {\bibinfo
  {title} {Acceleration of 60 {MeV} proton beams in the commissioning
  experiment of the sulf-10 {PW} laser},\ }\href
  {https://doi.org/10.1017/hpl.2022.17} {\bibfield  {journal} {\bibinfo
  {journal} {High Power Laser Science and Engineering}\ }\textbf {\bibinfo
  {volume} {10}},\ \bibinfo {pages} {e26} (\bibinfo {year} {2022})}\BibitemShut
  {NoStop}%
\bibitem [{\citenamefont {ORTEC}(2025{\natexlab{a}})}]{b39}%
  \BibitemOpen
  \bibfield  {author} {\bibinfo {author} {\bibnamefont {ORTEC}},\ }\href
  {https://www.ortec-online.com/products/radiation-detectors/high-purity-germanium-hpge-radiation-detectors/hpge-radiation-detector-types-how-choose/gmx-n-type-coaxial-hpge-radiation-detectors}
  {\emph {\bibinfo {title} {Gamma-X (GMX) N-type High Purity Germanium (HPGe)
  Coaxial Radiation Detectors}}} (\bibinfo {year}
  {2025}{\natexlab{a}})\BibitemShut {NoStop}%
\bibitem [{\citenamefont {{IEEE}}(1997)}]{b40}%
  \BibitemOpen
  \bibfield  {author} {\bibinfo {author} {\bibnamefont {{IEEE}}},\ }\bibfield
  {title} {\bibinfo {title} {{IEEE} standard test procedures for germanium
  gamma-ray detectors},\ }\href
  {https://doi.org/https://doi.org/10.1109/IEEESTD.1997.82400} {\bibfield
  {journal} {\bibinfo  {journal} {IEEE Std 325-1996}\ } (\bibinfo {year}
  {1997})}\BibitemShut {NoStop}%
\bibitem [{\citenamefont {ORTEC}(2025{\natexlab{b}})}]{b41}%
  \BibitemOpen
  \bibfield  {author} {\bibinfo {author} {\bibnamefont {ORTEC}},\ }\href
  {https://www.ortec-online.com/products/electronic-instruments/multi-channel-analyzers/workstation/dspec-50}
  {\emph {\bibinfo {title} {DSPEC 50/50A and DSPEC 502/502A Digital Signal
  Processing Gamma Spectrometers}}} (\bibinfo {year}
  {2025}{\natexlab{b}})\BibitemShut {NoStop}%
\bibitem [{\citenamefont {ORTEC}(2025{\natexlab{c}})}]{b42}%
  \BibitemOpen
  \bibfield  {author} {\bibinfo {author} {\bibnamefont {ORTEC}},\ }\href
  {https://www.ortec-online.com/products/software/gammavision} {\emph {\bibinfo
  {title} {GammaVision Gamma Spectroscopy}}} (\bibinfo {year}
  {2025}{\natexlab{c}})\BibitemShut {NoStop}%
\bibitem [{\citenamefont {Sebastián}\ \emph {et~al.}(2022)\citenamefont
  {Sebastián}, \citenamefont {Roque},\ and\ \citenamefont {Tania}}]{b80}%
  \BibitemOpen
  \bibfield  {author} {\bibinfo {author} {\bibfnamefont {S.}~\bibnamefont
  {Sebastián}}, \bibinfo {author} {\bibfnamefont {S.}~\bibnamefont {Roque}},\
  and\ \bibinfo {author} {\bibfnamefont {B.}~\bibnamefont {Tania}},\ }\bibfield
   {title} {\bibinfo {title} {Simulation of a hpge detector with geant4},\
  }\href {https://doi.org/http://dx.doi.org/10.33333/rp.vol50n2.01} {\bibfield
  {journal} {\bibinfo  {journal} {Revista Politécnica}\ }\textbf {\bibinfo
  {volume} {50}},\ \bibinfo {pages} {7} (\bibinfo {year} {2022})}\BibitemShut
  {NoStop}%
\bibitem [{\citenamefont {Agostinelli}\ \emph {et~al.}(2003)\citenamefont
  {Agostinelli}, \citenamefont {{Allison}}, \citenamefont {Amako},
  \citenamefont {Apostolakis}, \citenamefont {Araujo}, \citenamefont {Arce},
  \citenamefont {Asai}, \citenamefont {Axen}, \citenamefont {Banerjee},
  \citenamefont {Barrand}, \citenamefont {Behner}, \citenamefont {Bellagamba},
  \citenamefont {Boudreau}, \citenamefont {Broglia}, \citenamefont {Brunengo},
  \citenamefont {Burkhardt}, \citenamefont {Chauvie}, \citenamefont {Chuma},
  \citenamefont {Chytracek}, \citenamefont {Cooperman}, \citenamefont {Cosmo},
  \citenamefont {Degtyarenko}, \citenamefont {Dell'Acqua}, \citenamefont
  {Depaola}, \citenamefont {Dietrich}, \citenamefont {Enami}, \citenamefont
  {Feliciello}, \citenamefont {Ferguson}, \citenamefont {Fesefeldt},
  \citenamefont {Folger}, \citenamefont {Foppiano}, \citenamefont {Forti},
  \citenamefont {Garelli}, \citenamefont {Giani}, \citenamefont
  {Giannitrapani}, \citenamefont {Gibin}, \citenamefont {{Gómez Cadenas}},
  \citenamefont {González}, \citenamefont {{Gracia Abril}}, \citenamefont
  {Greeniaus}, \citenamefont {Greiner}, \citenamefont {Grichine}, \citenamefont
  {Grossheim}, \citenamefont {Guatelli}, \citenamefont {Gumplinger},
  \citenamefont {Hamatsu}, \citenamefont {Hashimoto}, \citenamefont {Hasui},
  \citenamefont {Heikkinen}, \citenamefont {Howard}, \citenamefont
  {Ivanchenko}, \citenamefont {Johnson}, \citenamefont {Jones}, \citenamefont
  {Kallenbach}, \citenamefont {Kanaya}, \citenamefont {Kawabata}, \citenamefont
  {Kawabata}, \citenamefont {Kawaguti}, \citenamefont {Kelner}, \citenamefont
  {Kent}, \citenamefont {Kimura}, \citenamefont {Kodama}, \citenamefont
  {Kokoulin}, \citenamefont {Kossov}, \citenamefont {Kurashige}, \citenamefont
  {Lamanna}, \citenamefont {Lampén}, \citenamefont {Lara}, \citenamefont
  {Lefebure}, \citenamefont {Lei}, \citenamefont {Liendl}, \citenamefont
  {Lockman}, \citenamefont {Longo}, \citenamefont {Magni}, \citenamefont
  {Maire}, \citenamefont {Medernach}, \citenamefont {Minamimoto}, \citenamefont
  {{Mora de Freitas}}, \citenamefont {Morita}, \citenamefont {Murakami},
  \citenamefont {Nagamatu}, \citenamefont {Nartallo}, \citenamefont {Nieminen},
  \citenamefont {Nishimura}, \citenamefont {Ohtsubo}, \citenamefont {Okamura},
  \citenamefont {O'Neale}, \citenamefont {Oohata}, \citenamefont {Paech},
  \citenamefont {Perl}, \citenamefont {Pfeiffer}, \citenamefont {Pia},
  \citenamefont {Ranjard}, \citenamefont {Rybin}, \citenamefont {Sadilov},
  \citenamefont {{Di Salvo}}, \citenamefont {Santin}, \citenamefont {Sasaki},
  \citenamefont {Savvas}, \citenamefont {Sawada}, \citenamefont {Scherer},
  \citenamefont {Sei}, \citenamefont {Sirotenko}, \citenamefont {Smith},
  \citenamefont {Starkov}, \citenamefont {Stoecker}, \citenamefont {Sulkimo},
  \citenamefont {Takahata}, \citenamefont {Tanaka}, \citenamefont {Tcherniaev},
  \citenamefont {{Safai Tehrani}}, \citenamefont {Tropeano}, \citenamefont
  {Truscott}, \citenamefont {Uno}, \citenamefont {Urban}, \citenamefont
  {Urban}, \citenamefont {Verderi}, \citenamefont {Walkden}, \citenamefont
  {Wander}, \citenamefont {Weber}, \citenamefont {Wellisch}, \citenamefont
  {Wenaus}, \citenamefont {Williams}, \citenamefont {Wright}, \citenamefont
  {Yamada}, \citenamefont {Yoshida},\ and\ \citenamefont {Zschiesche}}]{b37}%
  \BibitemOpen
  \bibfield  {author} {\bibinfo {author} {\bibfnamefont {S.}~\bibnamefont
  {Agostinelli}}, \bibinfo {author} {\bibfnamefont {J.}~\bibnamefont
  {{Allison}}}, \bibinfo {author} {\bibfnamefont {K.}~\bibnamefont {Amako}},
  \bibinfo {author} {\bibfnamefont {J.}~\bibnamefont {Apostolakis}}, \bibinfo
  {author} {\bibfnamefont {H.}~\bibnamefont {Araujo}}, \bibinfo {author}
  {\bibfnamefont {P.}~\bibnamefont {Arce}}, \bibinfo {author} {\bibfnamefont
  {M.}~\bibnamefont {Asai}}, \bibinfo {author} {\bibfnamefont {D.}~\bibnamefont
  {Axen}}, \bibinfo {author} {\bibfnamefont {S.}~\bibnamefont {Banerjee}},
  \bibinfo {author} {\bibfnamefont {G.}~\bibnamefont {Barrand}}, \bibinfo
  {author} {\bibfnamefont {F.}~\bibnamefont {Behner}}, \bibinfo {author}
  {\bibfnamefont {L.}~\bibnamefont {Bellagamba}}, \bibinfo {author}
  {\bibfnamefont {J.}~\bibnamefont {Boudreau}}, \bibinfo {author}
  {\bibfnamefont {L.}~\bibnamefont {Broglia}}, \bibinfo {author} {\bibfnamefont
  {A.}~\bibnamefont {Brunengo}}, \bibinfo {author} {\bibfnamefont
  {H.}~\bibnamefont {Burkhardt}}, \bibinfo {author} {\bibfnamefont
  {S.}~\bibnamefont {Chauvie}}, \bibinfo {author} {\bibfnamefont
  {J.}~\bibnamefont {Chuma}}, \bibinfo {author} {\bibfnamefont
  {R.}~\bibnamefont {Chytracek}}, \bibinfo {author} {\bibfnamefont
  {G.}~\bibnamefont {Cooperman}}, \bibinfo {author} {\bibfnamefont
  {G.}~\bibnamefont {Cosmo}}, \bibinfo {author} {\bibfnamefont
  {P.}~\bibnamefont {Degtyarenko}}, \bibinfo {author} {\bibfnamefont
  {A.}~\bibnamefont {Dell'Acqua}}, \bibinfo {author} {\bibfnamefont
  {G.}~\bibnamefont {Depaola}}, \bibinfo {author} {\bibfnamefont
  {D.}~\bibnamefont {Dietrich}}, \bibinfo {author} {\bibfnamefont
  {R.}~\bibnamefont {Enami}}, \bibinfo {author} {\bibfnamefont
  {A.}~\bibnamefont {Feliciello}}, \bibinfo {author} {\bibfnamefont
  {C.}~\bibnamefont {Ferguson}}, \bibinfo {author} {\bibfnamefont
  {H.}~\bibnamefont {Fesefeldt}}, \bibinfo {author} {\bibfnamefont
  {G.}~\bibnamefont {Folger}}, \bibinfo {author} {\bibfnamefont
  {F.}~\bibnamefont {Foppiano}}, \bibinfo {author} {\bibfnamefont
  {A.}~\bibnamefont {Forti}}, \bibinfo {author} {\bibfnamefont
  {S.}~\bibnamefont {Garelli}}, \bibinfo {author} {\bibfnamefont
  {S.}~\bibnamefont {Giani}}, \bibinfo {author} {\bibfnamefont
  {R.}~\bibnamefont {Giannitrapani}}, \bibinfo {author} {\bibfnamefont
  {D.}~\bibnamefont {Gibin}}, \bibinfo {author} {\bibfnamefont
  {J.}~\bibnamefont {{Gómez Cadenas}}}, \bibinfo {author} {\bibfnamefont
  {I.}~\bibnamefont {González}}, \bibinfo {author} {\bibfnamefont
  {G.}~\bibnamefont {{Gracia Abril}}}, \bibinfo {author} {\bibfnamefont
  {G.}~\bibnamefont {Greeniaus}}, \bibinfo {author} {\bibfnamefont
  {W.}~\bibnamefont {Greiner}}, \bibinfo {author} {\bibfnamefont
  {V.}~\bibnamefont {Grichine}}, \bibinfo {author} {\bibfnamefont
  {A.}~\bibnamefont {Grossheim}}, \bibinfo {author} {\bibfnamefont
  {S.}~\bibnamefont {Guatelli}}, \bibinfo {author} {\bibfnamefont
  {P.}~\bibnamefont {Gumplinger}}, \bibinfo {author} {\bibfnamefont
  {R.}~\bibnamefont {Hamatsu}}, \bibinfo {author} {\bibfnamefont
  {K.}~\bibnamefont {Hashimoto}}, \bibinfo {author} {\bibfnamefont
  {H.}~\bibnamefont {Hasui}}, \bibinfo {author} {\bibfnamefont
  {A.}~\bibnamefont {Heikkinen}}, \bibinfo {author} {\bibfnamefont
  {A.}~\bibnamefont {Howard}}, \bibinfo {author} {\bibfnamefont
  {V.}~\bibnamefont {Ivanchenko}}, \bibinfo {author} {\bibfnamefont
  {A.}~\bibnamefont {Johnson}}, \bibinfo {author} {\bibfnamefont {F.~W.}\
  \bibnamefont {Jones}}, \bibinfo {author} {\bibfnamefont {J.}~\bibnamefont
  {Kallenbach}}, \bibinfo {author} {\bibfnamefont {N.}~\bibnamefont {Kanaya}},
  \bibinfo {author} {\bibfnamefont {M.}~\bibnamefont {Kawabata}}, \bibinfo
  {author} {\bibfnamefont {Y.}~\bibnamefont {Kawabata}}, \bibinfo {author}
  {\bibfnamefont {M.}~\bibnamefont {Kawaguti}}, \bibinfo {author}
  {\bibfnamefont {S.}~\bibnamefont {Kelner}}, \bibinfo {author} {\bibfnamefont
  {P.}~\bibnamefont {Kent}}, \bibinfo {author} {\bibfnamefont {A.}~\bibnamefont
  {Kimura}}, \bibinfo {author} {\bibfnamefont {T.}~\bibnamefont {Kodama}},
  \bibinfo {author} {\bibfnamefont {R.}~\bibnamefont {Kokoulin}}, \bibinfo
  {author} {\bibfnamefont {M.}~\bibnamefont {Kossov}}, \bibinfo {author}
  {\bibfnamefont {H.}~\bibnamefont {Kurashige}}, \bibinfo {author}
  {\bibfnamefont {E.}~\bibnamefont {Lamanna}}, \bibinfo {author} {\bibfnamefont
  {T.}~\bibnamefont {Lampén}}, \bibinfo {author} {\bibfnamefont
  {V.}~\bibnamefont {Lara}}, \bibinfo {author} {\bibfnamefont {V.}~\bibnamefont
  {Lefebure}}, \bibinfo {author} {\bibfnamefont {F.}~\bibnamefont {Lei}},
  \bibinfo {author} {\bibfnamefont {M.}~\bibnamefont {Liendl}}, \bibinfo
  {author} {\bibfnamefont {W.}~\bibnamefont {Lockman}}, \bibinfo {author}
  {\bibfnamefont {F.}~\bibnamefont {Longo}}, \bibinfo {author} {\bibfnamefont
  {S.}~\bibnamefont {Magni}}, \bibinfo {author} {\bibfnamefont
  {M.}~\bibnamefont {Maire}}, \bibinfo {author} {\bibfnamefont
  {E.}~\bibnamefont {Medernach}}, \bibinfo {author} {\bibfnamefont
  {K.}~\bibnamefont {Minamimoto}}, \bibinfo {author} {\bibfnamefont
  {P.}~\bibnamefont {{Mora de Freitas}}}, \bibinfo {author} {\bibfnamefont
  {Y.}~\bibnamefont {Morita}}, \bibinfo {author} {\bibfnamefont
  {K.}~\bibnamefont {Murakami}}, \bibinfo {author} {\bibfnamefont
  {M.}~\bibnamefont {Nagamatu}}, \bibinfo {author} {\bibfnamefont
  {R.}~\bibnamefont {Nartallo}}, \bibinfo {author} {\bibfnamefont
  {P.}~\bibnamefont {Nieminen}}, \bibinfo {author} {\bibfnamefont
  {T.}~\bibnamefont {Nishimura}}, \bibinfo {author} {\bibfnamefont
  {K.}~\bibnamefont {Ohtsubo}}, \bibinfo {author} {\bibfnamefont
  {M.}~\bibnamefont {Okamura}}, \bibinfo {author} {\bibfnamefont
  {S.}~\bibnamefont {O'Neale}}, \bibinfo {author} {\bibfnamefont
  {Y.}~\bibnamefont {Oohata}}, \bibinfo {author} {\bibfnamefont
  {K.}~\bibnamefont {Paech}}, \bibinfo {author} {\bibfnamefont
  {J.}~\bibnamefont {Perl}}, \bibinfo {author} {\bibfnamefont {A.}~\bibnamefont
  {Pfeiffer}}, \bibinfo {author} {\bibfnamefont {M.~G.}\ \bibnamefont {Pia}},
  \bibinfo {author} {\bibfnamefont {F.}~\bibnamefont {Ranjard}}, \bibinfo
  {author} {\bibfnamefont {A.}~\bibnamefont {Rybin}}, \bibinfo {author}
  {\bibfnamefont {S.}~\bibnamefont {Sadilov}}, \bibinfo {author} {\bibfnamefont
  {E.}~\bibnamefont {{Di Salvo}}}, \bibinfo {author} {\bibfnamefont
  {G.}~\bibnamefont {Santin}}, \bibinfo {author} {\bibfnamefont
  {T.}~\bibnamefont {Sasaki}}, \bibinfo {author} {\bibfnamefont
  {N.}~\bibnamefont {Savvas}}, \bibinfo {author} {\bibfnamefont
  {Y.}~\bibnamefont {Sawada}}, \bibinfo {author} {\bibfnamefont
  {S.}~\bibnamefont {Scherer}}, \bibinfo {author} {\bibfnamefont
  {S.}~\bibnamefont {Sei}}, \bibinfo {author} {\bibfnamefont {V.}~\bibnamefont
  {Sirotenko}}, \bibinfo {author} {\bibfnamefont {D.}~\bibnamefont {Smith}},
  \bibinfo {author} {\bibfnamefont {N.}~\bibnamefont {Starkov}}, \bibinfo
  {author} {\bibfnamefont {H.}~\bibnamefont {Stoecker}}, \bibinfo {author}
  {\bibfnamefont {J.}~\bibnamefont {Sulkimo}}, \bibinfo {author} {\bibfnamefont
  {M.}~\bibnamefont {Takahata}}, \bibinfo {author} {\bibfnamefont
  {S.}~\bibnamefont {Tanaka}}, \bibinfo {author} {\bibfnamefont
  {E.}~\bibnamefont {Tcherniaev}}, \bibinfo {author} {\bibfnamefont
  {E.}~\bibnamefont {{Safai Tehrani}}}, \bibinfo {author} {\bibfnamefont
  {M.}~\bibnamefont {Tropeano}}, \bibinfo {author} {\bibfnamefont
  {P.}~\bibnamefont {Truscott}}, \bibinfo {author} {\bibfnamefont
  {H.}~\bibnamefont {Uno}}, \bibinfo {author} {\bibfnamefont {L.}~\bibnamefont
  {Urban}}, \bibinfo {author} {\bibfnamefont {P.}~\bibnamefont {Urban}},
  \bibinfo {author} {\bibfnamefont {M.}~\bibnamefont {Verderi}}, \bibinfo
  {author} {\bibfnamefont {A.}~\bibnamefont {Walkden}}, \bibinfo {author}
  {\bibfnamefont {W.}~\bibnamefont {Wander}}, \bibinfo {author} {\bibfnamefont
  {H.}~\bibnamefont {Weber}}, \bibinfo {author} {\bibfnamefont
  {J.}~\bibnamefont {Wellisch}}, \bibinfo {author} {\bibfnamefont
  {T.}~\bibnamefont {Wenaus}}, \bibinfo {author} {\bibfnamefont
  {D.}~\bibnamefont {Williams}}, \bibinfo {author} {\bibfnamefont
  {D.}~\bibnamefont {Wright}}, \bibinfo {author} {\bibfnamefont
  {T.}~\bibnamefont {Yamada}}, \bibinfo {author} {\bibfnamefont
  {H.}~\bibnamefont {Yoshida}},\ and\ \bibinfo {author} {\bibfnamefont
  {D.}~\bibnamefont {Zschiesche}},\ }\bibfield  {title} {\bibinfo {title}
  {Geant4—a simulation toolkit},\ }\href
  {https://doi.org/https://doi.org/10.1016/S0168-9002(03)01368-8} {\bibfield
  {journal} {\bibinfo  {journal} {Nuclear Instruments and Methods in Physics
  Research Section A: Accelerators, Spectrometers, Detectors and Associated
  Equipment}\ }\textbf {\bibinfo {volume} {506}},\ \bibinfo {pages} {250}
  (\bibinfo {year} {2003})}\BibitemShut {NoStop}%
\bibitem [{\citenamefont {{Allison}}\ \emph {et~al.}(2006)\citenamefont
  {{Allison}}, \citenamefont {Amako}, \citenamefont {Apostolakis},
  \citenamefont {Araujo}, \citenamefont {Arce~Dubois}, \citenamefont {Asai},
  \citenamefont {Barrand}, \citenamefont {Capra}, \citenamefont {Chauvie},
  \citenamefont {Chytracek}, \citenamefont {Cirrone}, \citenamefont
  {Cooperman}, \citenamefont {Cosmo}, \citenamefont {Cuttone}, \citenamefont
  {Daquino}, \citenamefont {Donszelmann}, \citenamefont {Dressel},
  \citenamefont {Folger}, \citenamefont {Foppiano}, \citenamefont {Generowicz},
  \citenamefont {Grichine}, \citenamefont {Guatelli}, \citenamefont
  {Gumplinger}, \citenamefont {Heikkinen}, \citenamefont {Hrivnacova},
  \citenamefont {Howard}, \citenamefont {Incerti}, \citenamefont {Ivanchenko},
  \citenamefont {Johnson}, \citenamefont {Jones}, \citenamefont {Koi},
  \citenamefont {Kokoulin}, \citenamefont {Kossov}, \citenamefont {Kurashige},
  \citenamefont {Lara}, \citenamefont {Larsson}, \citenamefont {Lei},
  \citenamefont {Link}, \citenamefont {Longo}, \citenamefont {Maire},
  \citenamefont {Mantero}, \citenamefont {Mascialino}, \citenamefont {McLaren},
  \citenamefont {Mendez~Lorenzo}, \citenamefont {Minamimoto}, \citenamefont
  {Murakami}, \citenamefont {Nieminen}, \citenamefont {Pandola}, \citenamefont
  {Parlati}, \citenamefont {Peralta}, \citenamefont {Perl}, \citenamefont
  {Pfeiffer}, \citenamefont {Pia}, \citenamefont {Ribon}, \citenamefont
  {Rodrigues}, \citenamefont {Russo}, \citenamefont {Sadilov}, \citenamefont
  {Santin}, \citenamefont {Sasaki}, \citenamefont {Smith}, \citenamefont
  {Starkov}, \citenamefont {Tanaka}, \citenamefont {Tcherniaev}, \citenamefont
  {Tome}, \citenamefont {Trindade}, \citenamefont {Truscott}, \citenamefont
  {Urban}, \citenamefont {Verderi}, \citenamefont {Walkden}, \citenamefont
  {Wellisch}, \citenamefont {Williams}, \citenamefont {Wright},\ and\
  \citenamefont {Yoshida}}]{b38}%
  \BibitemOpen
  \bibfield  {author} {\bibinfo {author} {\bibfnamefont {J.}~\bibnamefont
  {{Allison}}}, \bibinfo {author} {\bibfnamefont {K.}~\bibnamefont {Amako}},
  \bibinfo {author} {\bibfnamefont {J.}~\bibnamefont {Apostolakis}}, \bibinfo
  {author} {\bibfnamefont {H.}~\bibnamefont {Araujo}}, \bibinfo {author}
  {\bibfnamefont {P.}~\bibnamefont {Arce~Dubois}}, \bibinfo {author}
  {\bibfnamefont {M.}~\bibnamefont {Asai}}, \bibinfo {author} {\bibfnamefont
  {G.}~\bibnamefont {Barrand}}, \bibinfo {author} {\bibfnamefont
  {R.}~\bibnamefont {Capra}}, \bibinfo {author} {\bibfnamefont
  {S.}~\bibnamefont {Chauvie}}, \bibinfo {author} {\bibfnamefont
  {R.}~\bibnamefont {Chytracek}}, \bibinfo {author} {\bibfnamefont
  {G.}~\bibnamefont {Cirrone}}, \bibinfo {author} {\bibfnamefont
  {G.}~\bibnamefont {Cooperman}}, \bibinfo {author} {\bibfnamefont
  {G.}~\bibnamefont {Cosmo}}, \bibinfo {author} {\bibfnamefont
  {G.}~\bibnamefont {Cuttone}}, \bibinfo {author} {\bibfnamefont
  {G.}~\bibnamefont {Daquino}}, \bibinfo {author} {\bibfnamefont
  {M.}~\bibnamefont {Donszelmann}}, \bibinfo {author} {\bibfnamefont
  {M.}~\bibnamefont {Dressel}}, \bibinfo {author} {\bibfnamefont
  {G.}~\bibnamefont {Folger}}, \bibinfo {author} {\bibfnamefont
  {F.}~\bibnamefont {Foppiano}}, \bibinfo {author} {\bibfnamefont
  {J.}~\bibnamefont {Generowicz}}, \bibinfo {author} {\bibfnamefont
  {V.}~\bibnamefont {Grichine}}, \bibinfo {author} {\bibfnamefont
  {S.}~\bibnamefont {Guatelli}}, \bibinfo {author} {\bibfnamefont
  {P.}~\bibnamefont {Gumplinger}}, \bibinfo {author} {\bibfnamefont
  {A.}~\bibnamefont {Heikkinen}}, \bibinfo {author} {\bibfnamefont
  {I.}~\bibnamefont {Hrivnacova}}, \bibinfo {author} {\bibfnamefont
  {A.}~\bibnamefont {Howard}}, \bibinfo {author} {\bibfnamefont
  {S.}~\bibnamefont {Incerti}}, \bibinfo {author} {\bibfnamefont
  {V.}~\bibnamefont {Ivanchenko}}, \bibinfo {author} {\bibfnamefont
  {T.}~\bibnamefont {Johnson}}, \bibinfo {author} {\bibfnamefont
  {F.}~\bibnamefont {Jones}}, \bibinfo {author} {\bibfnamefont
  {T.}~\bibnamefont {Koi}}, \bibinfo {author} {\bibfnamefont {R.}~\bibnamefont
  {Kokoulin}}, \bibinfo {author} {\bibfnamefont {M.}~\bibnamefont {Kossov}},
  \bibinfo {author} {\bibfnamefont {H.}~\bibnamefont {Kurashige}}, \bibinfo
  {author} {\bibfnamefont {V.}~\bibnamefont {Lara}}, \bibinfo {author}
  {\bibfnamefont {S.}~\bibnamefont {Larsson}}, \bibinfo {author} {\bibfnamefont
  {F.}~\bibnamefont {Lei}}, \bibinfo {author} {\bibfnamefont {O.}~\bibnamefont
  {Link}}, \bibinfo {author} {\bibfnamefont {F.}~\bibnamefont {Longo}},
  \bibinfo {author} {\bibfnamefont {M.}~\bibnamefont {Maire}}, \bibinfo
  {author} {\bibfnamefont {A.}~\bibnamefont {Mantero}}, \bibinfo {author}
  {\bibfnamefont {B.}~\bibnamefont {Mascialino}}, \bibinfo {author}
  {\bibfnamefont {I.}~\bibnamefont {McLaren}}, \bibinfo {author} {\bibfnamefont
  {P.}~\bibnamefont {Mendez~Lorenzo}}, \bibinfo {author} {\bibfnamefont
  {K.}~\bibnamefont {Minamimoto}}, \bibinfo {author} {\bibfnamefont
  {K.}~\bibnamefont {Murakami}}, \bibinfo {author} {\bibfnamefont
  {P.}~\bibnamefont {Nieminen}}, \bibinfo {author} {\bibfnamefont
  {L.}~\bibnamefont {Pandola}}, \bibinfo {author} {\bibfnamefont
  {S.}~\bibnamefont {Parlati}}, \bibinfo {author} {\bibfnamefont
  {L.}~\bibnamefont {Peralta}}, \bibinfo {author} {\bibfnamefont
  {J.}~\bibnamefont {Perl}}, \bibinfo {author} {\bibfnamefont {A.}~\bibnamefont
  {Pfeiffer}}, \bibinfo {author} {\bibfnamefont {M.}~\bibnamefont {Pia}},
  \bibinfo {author} {\bibfnamefont {A.}~\bibnamefont {Ribon}}, \bibinfo
  {author} {\bibfnamefont {P.}~\bibnamefont {Rodrigues}}, \bibinfo {author}
  {\bibfnamefont {G.}~\bibnamefont {Russo}}, \bibinfo {author} {\bibfnamefont
  {S.}~\bibnamefont {Sadilov}}, \bibinfo {author} {\bibfnamefont
  {G.}~\bibnamefont {Santin}}, \bibinfo {author} {\bibfnamefont
  {T.}~\bibnamefont {Sasaki}}, \bibinfo {author} {\bibfnamefont
  {D.}~\bibnamefont {Smith}}, \bibinfo {author} {\bibfnamefont
  {N.}~\bibnamefont {Starkov}}, \bibinfo {author} {\bibfnamefont
  {S.}~\bibnamefont {Tanaka}}, \bibinfo {author} {\bibfnamefont
  {E.}~\bibnamefont {Tcherniaev}}, \bibinfo {author} {\bibfnamefont
  {B.}~\bibnamefont {Tome}}, \bibinfo {author} {\bibfnamefont {A.}~\bibnamefont
  {Trindade}}, \bibinfo {author} {\bibfnamefont {P.}~\bibnamefont {Truscott}},
  \bibinfo {author} {\bibfnamefont {L.}~\bibnamefont {Urban}}, \bibinfo
  {author} {\bibfnamefont {M.}~\bibnamefont {Verderi}}, \bibinfo {author}
  {\bibfnamefont {A.}~\bibnamefont {Walkden}}, \bibinfo {author} {\bibfnamefont
  {J.}~\bibnamefont {Wellisch}}, \bibinfo {author} {\bibfnamefont
  {D.}~\bibnamefont {Williams}}, \bibinfo {author} {\bibfnamefont
  {D.}~\bibnamefont {Wright}},\ and\ \bibinfo {author} {\bibfnamefont
  {H.}~\bibnamefont {Yoshida}},\ }\bibfield  {title} {\bibinfo {title} {Geant4
  developments and applications},\ }\href
  {https://doi.org/10.1109/TNS.2006.869826} {\bibfield  {journal} {\bibinfo
  {journal} {IEEE Transactions on Nuclear Science}\ }\textbf {\bibinfo {volume}
  {53}},\ \bibinfo {pages} {270} (\bibinfo {year} {2006})}\BibitemShut
  {NoStop}%
\bibitem [{b99(2025)}]{b99}%
  \BibitemOpen
  \href {https://authors.aps.org/Submissions/files/18714770} {\bibinfo {title}
  {{Supplemental Materials: Production of Iodine Isotopes via Ultra-intense
  Laser Driven Photonuclear Reactions}}} (\bibinfo {year} {2025})\BibitemShut
  {NoStop}%
\bibitem [{b84(2025)}]{b84}%
  \BibitemOpen
  \href {https://www.ansys.com/zh-cn/academic/students} {\bibinfo {title}
  {{ANSYS Students 2025r1.}}} (\bibinfo {year} {2025})\BibitemShut {NoStop}%
\bibitem [{\citenamefont {Henares}\ \emph {et~al.}(2019)\citenamefont
  {Henares}, \citenamefont {Puyuelo-Valdes}, \citenamefont {Hannachi},
  \citenamefont {Ceccotti}, \citenamefont {Ehret}, \citenamefont {Gobet},
  \citenamefont {Lancia}, \citenamefont {Marquès}, \citenamefont {Santos},
  \citenamefont {Versteegen},\ and\ \citenamefont {Tarisien}}]{b85}%
  \BibitemOpen
  \bibfield  {author} {\bibinfo {author} {\bibfnamefont {J.~L.}\ \bibnamefont
  {Henares}}, \bibinfo {author} {\bibfnamefont {P.}~\bibnamefont
  {Puyuelo-Valdes}}, \bibinfo {author} {\bibfnamefont {F.}~\bibnamefont
  {Hannachi}}, \bibinfo {author} {\bibfnamefont {T.}~\bibnamefont {Ceccotti}},
  \bibinfo {author} {\bibfnamefont {M.}~\bibnamefont {Ehret}}, \bibinfo
  {author} {\bibfnamefont {F.}~\bibnamefont {Gobet}}, \bibinfo {author}
  {\bibfnamefont {L.}~\bibnamefont {Lancia}}, \bibinfo {author} {\bibfnamefont
  {J.-R.}\ \bibnamefont {Marquès}}, \bibinfo {author} {\bibfnamefont {J.~J.}\
  \bibnamefont {Santos}}, \bibinfo {author} {\bibfnamefont {M.}~\bibnamefont
  {Versteegen}},\ and\ \bibinfo {author} {\bibfnamefont {M.}~\bibnamefont
  {Tarisien}},\ }\bibfield  {title} {\bibinfo {title} {Development of gas jet
  targets for laser-plasma experiments at near-critical density},\ }\href
  {https://doi.org/10.1063/1.5093613} {\bibfield  {journal} {\bibinfo
  {journal} {Review of Scientific Instruments}\ }\textbf {\bibinfo {volume}
  {90}},\ \bibinfo {pages} {063302} (\bibinfo {year} {2019})}\BibitemShut
  {NoStop}%
\bibitem [{\citenamefont {Liu}\ \emph {et~al.}(2021)\citenamefont {Liu},
  \citenamefont {Guo}, \citenamefont {Zhao}, \citenamefont {Ma}, \citenamefont
  {Ma}, \citenamefont {Lv}, \citenamefont {Meng}, \citenamefont {Zhang},
  \citenamefont {Ban}, \citenamefont {Wang}, \citenamefont {Xi}, \citenamefont
  {Tian},\ and\ \citenamefont {He}}]{b90}%
  \BibitemOpen
  \bibfield  {author} {\bibinfo {author} {\bibfnamefont {Q.~S.}\ \bibnamefont
  {Liu}}, \bibinfo {author} {\bibfnamefont {B.}~\bibnamefont {Guo}}, \bibinfo
  {author} {\bibfnamefont {B.~Z.}\ \bibnamefont {Zhao}}, \bibinfo {author}
  {\bibfnamefont {M.~J.}\ \bibnamefont {Ma}}, \bibinfo {author} {\bibfnamefont
  {X.~H.}\ \bibnamefont {Ma}}, \bibinfo {author} {\bibfnamefont
  {C.}~\bibnamefont {Lv}}, \bibinfo {author} {\bibfnamefont {X.~H.}\
  \bibnamefont {Meng}}, \bibinfo {author} {\bibfnamefont {J.}~\bibnamefont
  {Zhang}}, \bibinfo {author} {\bibfnamefont {X.~N.}\ \bibnamefont {Ban}},
  \bibinfo {author} {\bibfnamefont {Z.}~\bibnamefont {Wang}}, \bibinfo {author}
  {\bibfnamefont {X.~F.}\ \bibnamefont {Xi}}, \bibinfo {author} {\bibfnamefont
  {B.~X.}\ \bibnamefont {Tian}},\ and\ \bibinfo {author} {\bibfnamefont
  {C.~Y.}\ \bibnamefont {He}},\ }\bibfield  {title} {\bibinfo {title} {Effect
  of multiple parameters on the supersonic gas-jet target characteristics for
  laser wakefield acceleration},\ }\href
  {https://doi.org/10.1007/s41365-021-00910-1} {\bibfield  {journal} {\bibinfo
  {journal} {Nuclear Science and Techniques}\ }\textbf {\bibinfo {volume}
  {32}},\ \bibinfo {pages} {75} (\bibinfo {year} {2021})}\BibitemShut {NoStop}%
\bibitem [{\citenamefont {Derouillat}\ \emph {et~al.}(2018)\citenamefont
  {Derouillat}, \citenamefont {Beck}, \citenamefont {Pérez}, \citenamefont
  {Vinci}, \citenamefont {Chiaramello}, \citenamefont {Grassi}, \citenamefont
  {Flé}, \citenamefont {Bouchard}, \citenamefont {Plotnikov}, \citenamefont
  {Aunai}, \citenamefont {Dargent}, \citenamefont {Riconda},\ and\
  \citenamefont {Grech}}]{b81}%
  \BibitemOpen
  \bibfield  {author} {\bibinfo {author} {\bibfnamefont {J.}~\bibnamefont
  {Derouillat}}, \bibinfo {author} {\bibfnamefont {A.}~\bibnamefont {Beck}},
  \bibinfo {author} {\bibfnamefont {F.}~\bibnamefont {Pérez}}, \bibinfo
  {author} {\bibfnamefont {T.}~\bibnamefont {Vinci}}, \bibinfo {author}
  {\bibfnamefont {M.}~\bibnamefont {Chiaramello}}, \bibinfo {author}
  {\bibfnamefont {A.}~\bibnamefont {Grassi}}, \bibinfo {author} {\bibfnamefont
  {M.}~\bibnamefont {Flé}}, \bibinfo {author} {\bibfnamefont {G.}~\bibnamefont
  {Bouchard}}, \bibinfo {author} {\bibfnamefont {I.}~\bibnamefont {Plotnikov}},
  \bibinfo {author} {\bibfnamefont {N.}~\bibnamefont {Aunai}}, \bibinfo
  {author} {\bibfnamefont {J.}~\bibnamefont {Dargent}}, \bibinfo {author}
  {\bibfnamefont {C.}~\bibnamefont {Riconda}},\ and\ \bibinfo {author}
  {\bibfnamefont {M.}~\bibnamefont {Grech}},\ }\bibfield  {title} {\bibinfo
  {title} {Smilei : {A} collaborative, open-source, multi-purpose
  particle-in-cell code for plasma simulation},\ }\href
  {https://doi.org/https://doi.org/10.1016/j.cpc.2017.09.024} {\bibfield
  {journal} {\bibinfo  {journal} {Computer Physics Communications}\ }\textbf
  {\bibinfo {volume} {222}},\ \bibinfo {pages} {351} (\bibinfo {year}
  {2018})}\BibitemShut {NoStop}%
\bibitem [{\citenamefont {Lifschitz}\ \emph {et~al.}(2009)\citenamefont
  {Lifschitz}, \citenamefont {Davoine}, \citenamefont {Lefebvre}, \citenamefont
  {Faure}, \citenamefont {Rechatin},\ and\ \citenamefont {Malka}}]{b82}%
  \BibitemOpen
  \bibfield  {author} {\bibinfo {author} {\bibfnamefont {A.~F.}\ \bibnamefont
  {Lifschitz}}, \bibinfo {author} {\bibfnamefont {X.}~\bibnamefont {Davoine}},
  \bibinfo {author} {\bibfnamefont {E.}~\bibnamefont {Lefebvre}}, \bibinfo
  {author} {\bibfnamefont {J.}~\bibnamefont {Faure}}, \bibinfo {author}
  {\bibfnamefont {C.}~\bibnamefont {Rechatin}},\ and\ \bibinfo {author}
  {\bibfnamefont {V.}~\bibnamefont {Malka}},\ }\bibfield  {title} {\bibinfo
  {title} {Particle-in-cell modelling of laser–plasma interaction using
  fourier decomposition},\ }\href
  {https://doi.org/https://doi.org/10.1016/j.jcp.2008.11.017} {\bibfield
  {journal} {\bibinfo  {journal} {Journal of Computational Physics}\ }\textbf
  {\bibinfo {volume} {228}},\ \bibinfo {pages} {1803} (\bibinfo {year}
  {2009})}\BibitemShut {NoStop}%
\bibitem [{\citenamefont {Massimo}\ \emph {et~al.}(2019)\citenamefont
  {Massimo}, \citenamefont {Beck}, \citenamefont {Derouillat}, \citenamefont
  {Grech}, \citenamefont {Lobet}, \citenamefont {Pérez}, \citenamefont
  {Zemzemi},\ and\ \citenamefont {Specka}}]{b83}%
  \BibitemOpen
  \bibfield  {author} {\bibinfo {author} {\bibfnamefont {F.}~\bibnamefont
  {Massimo}}, \bibinfo {author} {\bibfnamefont {A.}~\bibnamefont {Beck}},
  \bibinfo {author} {\bibfnamefont {J.}~\bibnamefont {Derouillat}}, \bibinfo
  {author} {\bibfnamefont {M.}~\bibnamefont {Grech}}, \bibinfo {author}
  {\bibfnamefont {M.}~\bibnamefont {Lobet}}, \bibinfo {author} {\bibfnamefont
  {F.}~\bibnamefont {Pérez}}, \bibinfo {author} {\bibfnamefont
  {I.}~\bibnamefont {Zemzemi}},\ and\ \bibinfo {author} {\bibfnamefont
  {A.}~\bibnamefont {Specka}},\ }\bibfield  {title} {\bibinfo {title}
  {Efficient start-to-end {3D} envelope modeling for two-stage laser wakefield
  acceleration experiments},\ }\href {https://doi.org/10.1088/1361-6587/ab49cf}
  {\bibfield  {journal} {\bibinfo  {journal} {Plasma Physics and Controlled
  Fusion}\ }\textbf {\bibinfo {volume} {61}},\ \bibinfo {pages} {124001}
  (\bibinfo {year} {2019})}\BibitemShut {NoStop}%
\bibitem [{b93(2025)}]{b93}%
  \BibitemOpen
  \href {https://webbook.nist.gov/chemistry/fluid/} {\bibinfo {title}
  {{Thermophysical Properties of Fluid Systems}}} (\bibinfo {year}
  {2025})\BibitemShut {NoStop}%
\bibitem [{\citenamefont {Mangles}\ \emph {et~al.}(2005)\citenamefont
  {Mangles}, \citenamefont {Walton}, \citenamefont {Tzoufras}, \citenamefont
  {Najmudin}, \citenamefont {Clarke}, \citenamefont {Dangor}, \citenamefont
  {Evans}, \citenamefont {Fritzler}, \citenamefont {Gopal}, \citenamefont
  {Hernandez-Gomez}, \citenamefont {Mori}, \citenamefont {Rozmus},
  \citenamefont {Tatarakis}, \citenamefont {Thomas}, \citenamefont {Tsung},
  \citenamefont {Wei},\ and\ \citenamefont {Krushelnick}}]{b94}%
  \BibitemOpen
  \bibfield  {author} {\bibinfo {author} {\bibfnamefont {S.~P.~D.}\
  \bibnamefont {Mangles}}, \bibinfo {author} {\bibfnamefont {B.~R.}\
  \bibnamefont {Walton}}, \bibinfo {author} {\bibfnamefont {M.}~\bibnamefont
  {Tzoufras}}, \bibinfo {author} {\bibfnamefont {Z.}~\bibnamefont {Najmudin}},
  \bibinfo {author} {\bibfnamefont {R.~J.}\ \bibnamefont {Clarke}}, \bibinfo
  {author} {\bibfnamefont {A.~E.}\ \bibnamefont {Dangor}}, \bibinfo {author}
  {\bibfnamefont {R.~G.}\ \bibnamefont {Evans}}, \bibinfo {author}
  {\bibfnamefont {S.}~\bibnamefont {Fritzler}}, \bibinfo {author}
  {\bibfnamefont {A.}~\bibnamefont {Gopal}}, \bibinfo {author} {\bibfnamefont
  {C.}~\bibnamefont {Hernandez-Gomez}}, \bibinfo {author} {\bibfnamefont
  {W.~B.}\ \bibnamefont {Mori}}, \bibinfo {author} {\bibfnamefont
  {W.}~\bibnamefont {Rozmus}}, \bibinfo {author} {\bibfnamefont
  {M.}~\bibnamefont {Tatarakis}}, \bibinfo {author} {\bibfnamefont {A.~G.~R.}\
  \bibnamefont {Thomas}}, \bibinfo {author} {\bibfnamefont {F.~S.}\
  \bibnamefont {Tsung}}, \bibinfo {author} {\bibfnamefont {M.~S.}\ \bibnamefont
  {Wei}},\ and\ \bibinfo {author} {\bibfnamefont {K.}~\bibnamefont
  {Krushelnick}},\ }\bibfield  {title} {\bibinfo {title} {Electron acceleration
  in cavitated channels formed by a petawatt laser in low-density plasma},\
  }\href {https://doi.org/10.1103/PhysRevLett.94.245001} {\bibfield  {journal}
  {\bibinfo  {journal} {Phys. Rev. Lett.}\ }\textbf {\bibinfo {volume} {94}},\
  \bibinfo {pages} {245001} (\bibinfo {year} {2005})}\BibitemShut {NoStop}%
\bibitem [{\citenamefont {Najmudin}\ \emph {et~al.}(2003)\citenamefont
  {Najmudin}, \citenamefont {Krushelnick}, \citenamefont {Tatarakis},
  \citenamefont {Clark}, \citenamefont {Danson}, \citenamefont {Malka},
  \citenamefont {Neely}, \citenamefont {Santala},\ and\ \citenamefont
  {Dangor}}]{b95}%
  \BibitemOpen
  \bibfield  {author} {\bibinfo {author} {\bibfnamefont {Z.}~\bibnamefont
  {Najmudin}}, \bibinfo {author} {\bibfnamefont {K.}~\bibnamefont
  {Krushelnick}}, \bibinfo {author} {\bibfnamefont {M.}~\bibnamefont
  {Tatarakis}}, \bibinfo {author} {\bibfnamefont {E.~L.}\ \bibnamefont
  {Clark}}, \bibinfo {author} {\bibfnamefont {C.~N.}\ \bibnamefont {Danson}},
  \bibinfo {author} {\bibfnamefont {V.}~\bibnamefont {Malka}}, \bibinfo
  {author} {\bibfnamefont {D.}~\bibnamefont {Neely}}, \bibinfo {author}
  {\bibfnamefont {M.~I.~K.}\ \bibnamefont {Santala}},\ and\ \bibinfo {author}
  {\bibfnamefont {A.~E.}\ \bibnamefont {Dangor}},\ }\bibfield  {title}
  {\bibinfo {title} {The effect of high intensity laser propagation
  instabilities on channel formation in underdense plasmas},\ }\href
  {https://doi.org/10.1063/1.1534585} {\bibfield  {journal} {\bibinfo
  {journal} {Physics of Plasmas}\ }\textbf {\bibinfo {volume} {10}},\ \bibinfo
  {pages} {438} (\bibinfo {year} {2003})}\BibitemShut {NoStop}%
\bibitem [{\citenamefont {Boutoux}\ \emph {et~al.}(2015)\citenamefont
  {Boutoux}, \citenamefont {Rabhi}, \citenamefont {Batani}, \citenamefont
  {Binet}, \citenamefont {Ducret}, \citenamefont {Jakubowska}, \citenamefont
  {Nègre}, \citenamefont {Reverdin},\ and\ \citenamefont {Thfoin}}]{b88}%
  \BibitemOpen
  \bibfield  {author} {\bibinfo {author} {\bibfnamefont {G.}~\bibnamefont
  {Boutoux}}, \bibinfo {author} {\bibfnamefont {N.}~\bibnamefont {Rabhi}},
  \bibinfo {author} {\bibfnamefont {D.}~\bibnamefont {Batani}}, \bibinfo
  {author} {\bibfnamefont {A.}~\bibnamefont {Binet}}, \bibinfo {author}
  {\bibfnamefont {J.-E.}\ \bibnamefont {Ducret}}, \bibinfo {author}
  {\bibfnamefont {K.}~\bibnamefont {Jakubowska}}, \bibinfo {author}
  {\bibfnamefont {J.-P.}\ \bibnamefont {Nègre}}, \bibinfo {author}
  {\bibfnamefont {C.}~\bibnamefont {Reverdin}},\ and\ \bibinfo {author}
  {\bibfnamefont {I.}~\bibnamefont {Thfoin}},\ }\bibfield  {title} {\bibinfo
  {title} {Study of imaging plate detector sensitivity to 5-18 {MeV}
  electrons},\ }\href {https://doi.org/10.1063/1.4936141} {\bibfield  {journal}
  {\bibinfo  {journal} {Review of Scientific Instruments}\ }\textbf {\bibinfo
  {volume} {86}},\ \bibinfo {pages} {113304} (\bibinfo {year}
  {2015})}\BibitemShut {NoStop}%
\bibitem [{\citenamefont {Koning}\ and\ \citenamefont {Rochman}(2012)}]{b36}%
  \BibitemOpen
  \bibfield  {author} {\bibinfo {author} {\bibfnamefont {A.}~\bibnamefont
  {Koning}}\ and\ \bibinfo {author} {\bibfnamefont {D.}~\bibnamefont
  {Rochman}},\ }\bibfield  {title} {\bibinfo {title} {Modern nuclear data
  evaluation with the {TALYS} code system},\ }\href
  {https://doi.org/https://doi.org/10.1016/j.nds.2012.11.002} {\bibfield
  {journal} {\bibinfo  {journal} {Nuclear Data Sheets}\ }\textbf {\bibinfo
  {volume} {113}},\ \bibinfo {pages} {2841} (\bibinfo {year}
  {2012})}\BibitemShut {NoStop}%
\bibitem [{\citenamefont {Mohr}\ \emph {et~al.}(2000)\citenamefont {Mohr},
  \citenamefont {Vogt}, \citenamefont {Babilon}, \citenamefont {Enders},
  \citenamefont {Hartmann}, \citenamefont {Hutter}, \citenamefont {Rauscher},
  \citenamefont {Volz},\ and\ \citenamefont {Zilges}}]{b87}%
  \BibitemOpen
  \bibfield  {author} {\bibinfo {author} {\bibfnamefont {P.}~\bibnamefont
  {Mohr}}, \bibinfo {author} {\bibfnamefont {K.}~\bibnamefont {Vogt}}, \bibinfo
  {author} {\bibfnamefont {M.}~\bibnamefont {Babilon}}, \bibinfo {author}
  {\bibfnamefont {J.}~\bibnamefont {Enders}}, \bibinfo {author} {\bibfnamefont
  {T.}~\bibnamefont {Hartmann}}, \bibinfo {author} {\bibfnamefont
  {C.}~\bibnamefont {Hutter}}, \bibinfo {author} {\bibfnamefont
  {T.}~\bibnamefont {Rauscher}}, \bibinfo {author} {\bibfnamefont
  {S.}~\bibnamefont {Volz}},\ and\ \bibinfo {author} {\bibfnamefont
  {A.}~\bibnamefont {Zilges}},\ }\bibfield  {title} {\bibinfo {title}
  {Experimental simulation of a stellar photon bath by bremsstrahlung: the
  astrophysical $\gamma$-process},\ }\href
  {https://doi.org/https://doi.org/10.1016/S0370-2693(00)00862-5} {\bibfield
  {journal} {\bibinfo  {journal} {Physics Letters B}\ }\textbf {\bibinfo
  {volume} {488}},\ \bibinfo {pages} {127} (\bibinfo {year}
  {2000})}\BibitemShut {NoStop}%
\bibitem [{b43(2025)}]{b43}%
  \BibitemOpen
  \href {https://www.nndc.bnl.gov/nudat3/} {\bibinfo {title} {{National Nuclear
  Data Center, Brookhaven National Laboratory}}} (\bibinfo {year}
  {2025})\BibitemShut {NoStop}%
\bibitem [{\citenamefont {Jonsson}\ and\ \citenamefont {Forkman}(1968)}]{b8}%
  \BibitemOpen
  \bibfield  {author} {\bibinfo {author} {\bibfnamefont {G.~G.}\ \bibnamefont
  {Jonsson}}\ and\ \bibinfo {author} {\bibfnamefont {B.}~\bibnamefont
  {Forkman}},\ }\bibfield  {title} {\bibinfo {title} {($\gamma$, $xn$)
  reactions in \ce{^{127}I}},\ }\href
  {https://doi.org/https://doi.org/10.1016/0375-9474(68)90723-9} {\bibfield
  {journal} {\bibinfo  {journal} {Nuclear Physics A}\ }\textbf {\bibinfo
  {volume} {107}},\ \bibinfo {pages} {52} (\bibinfo {year} {1968})}\BibitemShut
  {NoStop}%
\bibitem [{\citenamefont {Debs}\ \emph {et~al.}(1955)\citenamefont {Debs},
  \citenamefont {Eisinger}, \citenamefont {Fairhall}, \citenamefont {Halpern},\
  and\ \citenamefont {Richter}}]{b47}%
  \BibitemOpen
  \bibfield  {author} {\bibinfo {author} {\bibfnamefont {R.~J.}\ \bibnamefont
  {Debs}}, \bibinfo {author} {\bibfnamefont {J.~T.}\ \bibnamefont {Eisinger}},
  \bibinfo {author} {\bibfnamefont {A.~W.}\ \bibnamefont {Fairhall}}, \bibinfo
  {author} {\bibfnamefont {I.}~\bibnamefont {Halpern}},\ and\ \bibinfo {author}
  {\bibfnamefont {H.~G.}\ \bibnamefont {Richter}},\ }\bibfield  {title}
  {\bibinfo {title} {Yields of photonuclear reactions with 320-{Mev} {X}-rays.
  {I}. experimental results},\ }\href {https://doi.org/10.1103/PhysRev.97.1325}
  {\bibfield  {journal} {\bibinfo  {journal} {Phys. Rev.}\ }\textbf {\bibinfo
  {volume} {97}},\ \bibinfo {pages} {1325} (\bibinfo {year}
  {1955})}\BibitemShut {NoStop}%
\end{thebibliography}%

\end{document}